\documentclass[preprints,article,accept,oneauthor,pdftex,10pt,a4paper]{mdpi} 

\firstpage{1} 
\pubvolume{xx}
\issuenum{1}
\articlenumber{5}
\pubyear{2018}
\copyrightyear{2018}
\history{Received: date; Accepted: date; Published: date}

\usepackage{bm}
\usepackage[cal=cm,scr=dutchcal]{mathalfa}
\usepackage[normalem]{ulem}

\makeatletter
\newcommand\reldoublebar{\mathrel{\smash=}}
\renewcommand{\Rightarrowfill@}[1]{%
\m@th \setboxz@h {$#1\reldoublebar $}\ht \z@ \z@
$#1\copy \z@
\mkern -6mu\cleaders \hbox {$#1\mkern -2mu\box \z@ \mkern -2mu$}\hfill
\mkern -6mu
\mathord \Rightarrow $}
\newcommand{\Overrightarrow}{\mathpalette{\overarrow@\Rightarrowfill@}}
\makeatother

\Title{Minkowski and Galilei/Newton Fluid Dynamics: A Geometric 3+1 Spacetime Perspective}

\Author{Christian Y. Cardall $^{1,2,\dagger}$\orcidA{}}

\AuthorNames{Christian Y. Cardall}

\address{%
$^{1}$ \quad Physics Division, Oak Ridge National Laboratory, Oak Ridge, TN 37831-6354, USA; cardallcy@ornl.gov \\
$^{2}$ \quad Department of Physics and Astronomy, University of Tennessee, Knoxville, TN 37996-1200, USA; ccardall@utk.edu}

\firstnote{This manuscript has been authored by UT-Battelle, LLC, under contract DE-AC05-00OR22725 with the US Department of Energy (DOE). The US government retains and the publisher, by accepting the article for publication, acknowledges that the US government retains a nonexclusive, paid-up, irrevocable, worldwide license to publish or reproduce the published form of this manuscript, or allow others to do so, for US government purposes. DOE will provide public access to these results of federally sponsored research in accordance with the DOE Public Access Plan (\texttt{http://energy.gov/downloads/doe-public-access-plan}).}

\abstract{A kinetic theory of classical particles serves as a unified basis for developing a geometric $3+1$ spacetime perspective on fluid dynamics capable of embracing both Minkowski and Galilei/Newton spacetimes.
Parallel treatment of these cases on as common a footing as possible reveals that the particle four-momentum is better regarded as comprising momentum and \textit{inertia} rather than momentum and energy; and consequently, that the object now known as the stress-energy or energy-momentum tensor is more properly understood as a stress-\textit{inertia} or \textit{inertia}-momentum tensor.
In dealing with both fiducial and comoving frames as fluid dynamics requires, tensor decompositions in terms of the four-velocities of observers associated with these frames render use of coordinate-free geometric notation not only fully viable, but conceptually simplifying.
A particle number four-vector, three-momentum $(1,1)$ tensor, and kinetic energy four-vector characterize a simple fluid and satisfy balance equations involving spacetime divergences on both Minkowski and Galilei/Newton spacetimes.
Reduced to a fully $3+1$ form, these equations yield the familiar conservative formulations of special relativistic and non-relativistic hydrodynamics as partial differential equations in inertial coordinates, and in geometric form will provide a useful conceptual bridge to arbitrary-Lagrange-Euler and general relativistic formulations.
}


\begin{document}
\section{Introduction}
\label{sec:Introduction}

In contemplating fluid dynamics---whether purely for deeper theoretical understanding and appreciation, or for the practical purpose of developing an extensible simulation code---it may be useful to abstract certain notions that are common to both the relativistic and non-relativistic cases.

Non-relativistic fluid dynamics is normally formulated as a set of evolution equations for time-dependent fields on position space.
Regarded as an application of mechanics and thermodynamics to an infinitesimal `fluid element,'
the equations express evolution of the element's particle density (due to its changing volume), its velocity (due to Newton's Second Law), and its internal energy density (due to the First Law of Thermodynamics).
Alternatively, the non-relativistic fluid equations can be expressed in conservative form as balance equations for particle number, momentum, and `total' (internal plus kinetic) energy in a fixed `control volume.'

In the case of relativistic fluid dynamics, spacetime allows a different perspective, as reflected in a formulation that is manifestly covariant with respect to Poincar\'e (inhomogeneous Lorentz) transformations in Minkowski (special relativistic) spacetime, or general coordinate transformations in Einstein (general relativistic) spacetime.
The velocity three-vector of the non-relativistic case, denoting the rate of change of a fluid element's position in space, is augmented and modified to a four-vector on spacetime, tangent to a fluid element's worldline.  
Thermodynamic state variables are scalar fields defined in a `comoving frame' (in which the First Law of Thermodynamics applies) riding along with a fluid element.
The spatial particle flux of the non-relativistic formulation is augmented and modified to a particle number four-vector whose spacetime four-divergence expresses conservation of particles.
The non-relativistic stress tensor expressing three-momentum flux is generalized to a stress-energy tensor on spacetime whose four-divergence expresses local conservation (or balance) of energy-momentum.

In the vast literature on fluid dynamics and continuum mechanics more generally (considering also elastic solids), it is virtually universal practice to adopt these separate viewpoints as distinct---even disjoint---conceptual points of departure.
In a non-relativistic treatment, time-dependent fields on position space are employed, with momentum and energy treated separately: energy conservation is obtained by an ad hoc combination of the First Law of Thermodynamics (applied to a mass element) with the equations for mass and momentum conservation.
In a relativistic formulation, a unified stress-energy tensor on spacetime is typically postulated from the outset, whose spacetime divergence gives the equations of motion (whose reduction to the non-relativistic equations is then argued post hoc in the limit of an infinite speed of light). 
These different perspectives are typically taken even when non-relativistic and relativistic approaches are discussed within the same work, for instance the classic text by Landau and Lifshitz \cite{Landau1987Fluid-Mechanics}, a recent very thorough monograph by Ferrase and Bini \cite{Ferrarese2008Introduction-to}, and an even more recent paper by the present author \cite{Cardall2017Relativistic-an}, to pluck to just a few examples from what is, again, a vast literature.
These disjoint conceptual points of departure are also taken by de Groot and collaborators in two separate works on kinetic theory, one non-relativistic \cite{de-Groot1962Non-equilibrium} and one relativistic \cite{Groot1980Relativistic-Ki}, in which the fluid dynamical limit is obtained in the course of the exposition.
Two examples that include a geometric and less familiar aspect---the deployment of exterior calculus (different from the usual tensor calculus more familiar to physicists), to more easily prove certain theorems such as conservation of vorticity and helicity---but that are still standard in their separate viewpoints on non-relativistic space and time or relativistic spacetime respectively, are a non-relativistic variational approach of Prix \cite{Prix2004Variational-des,Prix2005Variational-des};\footnote{
A series of papers by Carter and collaborators \cite{Carter2004Covariant-Analy,Carter2005Covariant-Analy,Carter2005-b-Covariant-Analy,Carter2006Covariant-Newto} addresses similar territory, but with the four-dimensional perspective on non-relativistic spacetime discussed below.
However, an object they call a `stress momentum energy tensor' seems in fact to be what is called below a `mass-momentum' or `stress-inertia' or `inertia-momentum' tensor, as its divergence provides only three momentum equations, with the fourth independent (`time') component of its divergence providing no new information.
In fact, unlike the non-covariant presentation of Prix \cite{Prix2004Variational-des,Prix2005Variational-des}, these papers do not present a conservative `total' energy equation.
Indeed they apparently see no need to discuss an energy equation at all, presumably not regarding as necessary for their purposes anything beyond the laws of thermodynamics applied to a fluid element. 
} 
and an introduction to relativistic hydrodynamics by Gourgoulhon \cite{Gourgoulhon2006An-introduction} that shows a reduction to the non-relativistic limit only after the relativistic equations have been discussed.

It seems widely underappreciated that the time and position space of non-relativistic physics also form a four-dimensional spacetime (e.g. \cite{Weyl1922Space---Time---,Trautman1965Foundations-and,Trautman1966Comparison-of-N,Penrose1968Structure-of-sp,Trautman1970Fibre-bundles-a,Ehlers1973The-Nature-and-,Arnold1978Mathematical-me,Geroch1978General-Relativ,Schutz1980Geometrical-Met,Penrose2004The-Road-to-Rea,Gourgoulhon2013Special-Relativ}))---call it, say, Galilei/Newton spacetime.\footnote{Projection of a conception of spacetime as a four-dimensional manifold back onto the thinking of Galileo and Newton is ahistorical and anachronistic, but useful for purposes of comparison with Minkowski and Einstein spacetimes.
The designation `Galilei/Newton' serves two purposes. 
First, it is a reminder that Newton, not having the spacetime and fiber bundle concepts available, assumed absolute position space (e.g.~\cite{Ehlers1973The-Nature-and-}), with Galilei invariance being an accidental consequence of the postulated Second Law rather than a symmetry of spacetime itself.
Second, this moniker allows Galilei/Newton spacetime with its global inertial frames to be distinguished from the phenomenologically equivalent Newton/Cartan spacetime---a low-velocity/low-energy limit of general relativity that features absolute time and Euclid geometry on each surface of simultaneity, but with the gravitational force eliminated, via the equivalence principle, in favor of space\textit{time} curvature (albeit via an affine connection not derivable from a spacetime metric; see e.g. Refs.~\cite{Trautman1965Foundations-and,Trautman1966Comparison-of-N,Ehlers1973Survey-of-Gener,Ehlers1973The-Nature-and-}).
}
There are probably several reasons for this.
First, the notion of spacetime as a four-dimensional manifold was conceived shortly after the publication of special relativity, and subsequently played a major role in the development of general relativity, so that the whole notion seems inextricably connected with these theories.
Conception of a non-relativistic spacetime was only retrospective, coming a few years later.
Second, when one approaches a problem in a non-relativistic context, the comfortable familiarity of separate time and position space is a mindset difficult to resist, as it accords more closely with both lived experience and early physics instruction. 
One feels as if one is avoiding various complications and counterintuitive notions.
Third---and ironically, in light of the previous point---in geometric terms Galilei/Newton spacetime is less elegant than Minkowski or Einstein spacetimes.
This is because there is no spacetime metric: while it is a differentiable manifold, Galilei/Newton spacetime is not a (pseudo)Riemann manifold, and must be given different structure. 
Each (hyper)surface of simultaneity \textit{is} a Riemann manifold, with a flat three-metric for position space; but 
without metric duality there is no raising or lowering of indices of spacetime tensors.
Fourth, especially when it comes to initial value problems, there are usually no practical reasons to consider a four-dimensional perspective on particular systems.
Once physical laws are determined (based in part on the postulated nature and symmetries of spacetime), most often solutions in particular cases are found by splitting spacetime into `space' and `time'---even in general relativity.

Nevertheless, formulating non-relativistic fluid dynamics in terms of tensors on four-dimensional Galilei/Newton spacetime---so that covariance with respect to inhomogeneous Galilei transformations is manifest---seems like a worthwhile enterprise, for it is as true of non-relativistic physics as it is of relativistic physics that contemplation of spacetime structure and symmetries gives a deeper feel for some of the constraints on the structure of physical law.
Such a program has been undertaken in more heavily mathematically-oriented literature beginning with Toupin and Truesdell \cite{Toupin1957World-invariant,Truesdell1960The-Classical-F}.
One might wish for an analogue of the relativistic stress-energy tensor on non-relativistic four-dimensional spacetime, with vanishing divergence in the absence of external forces; but the non-equivalence of non-relativistic mass and energy implies that the most straightforward non-relativistic analogue---a `mass-momentum' tensor \cite{Truesdell1960The-Classical-F,Dixon1975On-the-uniquene} (herein called the `stress-inertia' or `inertia-momentum tensor', see Sec.~\ref{sec:StressInertia})---cannot accommodate internal energy, and so the timelike projection of its spacetime divergence is merely redundant with conservation of mass (or, herein, baryon number).
An alternative rank-2 tensor on non-relativistic four-dimensional spacetime unifying stresses and energy flow (while excluding mass flow) is obtained by Duval and K\"unzle \cite{Duval1978Dynamics-of-con} with a variational method, but it does not have vanishing divergence in the absence of external fields.

An alternative view that somewhat unifies non-relativistic and relativistic spacetimes is opened by Duval et al. \cite{Duval1985Bargmann-struct}, who show that both non-relativistic and Minkowski four-dimensional spacetimes $\mathcal{M}$ can be regarded as embedded in a five-dimensional extended spacetime $\hat{\mathcal{M}}$ with a Lorentz metric. 
On this extended spacetime, the Bargmann group---an extension of the Galilei group---encodes the non-relativistic physics.
Recently, de Saxc\'e and Vall\'ee adopted this five-dimensional perspective in developing their geometric vision of non-relativistic continuum physics \cite{de-Saxce2012Bargmann-group-,de-Saxce2016Galilean-Mechan,de-Saxce20175-Dimensional-T}, defining a rank-2 `energy-momentum-mass tensor' $\hat{\bm{T}}$ that is a one-form with respect to five-dimensional $\hat{\mathcal{M}}$ and a vector with respect to four-dimensional $\mathcal{M}$, whose divergence on $\mathcal{M}$ vanishes in the absence of external forces.
Represented as a $4\times 5$ matrix $\hat{T}^\mu_{\hat{\nu}}$ in `Bargmann coordinates' on $\hat{\mathcal{M}}$, the columns represent flux vectors of energy, three-momentum, and mass on $\mathcal{M}$.
Despite the heavily geometric spacetime approach, they focus on the non-relativistic case and briefly mention only in passing the relationship of their formalism to relativistic physics.   

In pondering the difficulty of `geometrizing' non-relativistic fluid dynamics due to the inequivalence of the mass and internal energy of fluid elements, it may be useful to step back and consider three `levels' of geometrization, and how much is required for a given purpose.

The lowest level of geometrization is essentially none: one simply works with partial differential equations in a particular coordinate system expressing the time evolution of physical fields in position space. 
This is concrete, and ultimately necessary for numerical simulations on computers; but it can obscure the underlying structure of the theory, and the coordinate transformations required to compare Eulerian (static `lab' frame), Lagrangian (comoving `material' frame), and arbitrary-Lagrangian-Eulerian (something in between) formulations are messy and confusing.
In the general relativistic case, a blizzard of indices and connection coefficients contributes to the obfuscation of the central physical ideas.

At the opposite end, the highest level of geometrization involves the deployment of as few geometric objects (typically tensors) as possible on spacetime, fully embodying the underlying symmetries of the theory, and the use of coordinate-free expressions whenever possible. 
In the case of relativity this is quite appealing: for a simple fluid, a unified stress-energy tensor and number flux vector, each with vanishing divergence, express the conservation laws in a manifestly covariant way, making no commitments as to a particular time coordinate or slicing of spacetime.
In the non-relativistic case, however, absolute time and the associated absence of a full metric structure for four-dimensional spacetime mean that one is driven to the unfamiliar (to many day-to-day practitioners of fluid dynamics) and ugly (by comparison with the relativistic case) Bargmann group on a five-dimensional extended spacetime in order to obtain, as described above, an energy-momentum-mass tensor with vanishing divergence.

This paper contributes a happy medium, a `mid-level' geometrization of non-relativistic fluid dynamics, which might be called \textit{a geometric 3+1 spacetime perspective}.
The label `$3+1$' signals a primary concession made from the outset: a foliation of spacetime---into three-dimensional spacelike slices to be labeled by a single parameter (the time coordinate)---is to be chosen.
In non-relativistic spacetime the $3+1$ approach is not only natural, but essentially compulsory, given absolute time.
But for the computer simulation of relativistic systems the $3+1$ perspective is ultimately also natural, since time evolution from initial conditions is basically the only game in town.
In this sense, a geometric $3+1$ spacetime perspective provides for as much unification between the non-relativistic and relativistic cases as needed or desired for this particular purpose.

Modulo the `$3+1$' qualifier, however, the approach presented in this paper is otherwise a \textit{geometric spacetime perspective}.\footnote{An excellent monograph by Gourgoulhon \cite{Gourgoulhon201231-Formalism-in}, while focused on general relativity, provides the necessary background in differential geometry (including coordinate-free notation) and a conceptual understanding of the geometric $3+1$ spacetime approach that are helpful in back-porting these ideas to Galilei/Newton and Minkowski spacetimes.
A helpful treatment of Minkowski spacetime by the same author, also with a geometric approach, is Ref.~\cite{Gourgoulhon2013Special-Relativ}.
The `mathematical tools' chapters of the book by de Saxc\'e and Vall\'ee \cite{de-Saxce2016Galilean-Mechan} are also helpful.}
Particle momenta, and the four-velocities of `fiducial' (lab frame) and `comoving' (material) observers, are four-vectors on spacetime.
In a first sub-level of the $3+1$ perspective, spacetime divergences of spacetime tensors $\bm{N}$, $\bm{M}$, and $\bm{E}$ representing four-fluxes of particle number, three-momentum, and kinetic/internal energy are contemplated; see Eqs.~(\ref{eq:Divergence_4_N})-(\ref{eq:Divergence_4_E}).
[While spacetime divergences of spacetime tensors are dealt with here, this is partly $3+1$ because of the separation of three-momentum four-flux $\bm{M}$ from energy four-flux $\bm{E}$. 
This is the second primary accommodation to the non-relativistic case, related to the first: the absolute nature of inertia (mass) in non-relativistic physics is in a sense `conjugate' to absolute time, and requires separate treatment of energy and momentum.
The simple relationship between $\bm{E}$, $\bm{N}$, and the momentum-inertia tensor $\bm{T}$ in the relativistic case---thanks to the unification of mass and energy---is noted in this paper after the fact rather than taken as primary.]
In the second and final sub-level of the $3+1$ perpsective, the spacetime divergences of $\bm{N}$, $\bm{M}$, and $\bm{E}$ are decomposed into a Lie derivative of a density from one spacelike slice to the next (a geometric expression of a time derivative) and a covariant three-divergence of a flux within the spacelike slice (a geometric expression of position space derivatives); see Eqs.~(\ref{eq:Divergence_3_1_E})-(\ref{eq:Divergence_3_1_E}).
All of this is done to a large extent with a coordinate-free mode of expression.
Two payoffs of this $3+1$ geometric perspective are that it (1) treats the non-relativistic and relativistic cases on a similar conceptual footing, illuminating the relationship between them, and (2) will provide, in future work, a conceptual bridge to arbitrary-Lagrange-Euler and general relativistic formulations useful to numerical practitioners interested in going beyond static Eulerian coordinates and/or flat spacetime.

Two other somewhat unique aspects of this paper enhance one of its main contributions, the conceptually unified treatment of the Galilei/Newton (non-relativistic) and Minkowski (special relativistic) cases.
First, the presentation proceeds in an interleaved parallel manner, with sections or subsections generally treating both the Minkowski and Galilei/Newtonian cases (typically in that order, since Minkowski spacetime is the more familiar arena for four-dimensional thinking) so as to enhance comparison and contrast.
Second, kinetic theory on both Galilei/Newton and Minkowski spacetimes, based on the four-momenta of classical particles, is used as a unified basis for motivating the structure and interpretation of the geometric objects that characterize a fluid-dynamical approach common to both spacetimes.
Ultimately this must be regarded as something of a pedagogical device aimed at motivating intuition and interpretation: the kinetic theory of classical particles is not the fundamental theory of the world, and
more generally fluid dynamics may be regarded as a generic long-wavelength, low-frequency continuum approximation to any physical system (e.g. \cite{Schafer2014Fluid-Dynamics-}), regardless of the nature of its microscopic dynamics.
But a resort to kinetic theory serves its present purpose as a means of avoiding the usual disjoint conceptual perspectives on non-relativistic and relativistic fluid dynamics discussed early in this section. 

This paper is organized as follows.
Descriptions of Minkowski and Galilei/Newton spacetimes and the trajectories of classical particles thereon serve as a starting point in Sec.~\ref{sec:Spacetime}.
Fluid dynamics, obtained from kinetic theory on spacetime, is then discussed in Sec.~\ref{sec:FluidDynamics}.
Concluding remarks are given in Sec.~\ref{sec:Conclusion}, 
along with four tables summarizing the many entities appearing in the formalism, including brief descriptions and references to equations in the text.
For simplicity in demonstrating the Galilei/Newton spacetime perspective, Minkowski spacetime is the only relativistic case treated; only a single-component fluid is contemplated; and no discussions are included of specific microscopic models, or the closures via specific constitutive relations that might be obtained from them.\footnote{A primary example of a constitutive relation---the only one required of a simple perfect fluid---is an equation of state giving the pressure in terms of the particle number density and entropy, temperature, or internal energy. 
Another prominent example of a constitutive relation yields the celebrated Navier-Stokes equation in the non-relativistic case: dissipative stresses are assumed to be expressed as a viscosity parameter multiplying the gradient of velocity.
The relativistic analogue remains a subject of active research; further discussion is beyond the scope of this paper, but Refs.~\cite{Landau1987Fluid-Mechanics,Groot1980Relativistic-Ki,Schafer2014Fluid-Dynamics-} provide some classic and recent points of entry to the literature. 
}
Also by way of maintaining focus on the basic geometric objects characterizing fluid dynamics: while external gravitational and electromagnetic forces are considered as the paradigmatic examples consistent with the flat spacetimes of Galilei/Newton and Minkowski respectively, the present work includes no further discussion of the Poisson and Maxwell equations determining these fields, nor possible reformulations that would move the gravitational potential and electric and magnetic fields from source terms into the definitions of fluid densities and fluxes.

\section{Spacetime}
\label{sec:Spacetime}

Before discussing the particular features of Minkowski and Galilei/Newton spacetimes, consider the global inertial reference frames they both admit, as well as the causal structures that differentiate them.
While Ref.~\cite{Gourgoulhon201231-Formalism-in} focuses on Einstein spacetime rather than the Minkowski and Galilei/Newton spacetimes considered here, its notation and discussion of differential geometry as applied to spacetime provide useful background. 
A helpful treatment of Minkowski spacetime by the same author, also with a geometric approach, is Ref.~\cite{Gourgoulhon2013Special-Relativ}.

\subsection{Global Inertial Frames}
\label{sec:GlobalInertialFrames}

Spacetime $\mathcal{M}$ is a four-dimensional differentiable manifold whose points are called `events.'
Any four-dimensional differentiable manifold can be locally charted by coordinates $\left( x^\mu \right)  \in \mathbb{R}^4$, with Greek indices running over $0,1,2,3$; call this a reference frame $\mathscr{O}$.
Thus $\mathscr{O}$ labels each event by four real coordinate values. 
Call $t = x^0$ the time coordinate.
It can be regarded as a scalar field on $\mathcal{M}$, whose non-intersecting level surfaces $\mathcal{S}_t$ are three-dimensional hypersurfaces.
Let the remaining coordinates $\left( x^i \right)$---with Latin indices running over $1,2,3$---chart the hypersurfaces $\mathcal{S}_t$, and call them position space coordinates.
The hypersurfaces $\mathcal{S}_t$ constitute a foliation of $\mathcal{M}$, and the coordinates as described are said to be adapted to the foliation.

Define a time one-form 
\begin{equation}
\bm{t} = \bm{\nabla} t = \bm{\mathrm{d}}t
\label{eq:TimeForm_0}
\end{equation}
as the spacetime gradient of the time coordinate $t$ (which coincides with the natural basis one-form $\bm{\mathrm{d}}t$ associated with $t$), exploiting the fact that no metric is needed to take the gradient of a scalar field.
[Tensors are denoted with a boldface symbol; without indices, their type---e.g. vector $(1,0)$, bilinear form $(0,2)$, etc.---should be stated when the symbol is introduced.]
This one-form $\bm{t}$ has components
\begin{equation}
\left[ t_\mu \right]_\mathscr{O} = \left[ \frac{\partial t}{\partial x^\mu} \right]_\mathscr{O} = \begin{bmatrix} 1 && 0 && 0 && 0
\label{eq:TimeForm}
\end{bmatrix}_\mathscr{O}
\end{equation}
with respect to $\mathscr{O}$, i.e. the above-discussed coordinates adapted to the foliation.

A metric is also not needed to define curves on $\mathcal{M}$, and tangent vectors in terms of them; exploit this to represent particles in spacetime by their worldlines, which are curves on $\mathcal{M}$ that intersect each $\mathcal{S}_t$ once.
Parametrize a particle worldline $\mathcal{X}(\lambda)$ by an affine parameter $\lambda$, and consider the four-vector
\begin{equation}
\bm{p} = \frac{\mathrm{d}\mathcal{X}(\lambda)}{\mathrm{d}\lambda}
\label{eq:FourMomentum}
\end{equation}
tangent to $\mathcal{X}(\lambda)$.
Denote by the four functions $X^\mu( \lambda)$ the coordinates assigned by $\mathscr{O}$ to the event on $\mathcal{X}(\lambda)$ specified by a particular value of $\lambda$; then
\begin{equation}
p^\mu = \frac{\mathrm{d}X^\mu}{\mathrm{d}\lambda}
\label{eq:FourMomentumCoordinates}
\end{equation}
are the components of $\bm{p}$ according to $\mathscr{O}$.
Assuming the coordinate $t$ has units of time, scale $\lambda$ such that it has units of $\mathrm{time} / \mathrm{mass}$; then $\bm{t}( \bm{p} ) = \bm{t} \cdot \bm{p} = t_\alpha \,p^\alpha = p^0$ has units of mass, and is interpreted as the particle inertia as measured by $\mathscr{O}$. 
[In this paper the dot operator ($\cdot$) only represents a contraction of indices in which one is already up and the other already down.
Beware it does not represent a scalar product, which is not available on Galilei/Newton spacetime; a scalar product will always be expressed explicitly in terms of a metric tensor.
Summation over repeated indices is implied where one is up and one is down.]
Call $\bm{p}$ the particle four-momentum.

For a particle of mass $m > 0$,
define also its four-velocity $\bm{u}$ in terms of its four-momentum $\bm{p}$ by
\begin{equation}
\bm{u} = \frac{\bm{p}}{m}.
\label{eq:FourVelocity_0}
\end{equation} 
A particle has mass $m > 0$ if at any point along its worldline there exist reference frames in which it is at rest; in such a frame, the particle's inertia is its mass $m$.
Since $\bm{p}$ is a four-vector and $m$ is a scalar, $\bm{u}$ is also a four-vector.
Reparametrize the worldline in terms of a new affine parameter such that $\mathrm{d}\tau = m\, \mathrm{d}\lambda$, where the new parameter $\tau$ has units of time;
then
\begin{equation}
\left[ u^\mu \right]_\mathscr{O} = \left[ \frac{\mathrm{d}X^\mu}{\mathrm{d}\tau} \right]_\mathscr{O} = \frac{\mathrm{d}t}{\mathrm{d}\tau} \begin{bmatrix} 1 \\ v^i \end{bmatrix}_\mathscr{O}
\label{eq:FourVelocity}
\end{equation}
are the components of $\bm{u}$ reckoned by $\mathscr{O}$, where
\begin{equation}
v^i = \frac{\mathrm{d}X^i}{\mathrm{d}t}
\end{equation}
are the components of a velocity three-vector $\bm{v}$ on $\mathcal{S}_t$.
The push-forward of $\bm{v}$ onto $\mathcal{M}$, denoted with the same symbol, has components
\begin{equation}
\left[ v^\mu \right]_\mathscr{O} = \begin{bmatrix} 0 \\ v^i \end{bmatrix}_\mathscr{O}
\label{eq:ThreeVelocity}
\end{equation}
in coordinates (such as $\mathscr{O}$) adapted to the foliation; it is a four-vector on $\mathcal{M}$ that is tangent to $\mathcal{S}_t$. 
The leading factor $\mathrm{d}t/\mathrm{d}\tau$ in Eq.~(\ref{eq:FourVelocity}) cannot be specified further until additional details are given about a particular spacetime.

The coordinate curves of a reference frame $\mathscr{O}$ can be used to define the worldlines of a family of observers, with four-velocity field $\bm{w}$, associated with $\mathscr{O}$.
The coordinate curves for coordinate $x^\mu$ are those curves parametrized by $x^\mu$, with the other coordinates $\left(x^\nu, \ \nu \ne \mu\right)$ held fixed.
Given the coordinates $\left(t, x^i \right)$ of reference frame $\mathscr{O}$, the $t$ coordinate curves---those of fixed $\left(x^i \right)$---are the worldlines of the $\mathscr{O}$ observers. 
The four-velocity $\bm{w}$ of these observers is proportional, by some factor $a$, to the natural basis vector $\bm{\partial}_t$ tangent to the $t$ coordinate curves, so that $\bm{w}$ has components
\begin{equation}
\left[ w^\mu \right]_\mathscr{O} = a \begin{bmatrix} 1 \\ 0 \\ 0 \\ 0 \end{bmatrix}_\mathscr{O}
\label{eq:O_Worldlines}
\end{equation}
according to $\mathscr{O}$.
Comparing with Eq.~(\ref{eq:FourVelocity}), it is apparent that the $\mathscr{O}$ observers are---rather obviously---at rest with respect to themselves.

While what has been said so far pertains to spacetimes based on any four-dimensional differentiable manifold, Minkowski and Galilei/Newton spacetimes are based on a special type of differentiable manifold---an affine space---that provides some additional features.
Prior to being endowed with additional structure, an affine space is invariant under general linear transformations and translations, admitting infinitely extended parallel lines.
With additional structure to be discussed in Secs.~\ref{sec:CausalStructure}, \ref{sec:MinkowskiSpacetime} and \ref{sec:GalileiNewtonSpacetime}, the classes of linear transformations under which Minkowski and Galilei/Newton spacetimes are invariant are more restricted; but they retain invariance under spacetime translations, and admit infinitely extended parallel lines in the four dimensions of spacetime.
This means that coordinates $\left(t, x^i \right)$ of a reference frame $\mathscr{O}$ can be chosen such that the entire spacetime is covered by a global coordinate chart, with the worldlines of $\mathscr{O}$ observers being infinitely extended parallel lines.
If the coordinates $\left(x^i \right)$ are adapted to the foliation defined by level surfaces of $t$, the fact that their coordinate curves are both tangent to the hypersurfaces $\mathcal{S}_t$ and can be chosen to be infinitely extended parallel lines implies that the $\mathcal{S}_t$ can be chosen to be infinite parallel affine hyperplanes.

Begin the construction of what shall be called a `global inertial' reference frame $\mathscr{O}$ in Minkowski or Galilei/Newton spacetime by choosing its coordinate curves to be infinitely extended parallel lines, and the hypersurfaces $\mathcal{S}_t$ to be parallel hyperplanes.
The worldlines of the $\mathscr{O}$ observers are `straight,' and are therefore geodesics.
Physically, their straightness corresponds to an absence of acceleration; hence the designation of $\mathscr{O}$ as a global inertial reference frame.
Mathematically, the tangent vectors $\bm{w}$ of the $\mathscr{O}$ observer worldlines, with components displayed in Eq.~(\ref{eq:O_Worldlines}), satisfy the geodesic equation
\begin{equation}
\bm{w} \cdot \bm{\nabla} \bm{w} = 0.
\end{equation}
The infinite parallelism of the coordinate curves also implies that the partial derivatives with respect to these coordinates constitute components of a covariant derivative with vanishing connection coefficients (since the natural basis vectors do not vary from point to point), so that
\begin{equation}
 w^\alpha\, \nabla_\alpha w^\mu = w^\alpha\, \frac{\partial w^\mu}{\partial x^\alpha} = \frac{\mathrm{d} w^\mu}{\mathrm{d}t} = 0
\end{equation}
in $\mathscr{O}$, implying that 
$a = \mathrm{constant}$ in Eq.~(\ref{eq:O_Worldlines}).

Once additional structure associated with causality is specified as discussed below, it turns out that a last element of structure that Minkowski and Galilei/Newton spacetimes share is that the parallel hyperplanes $\mathcal{S}_t$ of a global inertial frame $\mathscr{O}$ have distances defined by Euclid geometry.
That is, each $\mathcal{S}_t$ is a flat three-dimensional Riemann manifold with metric $\bm{\gamma}$, charted for instance in $\mathscr{O}$ by the position space coordinates $\left( x^i \right)$.
To (almost) complete the specification of a global inertial frame $\mathscr{O}$, require that the 
three-metric $\bm{\gamma}$ have components
\begin{equation}
\left[ \gamma_{ij} \right]_\mathscr{O} = \begin{bmatrix} 1 & 0 & 0 \\ 0 & 1 & 0 \\ 0 & 0 & 1 \end{bmatrix}_\mathscr{O},
\label{eq:ThreeMetric}
\end{equation}
that is, that $\left( x^i \right)$ are orthonormal rectangular coordinates. 
(Full specification entails also a choice of orientation, for instance `right-handed.')

\subsection{Causal Structure}
\label{sec:CausalStructure}

While Minkowski and Galilei/Newton spacetimes are both affine spaces allowing global inertial observers, they are distinguished by additional structure---in particular, structure related to causality---that constrains the relationships between global inertial reference frames.
In particular, the symmetry of an affine space under general linear transformations is restricted to symmetry under Lorentz transformations (rotations and Lorentz boosts) in Minkowski spacetime and Galilei transformations (rotations and Galilei boosts) in Galilei/Newton spacetime.
(As previously noted, the symmetry of affine spaces under translations is preserved in both cases.)

However, there must be enough remaining symmetry to respect the principle of special relativity, anticipated already by Galileo in his famous description of the goings-on below decks in a uniformly moving ship (e.g. \cite{Penrose2004The-Road-to-Rea}): physical law takes the same mathematical form in all global inertial reference frames. 

At the very least this requires that position space be relative: one cannot say whether two events at different times occur in the same place.
That is, two families of global inertial observers in relative uniform motion, obeying the same laws of physics expressed in the same mathematical form, assign different position space coordinates to the two events.
Upon reflection with Galileo, this seems persuasive enough, at least to us, from common experience; but it is also subtle enough to apparently have eluded Aristotle, and to have made controversial (even setting aside religious controversy) the revolutionary perspective of Copernicus.

As discussed further in Sec.~\ref{sec:GalileiNewtonSpacetime}, the causal structure of Galilei/Newton spacetime $\mathbb{G}$ accepts only this much relativity, combining relative position space with an absolute time on which all observers agree.
This accords with observed mechanical phenomena at low speed (much smaller than the speed of light $c$) and low energy per particle (much smaller than $m c^2$).
Mathematically, absolute time is implemented by a projection that maps each point of $\mathbb{G}$ onto a one-dimensional manifold $\mathbb{T}$---`time.' 
The time manifold $\mathbb{T}$ is homogeneous, so that physical law is invariant under time translations.
Physical time intervals are simply differences of a time coordinate $t$, that is, a one-dimensional Euclid `distance' on $\mathbb{T}$. 
The set of events in $\mathbb{G}$ mapped to a particular value of $t$ is a hyperplane of simultaneity $\mathcal{S}_t$, a three-dimensional Riemann manifold $\mathbb{E}^3$ governed by Euclid geometry, constituting `position space' at time $t$.
Importantly, spacetime $\mathbb{G}$ is \textit{not} simply the product space $\mathbb{T} \times \mathbb{E}^3$.
Instead, Galilei relativity is implemented by allowing position space at different times to be completely independent: $\mathcal{S}_{t_1}$ at time $t_1$ is \textit{not the same space} as $\mathcal{S}_{t_2}$ at time $t_2$.
While they are both identical instantiations of the three-dimensional flat Euclid manifold $\mathbb{E}^3$, there is no \textit{a priori} relationship between them prior to selection of a particular global inertial reference frame as `fiducial.' 
The mathematical structure that combines absolute time with relative space is comparatively exotic: $\mathbb{G}$ is a fiber bundle (e.g. \cite{Penrose1968Structure-of-sp,Trautman1970Fibre-bundles-a,Schutz1980Geometrical-Met,Penrose2004The-Road-to-Rea}), with one-dimensional base space $\mathbb{T}$ and three-dimensional typical fiber $\mathbb{E}^3$.
As for causality: the time ordering of events is absolute; but there is no upper speed limit, and indeed forces act instantaneously at any distance within a single time slice $\mathcal{S}_t$.

However, absolute time---or more specifically, absolute simultaneity---is given up in Minkowski spacetime $\mathbb{M}$, as discussed in Sec.~\ref{sec:MinkowskiSpacetime}: one cannot say whether two events in different places occur at the same time.
That is, two families of global inertial observers in relative uniform motion, obeying the same laws of physics expressed in the same mathematical form, assign different position space \textit{and} time coordinates to the two events.
This is required in order to accept at face value an absolute speed of light independent of the motions of the source and/or detector, as predicted by the Maxwell equations of electrodynamics.
At each event $\mathcal{A}$ in $\mathbb{M}$ there are past and future light cones (called null cones in Sec.~\ref{sec:MinkowskiSpacetime}), on which all observers agree, consisting of $\mathcal{A}$ together with all events displaced from $\mathcal{A}$ by light rays.
The mathematical structure that combines absolute light cones with relative space and time is comparatively simple: $\mathbb{M}$ is a pseudo-Riemann manifold, in particular a flat Lorentz manifold, with a spacetime metric tensor embodying the light cone structure.
As for causality: the mathematical form of the Lorentz transformations that preserve the metric (and therefore the light cone structure) indicates that $c$ is an upper speed limit.
Moreover, under Lorentz transformations, for events $\mathcal{B}$ inside the past and future light cones of $\mathcal{A}$, all observers agree on the time ordering of $\mathcal{A}$ and $\mathcal{B}$; but for events $\mathcal{B}$ outside the light cones, the time ordering of $\mathcal{A}$ and $\mathcal{B}$ depends on the observer.
Therefore causality demands that the worldlines of physical particles or signals transmitted by field disturbances passing through $\mathcal{A}$ must lie within the past and future light cones of $\mathcal{A}$. 

\subsection{Minkowski Spacetime}
\label{sec:MinkowskiSpacetime}

As just described in Sec.~\ref{sec:CausalStructure}, Minkowski spacetime $\mathbb{M}$ is a flat pseudo-Riemann manifold, adding to its affine structure an indefinite non-degenerate bilinear form---the Minkowski metric $\bm{g}$. 
The metric defines a scalar product $\bm{g}\left( \bm{a}, \bm{b} \right) = \bm{g}\left( \bm{b}, \bm{a} \right)$ of two vectors $\bm{a}$ and $\bm{b}$.
The metric $\bm{g}$ is called indefinite because the 
scalar product of a vector with itself need not be positive: a vector $\bm{a}$ is spacelike if $\bm{g}\left( \bm{a}, \bm{a} \right) > 0$, timelike if $\bm{g}\left( \bm{a}, \bm{a} \right) < 0$, and null if $\bm{g}\left( \bm{a}, \bm{a} \right) = 0$.
This nomenclature arises by comparing the relative contributions of the time and position space components of a vector to its scalar product with itself. 

The canonical form of the Minkowski metric is displayed with respect to the Minkowski coordinates of a global inertial reference frame.
The scalar product of an infinitesimal spacetime displacement $\mathrm{d}\bm{\ell} = \mathrm{d}x^\alpha \, \bm{\partial}_\alpha$ with itself is called the line element. 
In particular, let $\left(x^\mu \right) = \left( t, x, y, z \right) \in \mathbb{R}^4$ be the coordinates of a global inertial frame $\mathscr{O}$ on $\mathbb{M}$; these are Minkowski coordinates if at every event 
the line element
is expressed as
\begin{eqnarray}
\mathrm{d}\ell^2 &=& \bm{g}\left( \mathrm{d}\bm{\ell}, \mathrm{d}\bm{\ell} \right) \nonumber \\
	&=& g_{\alpha\beta}\, \mathrm{d}x^\alpha \mathrm{d}x^\beta \nonumber \\
	&=& - c^2 \mathrm{d}t^2 + \mathrm{d}x^2 + \mathrm{d}y^2 + \mathrm{d}z^2, \label{eq:LineElement} 
\end{eqnarray}
where $c$ is the speed of light, so that the matrix of components of $\bm{g}$ has the canonical form 
\begin{equation}
\left[ g_{\mu\nu} \right]_\mathscr{O} = 
\begin{bmatrix}
-c^2 & 0_j \\
0_i & \gamma_{ij}
\end{bmatrix}_\mathscr{O} = \left[ \eta_{\mu\nu} \right],
\label{eq:MinkowskiMetric}
\end{equation}
where
\begin{equation}
\left[ \eta_{\mu\nu} \right] = 
\begin{bmatrix}
-c^2 & 0 & 0 & 0 \\
0 & 1 & 0 & 0 \\
0 & 0 & 1 & 0 \\
0 & 0 & 0 & 1
\end{bmatrix}
\label{eq:MinkowskiMatrix}
\end{equation}
is the Minkowski matrix.
Here $t$ is the time coordinate, corresponding to physical time according to $\mathscr{O}$.
The position space coordinates $\left(x^i \right) = \left(x,y,z \right)$ chart the parallel hyperplanes $\mathcal{S}_t$; these are hypersurfaces of simultaneity in inertial frame $\mathscr{O}$.
The Minkowski metric $\bm{g}$ is manifestly compatible with the three-metric $\bm{\gamma}$, previewed in Eq.~(\ref{eq:ThreeMetric}), that provides Euclid geometry on $\mathcal{S}_t$; in geometric terms, $\bm{\gamma}$ is the pull-back of $\bm{g}$ onto $\mathcal{S}_t$.

The metric provides a duality between vectors and linear forms, which allows the raising and lowering of indices of tensors on $\mathbb{M}$.
This is made possible by the fact that $\bm{g}$ is non-degenerate, implying it has an inverse, whose components in Minkowski coordinates are
\begin{equation}
\left[ g^{\mu\nu} \right]_\mathscr{O} = 
\begin{bmatrix}
-1/c^2 & 0^j \\
0^i & \gamma^{ij}
\end{bmatrix}_\mathscr{O} =
\begin{bmatrix}
-1/c^2 & 0 & 0 & 0 \\
0 & 1 & 0 & 0 \\
0 & 0 & 1 & 0 \\
0 & 0 & 0 & 1
\end{bmatrix}_\mathscr{O}.
\label{eq:MinkowskiInverseMetric}
\end{equation}
In coordinate-free notation, denote by an underbar the linear form $\underline{\bm{a}}$ associated by metric duality with a vector $\bm{a}$.
However this is not necessary when indices are used, as the index position in $a^\mu$ or $a_\mu = g_{\mu\alpha} a^\alpha$ indicates whether the components are those of a vector or linear form respectively.
In a similar spirit, denote by $\overrightarrow{\bm{A}}$ the tensor with first index raised associated by metric duality with a multilinear form $\bm{A}$, and by $\Overrightarrow{\bm{A}}$ the tensor with the first and second indices raised.
Again, the arrow is not necessary when indices are used, as the index position is sufficient to distinguish, for example, $A_{\mu\nu}$ from ${A^\mu}_\nu = g^{\mu\alpha} A_{\alpha\nu}$ and from $A^{\mu\nu} = g^{\mu\alpha} g^{\nu\beta}A_{\alpha\beta}$. 
Note that in this notation, the inverse metric is denoted $\Overrightarrow{\bm{g}}$.
Moreover, the metric with a single index raised is $\overrightarrow{\bm{g}} = \bm{\delta}$, that is, the $(1,1)$ identity tensor $\bm{\delta}$ with components ${\delta^\mu}_\nu$ of the Kronecker delta in any basis.
Finally, denote by $\overleftarrow{\bm{A}}$ the tensor with last index raised associated by metric duality with a multilinear form $\bm{A}$, for instance ${A_\mu}^\nu = g^{\nu\alpha} A_{\mu\alpha}$.

The Minkowski metric allows an interpretation of the affine parameter $\tau$ of the particle four-velocity discussed in Sec.~\ref{sec:GlobalInertialFrames}.
Consider an infinitesimal displacement tangent to a timelike particle worldline (i.e. one whose tangent vectors are timelike at each point on the curve).
On the one hand, Eq.~(\ref{eq:LineElement}) yields
\begin{equation}
\sqrt{-\mathrm{d}\ell^2} = \frac{c\, \mathrm{d}t}{\Lambda_{\bm{v}}},
\end{equation}
where
\begin{equation}
\frac{1}{\Lambda_{\bm{v}}} = \sqrt{1 - \frac{\bm{\gamma}\left( \bm{v}, \bm{v} \right)}{c^2}} = \sqrt{1 - \frac{\gamma_{ij} v^i v^j }{c^2}}
\label{eq:LorentzFactor}
\end{equation}
defines the Lorentz factor $\Lambda_{\bm{v}}$ in terms of the instantaneous particle three-velocity.
On the other hand, in a (possibly accelerated) `comoving' reference frame carried along with the particle, this displacement involves no change in the spatial coordinates; it is purely an interval in the comoving time coordinate.
Let the $\tau$ affinely parametrizing the particle worldline be this `proper time' coordinate; then, since $\mathrm{d}\ell^2$ is a scalar, $\sqrt{-\mathrm{d}\ell^2} = c \,\mathrm{d}\tau$, and
\begin{equation}
\mathrm{d}t = \Lambda_{\bm{v}} \,\mathrm{d}\tau.
\label{eq:ProperTime_M}
\end{equation}
Thus the particle four-velocity of Eq.~(\ref{eq:FourVelocity}) becomes
\begin{equation}
\left[ u^\mu \right]_\mathscr{O} = \Lambda_{\bm{v}} \begin{bmatrix} 1 \\ v^i \end{bmatrix}_\mathscr{O}
\label{eq:FourVelocity_M}
\end{equation}
as reckoned by $\mathscr{O}$.
The scalar product of a four-velocity with itself is readily seen to be $\bm{g}\left( \bm{u}, \bm{u} \right) = -c^2$; in units with $c=1$, it is a timelike unit vector.
That $c$ is a maximum speed is indicated by the expression for the Lorentz factor in Eq.~(\ref{eq:LorentzFactor}).

Turn next to the particle four-momentum $\bm{p}$.
For a particle of mass $m > 0$, 
$\bm{p} = m \bm{u}$ satisfies $\bm{g}\left( \bm{p}, \bm{p} \right) = -m^2 c^2$ and has components
\begin{equation} 
\left[ p^\mu \right]_\mathscr{O} = \begin{bmatrix} \epsilon_{\bm{q}} / c^2 \\ q^i \end{bmatrix}_\mathscr{O},
\label{eq:FourMomentum_M}
\end{equation}
with particle energy $\epsilon_{\bm{q}} = \Lambda_{\bm{v}}\, m c^2$ and three-momentum $q^i = \Lambda_{\bm{v}}\, m v^i$ measured by $\mathscr{O}$.
Note that $\epsilon_{\bm{q}} = c \sqrt{ m^2 c^2 + \bm{\gamma}\left( \bm{q}, \bm{q} \right)}$, and that $\bm{q}$ is a three-vector on $\mathcal{S}_t$, whose push-forward onto $\mathbb{M}$ (denoted by the same symbol $\bm{q}$) has components
\begin{equation}
\left[ q^\mu \right]_\mathscr{O} = \begin{bmatrix} 0 \\ q^i \end{bmatrix}_\mathscr{O}
\label{eq:ThreeMomentum_M}
\end{equation}
according to $\mathscr{O}$.
The four-momentum of a massless particle is null, i.e. $\bm{g}\left( \bm{p}, \bm{p} \right) = 0$.
Its worldline cannot be parametrized by a proper time; instead its affine parameter $\lambda$ must be scaled such that
\begin{equation} 
\left[ p^\mu \right]_\mathscr{O} = \left[ \frac{\mathrm{d}X^\mu}{\mathrm{d}\lambda} \right]_\mathscr{O} = \begin{bmatrix} \epsilon_{\bm{q}} / c^2 \\ q^i \end{bmatrix}_\mathscr{O},
\end{equation}
with particle energy $\epsilon_{\bm{q}} = c \sqrt{ \bm{\gamma}\left( \bm{q}, \bm{q} \right)}$.
For later comparison with Galilei/Newton spacetime, it is worth remembering that the component $p^0$ is not the energy itself, but the inertia $\epsilon_{\bm{q}} / c^2$, which in Minkowski spacetime varies with the particle momentum.

\begin{figure}[t!]
\centering
\includegraphics[width=3.3in]{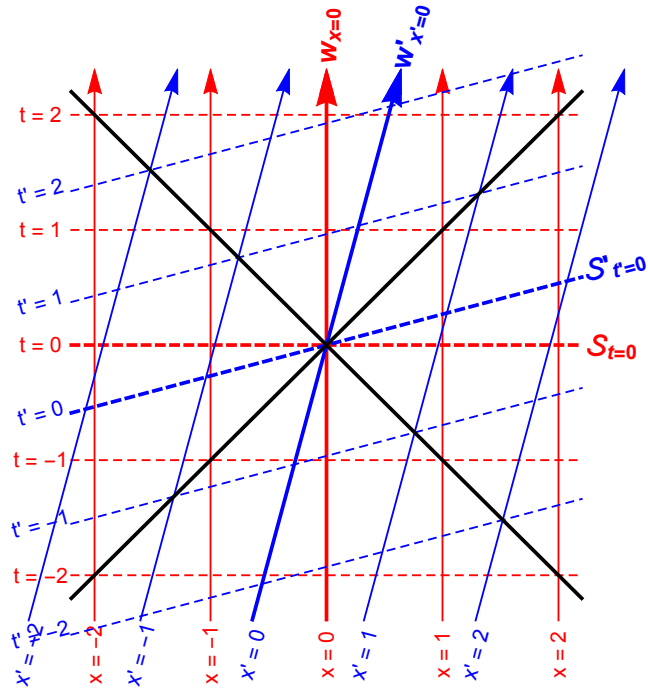}
\caption{Minkowski spacetime $\mathbb{M}$ as charted in time coordinate $t$ and position space coordinate $x$ (other position space dimensions suppressed) by a global inertial frame $\mathscr{O}$ (red), in units with $c=1$.
A second inertial frame $\mathscr{O}'$ (blue) with coordinates $t', x'$ moving with positive velocity in the $x$ direction relative to $\mathscr{O}$ is also shown,
as are the past and future null cones (black) emanating from the common origin of $\mathscr{O}$ and $\mathscr{O}'$.
Curves of constant $x,y,z$ and $x',y',z'$ can be regarded as the worldlines of families of inertial observers (solid lines with arrows; those passing through the origin are labeled and distinguished with heavier line weight), whose four-velocities are given by timelike tangent vectors $\bm{w}$ and $\bm{w'}$ to the $t$ and $t'$ coordinate curves.
Hyperplanes of simultaneity $\mathcal{S}$ and $\mathcal{S}'$ of $\mathscr{O}$ and $\mathscr{O'}$ (dashed lines; those passing through the origin are labeled and distinguished with heavier line weight) are defined by constant values of $t$ and $t'$ respectively, and are submanifolds with Euclid geometry charted by the position space coordinates. 
}
\label{fig:Minkowski_1}
\end{figure}

A spacetime diagram of $\mathbb{M}$ as charted by the inertial frame $\mathscr{O}$ (red), with $c=1$, is shown in Fig.~\ref{fig:Minkowski_1}.
By convention, the worldlines of the fiducial observers---relative to whose reference frame, in this case $\mathscr{O}$, measurements are made---are vertical, and the hyperplanes of simultaneity of $\mathscr{O}$ are horizontal.
According to this convention, a speed measured by $\mathscr{O}$ corresponds to the slope of a worldline relative to vertical.
With $c=1$, light rays are at $45^\circ$, midway between the $\mathscr{O}$ observer worldlines and their hyperplanes of simultaneity.
The future null cone of an event $\mathcal{A}$ consists of all null vectors $\bm{a}$ with time component $a^0 > 0$;
it bounds (from `below') all events that can be affected by $\mathcal{A}$.
Similarly, a past null cone bounds (from `above') all events that can influence $\mathcal{A}$.
The future and past of $\mathcal{A}$ are separated by a region of spacelike displacements from $\mathcal{A}$, which can neither influence nor be influenced by $\mathcal{A}$.
A particle with mass $m > 0$ has a timelike worldline; that $c$ is a maximum speed corresponds geometrically to the requirement that a worldline passing through $\mathcal{A}$ cannot pass outside the past and future null cones of $\mathcal{A}$.
The future and past null cones of the origin are also shown in Fig.~\ref{fig:Minkowski_1} (black).

Minkowski spacetime is invariant under Poincar\'e transformations.
These leave Eq.~(\ref{eq:LineElement}) in the same mathematical form, including the metric components as given in Eqs.~(\ref{eq:MinkowskiMetric})-(\ref{eq:MinkowskiInverseMetric}), and therefore preserve the causal structure.
The Poincar\'e group includes 10 continuous symmetries: 4 spacetime translations, 3 position space rotations, and 3 pseudo-rotations---the familiar Lorentz boosts that mix time and space.
The discrete symmetries are spatial inversion (mirror reflection or parity reversal) and time reversal.

\begin{figure}[t!]
\centering
\includegraphics[width=3.3in]{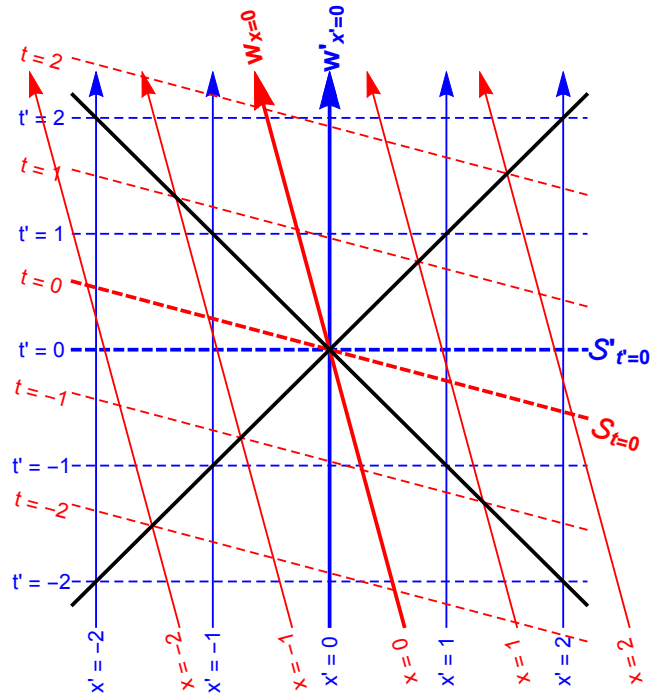}
\caption{Minkowski spacetime $\mathbb{M}$ as charted in time coordinate $t'$ and position space coordinate $x'$ (other position space dimensions suppressed) by an inertial reference frame $\mathscr{O}'$ (blue), in units with $c=1$, showing also $\mathscr{O}$ (red) moving with negative velocity in the $x'$ direction relative to $\mathscr{O}'$.
This figure is a Lorentz boost or pseudo-rotation of Fig.~\ref{fig:Minkowski_1}, under which the null cones remain unchanged.
}
\label{fig:Minkowski_2}
\end{figure}
 
Lorentz boosts are transformations between inertial frames in uniform relative motion, with different families of inertial observers.
An inertial frame $\mathscr{O}'$ (blue) in relative motion to $\mathscr{O}$ is also shown in Fig.~\ref{fig:Minkowski_1}, in particular the worldlines of a few of its observers and a few of its surfaces of simultaneity.
Some of the remarkable aspects of Minkowski spacetime are discernible.
For instance, the relativity of simultaneity is obvious in the non-coincidence of the hyperplanes of simultaneity $\mathcal{S}$ and $\mathcal{S}'$.
Length contraction can be seen, for example, in the fact that in the hyperplane $\mathcal{S}_{t=0}$ of $\mathscr{O}$, the event labeled $x' = 2$ by $\mathscr{O}'$ has a coordinate $x < 2$ according to $\mathscr{O}$.
An example of time dilation can be seen by noting that the time interval experienced by an $\mathscr{O}'$ observer traveling along $x' = 0$ from the origin to $t' = 2$ corresponds to a time interval $t > 2$ as reckoned in $\mathscr{O}$.

The principle of special relativity
is that the laws of physics must take the same mathematical form in any global inertial frame.
As applied to $\mathbb{M}$, this means invariance under Poincar\'e transformations. 
For instance, the laws of physics as expressed in $\mathscr{O}'$, whose charting of $\mathbb{M}$ is shown in Fig.~\ref{fig:Minkowski_2}---a Lorentz boost or pseudo-rotation of Fig.~\ref{fig:Minkowski_1}---must have the same form as in $\mathscr{O}$.
(That the null cones remain unchanged under this transformation reflects the invariance of the Maxwell equations under Lorentz boosts, so that the speed of light is the same for all inertial observers.)
This requirement is satisfied mathematically by expressing the laws of physics in terms of geometric objects, e.g. tensors on $\mathbb{M}$.
For instance, in describing the acceleration of a particle as
\begin{equation}
m\, \bm{u} \cdot \bm{\nabla} \bm{u} = m \frac{\mathrm{d}\bm{u}}{\mathrm{d}\tau} = \frac{\mathrm{d}\bm{p}}{\mathrm{d}\tau} = \bm{f},
\label{eq:Force_M}
\end{equation}
the force $\bm{f}$ must be a four-vector on $\mathbb{M}$ in order that the the equation be covariant under Poincar\'e transformations.
Moreover, because $\bm{g}\left( \bm{p}, \bm{p} \right) = -m^2 c^2$ is a constant, Eq.~(\ref{eq:Force_M}) requires $\bm{g}\left( \bm{p}, \bm{f} \right) = 0$. 
The Lorentz force law on a charged particle due to the electromagnetic field is the paradigmatic example satisfying these requirements.

According to the relation between time coordinate $t$ and worldline affine parameter $\tau$ discussed above, the time coordinate $t$ of a global inertial frame $\mathscr{O}$ is equal to $\tau$ for the $\mathscr{O}$ observers.
Thus the components of the $\mathscr{O}$ observer four-velocity $\bm{w}$ of Eq.~(\ref{eq:O_Worldlines}) and its associated linear form $\underline{\bm{w}}$ are 
\begin{equation}
\left[ w^\mu \right]_\mathscr{O} = \begin{bmatrix} 1 \\ 0 \\ 0 \\ 0 \end{bmatrix}_\mathscr{O}, 
\ \ \ \left[ w_\mu \right]_\mathscr{O} = \begin{bmatrix} -c^2 && 0 && 0 && 0 \end{bmatrix}_\mathscr{O}
\label{eq:O_Worldlines_M}
\end{equation}
as measured by $\mathscr{O}$.
Comparing with Eq.~(\ref{eq:TimeForm}) for the time one-form reveals that
\begin{equation}
\underline{\bm{w}} = -c^2\, \bm{t},
\end{equation}
an association made possible by the Minkowski metric.

The $\mathscr{O}$ observer four-velocity $\bm{w}$ is normal to $\mathcal{S}_t$, in the sense that it has a vanishing scalar product with the natural basis vectors $\bm{\partial}_x$, $\bm{\partial}_y$, and $\bm{\partial}_z$ tangent to the coordinate curves charting $\mathcal{S}_t$. 
However, that visually $\bm{w}$ makes a right angle with $\mathcal{S}_t$ in Fig.~\ref{fig:Minkowski_1}, looking like a normal in Euclid geometry, is an accident of convention.
The $\mathscr{O}'$ observer four-velocity $\bm{w}'$ is normal to $\mathcal{S}'_{t'}$, but visually they do not make a right angle in Fig.~\ref{fig:Minkowski_1}.
For both $\mathscr{O}$ and $\mathscr{O}'$, the normality according to the Minkowski metric is manifest graphically in the positioning of the null cone at an angle midway between the observer four-velocity and the hyperplanes of simultaneity.

Consider the projection tensor
\begin{equation}
\bm{\gamma} = \bm{g} + c^2 \, \bm{t} \otimes \bm{t} = \bm{g} + \frac{1}{c^2} \, \underline{\bm{w}} \otimes \underline{\bm{w}},
\label{eq:Projection_0_2_M}
\end{equation}
with components
\begin{equation}
\left[ \gamma_{\mu\nu} \right]_\mathscr{O} = \begin{bmatrix} 0 & 0_j \\ 0_i & \gamma_{ij} \end{bmatrix}_\mathscr{O} 
= \begin{bmatrix} 0 & 0 & 0 & 0 \\ 0 & 1 & 0 & 0 \\ 0 & 0 & 1 & 0 \\ 0 & 0 & 0 & 1 \end{bmatrix}_\mathscr{O}
\label{eq:Projection_0_2_M_Components}
\end{equation}
in inertial frame $\mathscr{O}$.
Use of the same symbol $\bm{\gamma}$ as for the three-metric is justified by the fact that for vectors $\bm{a}$ and $\bm{b}$ on $\mathbb{M}$ but tangent to $\mathcal{S}_t$, the result $\bm{\gamma}\left( \bm{a}, \bm{b} \right)$ is the same as if $\bm{\gamma}$, $\bm{a}$, and $\bm{b}$ were all tensors on $\mathcal{S}_t$.
With first index raised, this projection tensor reads
\begin{equation}
\overrightarrow{\bm{\gamma}} = \bm{\delta} + c^2 \, \overrightarrow{\bm{t}} \otimes \bm{t} = \bm{\delta} + \frac{1}{c^2} \, \bm{w} \otimes \underline{\bm{w}} = \bm{\delta} - \bm{w} \otimes \bm{t},
\label{eq:Projection_1_1_M}
\end{equation}
also with components
\begin{equation}
\left[ {\gamma^\mu}_\nu \right]_\mathscr{O} = \begin{bmatrix} 0 & 0_j \\ 0^i & {\delta^i}_j \end{bmatrix}_\mathscr{O} 
= \begin{bmatrix} 0 & 0 & 0 & 0 \\ 0 & 1 & 0 & 0 \\ 0 & 0 & 1 & 0 \\ 0 & 0 & 0 & 1 \end{bmatrix}_\mathscr{O}
\label{eq:Projection_1_1_M_Components}
\end{equation}
with respect to $\mathscr{O}$.
This raised version $\overrightarrow{\bm{\gamma}}$ projects a vector on $\mathbb{M}$ into a vector tangent to $\mathcal{S}_t$.
The other mixed version,
\begin{equation}
\overleftarrow{\bm{\gamma}} = \bm{\delta} + c^2 \,  \bm{t} \otimes \overrightarrow{\bm{t}} = \bm{\delta} + \frac{1}{c^2} \, \underline{\bm{w}} \otimes \bm{w} = \bm{\delta} - \bm{t} \otimes \bm{w}
\label{eq:Projection_1_1_M_L}
\end{equation} 
with its second index raised, i.e. ${\gamma_\mu}^\nu$, also has the same components according to $\mathscr{O}$.
Finally, the projection tensor with both indices raised is
\begin{equation}
\Overrightarrow{\bm{\gamma}} = \Overrightarrow{\bm{g}} + c^2 \, \overrightarrow{\bm{t}} \otimes \overrightarrow{\bm{t}} = \Overrightarrow{\bm{g}} + \frac{1}{c^2} \, \bm{w} \otimes \bm{w},
\label{eq:Projection_2_0_M}
\end{equation}
yet again with components
\begin{equation}
\left[ \gamma^{\mu\nu} \right]_\mathscr{O} = \begin{bmatrix} 0 & 0^j \\ 0^i & \gamma^{ij} \end{bmatrix}_\mathscr{O} 
= \begin{bmatrix} 0 & 0 & 0 & 0 \\ 0 & 1 & 0 & 0 \\ 0 & 0 & 1 & 0 \\ 0 & 0 & 0 & 1 \end{bmatrix}_\mathscr{O}
\label{eq:Projection_2_0_M_Components}
\end{equation}
with respect to $\mathscr{O}$.

The particle four-velocity $\bm{u}$ and four-momentum $\bm{p}$ can be decomposed in terms of the $\mathscr{O}$ observer four-velocity $\bm{w}$.
Comparison of Eqs.~(\ref{eq:ThreeVelocity}), (\ref{eq:FourVelocity_M}), and (\ref{eq:O_Worldlines_M}) reveals
\begin{equation}
\bm{u} = \Lambda_{\bm{v}} \left( \bm{w} + \bm{v} \right).
\label{eq:FourVelocityDecomposition_M}
\end{equation}
As for the four-momentum,
\begin{equation}
\bm{p} = \frac{\epsilon_{\bm{q}} }{c^2} \,\bm{w} + \bm{q},
\label{eq:FourMomentumDecomposition_M}
\end{equation}
thanks to Eqs.~(\ref{eq:FourMomentum_M}), (\ref{eq:ThreeMomentum_M}), and (\ref{eq:O_Worldlines_M}).
In each case, contraction with $\bm{t}$ or $\underline{\bm{w}}$ projects out the timelike component proportional to $\bm{w}$ (i.e. $\Lambda_{\bm{v}}$ and $\epsilon_{\bm{q}}$, up to a sign and factors of $c^2$), and contraction with $\overrightarrow{\bm{\gamma}}$ projects out the portion tangent to $\mathcal{S}_t$ (i.e. $\Lambda_{\bm{v}} \bm{v}$ and $\bm{q}$).

A couple of final comments on the particle energy are in order.
First, define a kinetic energy
\begin{equation}
e_{\bm{q}} = \epsilon_{\bm{q}} - m c^2 = \left( \Lambda_{\bm{v}} - 1 \right) m c^2
\label{eq:KineticEnergy_M}
\end{equation}
by subtracting from the total particle energy the portion associated with the rest mass.
Second, determine the evolution of the kinetic energy from Eq.~(\ref{eq:Force_M}).
Note that
\begin{equation}
\frac{\mathrm{d} e_{\bm{q}}}{\mathrm{d}\tau} = \frac{\mathrm{d} \epsilon_{\bm{q}}}{\mathrm{d}\tau} = \frac{\mathrm{d} }{\mathrm{d}\tau}\left[ -\bm{g}\left( \bm{w}, \bm{p} \right) \right] = -\bm{g}\left( \bm{w}, \bm{f} \right),
\end{equation}
in which $\bm{w}$, as the constant four-velocity of a global inertial observer, is not affected by the derivative.
Using Eq.~(\ref{eq:FourVelocityDecomposition_M}) and the fact that $\bm{u}$ is orthogonal to the acceleration (and hence to $\bm{f}$) results in
\begin{equation}
\frac{\mathrm{d} e_{\bm{q}}}{\mathrm{d}\tau} = \bm{g} \left( \bm{v}, \bm{f} \right) = \bm{\gamma} \left( \bm{v}, \bm{f} \right)
\label{eq:Power_M}
\end{equation}
for the evolution of kinetic energy, where the last equality follows from the fact that $\bm{v}$ is tangent to $\mathcal{S}_t$.

\subsection{Galilei/Newton Spacetime}
\label{sec:GalileiNewtonSpacetime}

As described in Sec.~\ref{sec:CausalStructure}, Galilei/Newton spacetime $\mathbb{G}$ is a fiber bundle. 
This bundle has base space $\mathbb{T}$, `time,' a one-dimensional flat manifold with Euclid distance expressing physical time intervals in terms of differences of a time coordinate $t$.
The bundle has typical fiber $\mathbb{E}^3$, `position space,' a three-dimensional flat manifold with metric $\bm{\gamma}$; in rectangular coordinates $\left( x, y, z \right)$ on $\mathbb{E}^3$, $\bm{\gamma}$ has the components displayed in Eq.~(\ref{eq:ThreeMetric}), giving Euclid geometry.
Each fiber is a hyperplane of simultaneity $\mathcal{S}_t$ corresponding to time $t$.
The position space coordinates can be set independently on each $\mathcal{S}_t$; but because of the affine structure of $\mathbb{G}$, they can be chosen such that the coordinate curves of $x$, $y$, and $z$---considered as scalar functions of the events in $\mathbb{G}$---are infinitely extended and mutually orthogonal parallel lines, as discussed in Sec.~\ref{sec:GlobalInertialFrames}.
In this case $\left(x^\mu \right) = \left( t, x, y, z \right) \in \mathbb{R}^4$ are Galilei coordinates constituting a global inertial frame on $\mathscr{O}$.

There is no spacetime metric on $\mathbb{G}$.
There is no scalar product between four-vectors;
physical time and distance are measured separately by the Euclid geometries of $\mathbb{T}$ and $\mathbb{E}^3$.
There is no metric duality between vectors and linear forms, and more generally no raising and lowering of indices of tensors on $\mathbb{G}$.
However, raising and lowering of indices of tensors on $\mathcal{S}_t$ \textit{are} allowed with respect to the three-metric $\bm{\gamma}$.
Considering also that on $\mathbb{M}$ the actions of $\bm{g}$ and $\bm{\gamma}$ on tensors tangent to $\mathcal{S}_t$ coincide, allow, for notational convenience---albeit with some trepidation---a limited use of the underbar and overarrow notation with tensors on $\mathbb{G}$, but \textit{strictly confined to tensors tangent to $\mathcal{S}_t$}, in which the associated raising and lowering is with respect to $\bm{\gamma}$. 
As discussed further below, primary examples are what might loosely be regarded as variants of $\bm{\gamma}$ itself: $\overrightarrow{\bm{\gamma}}$, $\overleftarrow{\bm{\gamma}}$, and $\Overrightarrow{\bm{\gamma}}$.
These are related by metric duality only on $\mathcal{S}_t$; their four-dimensional extensions must be understood as independent tensors on $\mathbb{G}$, as if they were represented by different symbols altogether.

Let the universal time coordinate $t$ serve as the affine parameter $\tau$ of the particle four-velocity discussed in Sec.~\ref{sec:GlobalInertialFrames}.
That is,
\begin{equation}
\mathrm{d}t = \mathrm{d}\tau,
\label{eq:ProperTime_G}
\end{equation}
and the particle four-velocity of Eq.~(\ref{eq:FourVelocity}) is simply
\begin{equation}
\left[ u^\mu \right]_\mathscr{O} = \begin{bmatrix} 1 \\ v^i \end{bmatrix}_\mathscr{O}
\label{eq:FourVelocity_G}
\end{equation}
as reckoned by $\mathscr{O}$.

The four-momentum $\bm{p} = m \bm{u}$ has components
\begin{equation} 
\left[ p^\mu \right]_\mathscr{O} = \begin{bmatrix} m \\ q^i \end{bmatrix}_\mathscr{O},
\label{eq:FourMomentum_G}
\end{equation}
with three-momentum $q^i = m v^i$ measured by $\mathscr{O}$.
That is, once again $\bm{q}$ is a three-vector on $\mathcal{S}_t$, whose push-forward onto $\mathbb{G}$ (denoted by the same symbol $\bm{q}$) has components
\begin{equation}
\left[ q^\mu \right]_\mathscr{O} = \begin{bmatrix} 0 \\ q^i \end{bmatrix}_\mathscr{O}
\label{eq:ThreeMomentum_G}
\end{equation}
according to $\mathscr{O}$.
In contrast to the relative time and relative particle energy---or rather, inertia $\epsilon_{\bm{q}} / c^2$---of Minkowski spacetime, the absolute time of Galilei/Newton spacetime induces an absolute particle inertia $p^0 = m$.

\begin{figure}[t!]
\centering
\includegraphics[width=3.3in]{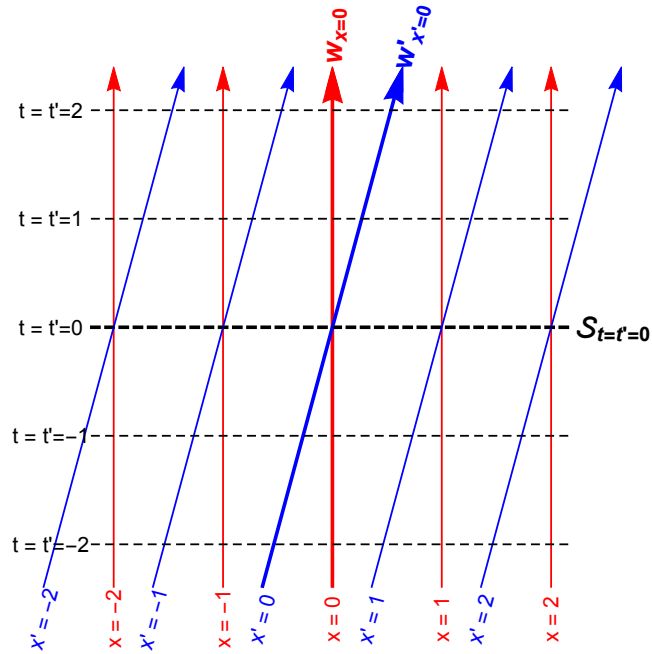}
\caption{Galilei/Newton spacetime $\mathbb{G}$ as charted in time coordinate $t$ and position space coordinate $x$ (other position space dimensions suppressed) by a global inertial frame $\mathscr{O}$ (observer worldlines, red; surfaces of simultaneity, black).
A second inertial frame $\mathscr{O}'$ (observer worldlines, blue; surfaces of simultaneity, black) with coordinates $t', x'$ moving with positive velocity in the $x$ direction relative to $\mathscr{O}$ is also shown.
In accordance with the absolute nature of time in Galilei/Newton spacetime, the surfaces of simultaneity of $\mathscr{O}$ and $\mathscr{O}'$ coincide.
}
\label{fig:GalileiNewton_1}
\end{figure}

A spacetime diagram of $\mathbb{G}$ as charted by the inertial frame $\mathscr{O}$ (red) is shown in Fig.~\ref{fig:GalileiNewton_1}.
As with Figs.~\ref{fig:Minkowski_1} and \ref{fig:Minkowski_2}, by convention, the worldlines of the fiducial observers---relative to whose reference frame, in this case $\mathscr{O}$, measurements are made---are vertical.
The hyperplanes of simultaneity, universal for all observers, are horizontal.
According to this convention, a speed measured by $\mathscr{O}$ corresponds to the slope of a worldline relative to vertical, as was also the case in Figs.~\ref{fig:Minkowski_1} and \ref{fig:Minkowski_2}.

Galilei/Newton spacetime is invariant under inhomogeneous Galilei transformations.
These leave the foliation of $\mathbb{G}$ into hyperplanes of absolute simultaneity $\mathcal{S}_t$ intact, and therefore preserve the causal structure.
Inhomogeneous Galilei transformations also leave the position space line element within each $\mathcal{S}_t$ unchanged, so that the components of $\bm{\gamma}$ are still given by Eq.~(\ref{eq:ThreeMetric}).
Like the Poincar\'e group, the inhomogeneous Galilei group includes 10 continuous symmetries: 4 spacetime translations; 3 position space rotations; and 3 Galilei boosts that leave time unchanged, but mix time into the position space coordinates.
Again the discrete symmetries are spatial inversion and time reversal.

\begin{figure}[t!]
\centering
\includegraphics[width=3.3in]{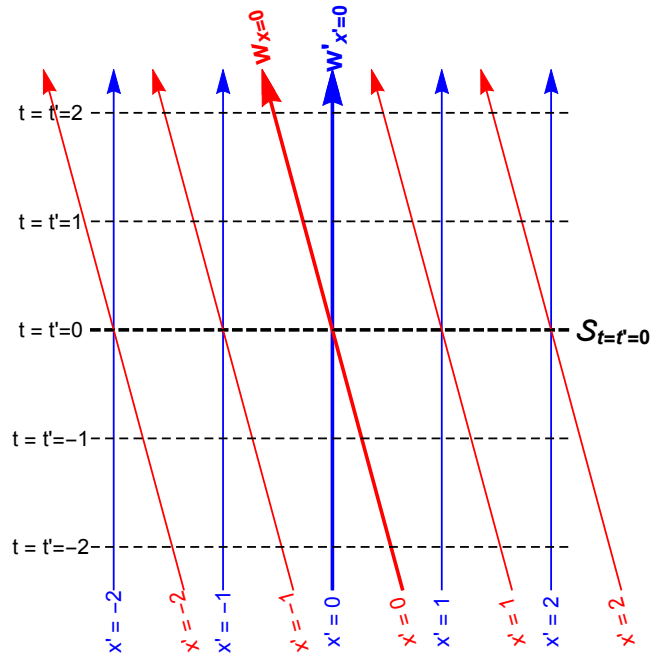}
\caption{Galilei/Newton spacetime $\mathbb{G}$ as charted in time coordinate $t'$ and position space coordinate $x'$ (other position space dimensions suppressed) by an inertial reference frame $\mathscr{O}'$ (observer worldlines, blue), showing also $\mathscr{O}$ (observer worldlines, red) moving with negative velocity in the $x'$ direction relative to $\mathscr{O}'$.
This figure is a Galilei boost or `beveling' of Fig.~\ref{fig:GalileiNewton_1}, under which the absolute hyperplanes of simultaneity remain unchanged.
}
\label{fig:GalileiNewton_2}
\end{figure}

Galilei boosts are transformations between inertial frames in uniform relative motion, with different families of inertial observers.
An inertial frame $\mathscr{O}'$ (blue) in relative motion to $\mathscr{O}$ is also shown in Fig.~\ref{fig:GalileiNewton_1}, in particular the worldlines of a few of its observers.
The relativity of simultaneity, length contraction, and time dilation discernible in Figs.~\ref{fig:Minkowski_1} and \ref{fig:Minkowski_2} are absent from Fig.~\ref{fig:GalileiNewton_1}. 

As applied to $\mathbb{G}$, the principle of special relativity means invariance under inhomogeneous Galilei transformations. 
For instance, the laws of physics as expressed in $\mathscr{O}'$, whose charting of $\mathbb{G}$ is shown in Fig.~\ref{fig:GalileiNewton_2}, must have the same form as in $\mathscr{O}$.
(Under a Galilei boost the hyperplanes of simultaneity `freely slide' past one another, like a beveling of a deck of cards, such that a `tilted' family of worldlines can be brought to point `straight up.')
As with Minkowski spacetime, this requirement is satisfied mathematically in Galilei/Newton spacetime by expressing the laws of physics in terms of geometric objects, e.g. tensors on $\mathbb{G}$.
For instance, in describing the acceleration of a particle as
\begin{equation}
m \,\bm{u} \cdot \bm{\nabla} \bm{u} = m \frac{\mathrm{d}\bm{u}}{\mathrm{d}\tau} = \frac{\mathrm{d}\bm{p}}{\mathrm{d}\tau} = \bm{f},
\label{eq:Force_G}
\end{equation}
the force $\bm{f}$ must be a four-vector on $\mathbb{G}$ in order that the the equation be covariant under inhomogeneous Galilei transformations. 
Because the derivative of the constant particle inertia $m = p^0$ vanishes, $\bm{t}\left( \bm{f} \right) = \bm{t}\cdot \bm{f}  = t_\alpha f^\alpha = 0$, that is, $\bm{f}$ must be tangent to $\mathcal{S}_t$---it is an instantaneous force---and it can depend only on the \textit{differences} of particles' positions and velocities on $\mathcal{S}_t$. 
Newton's gravitational force law is the paradigmatic example satisfying this requirement.

With use of the Galilei/Newton absolute time $t$ as the affine parameter $\tau$ of particle worldlines, as discussed above, the components of the $\mathscr{O}$ observer four-velocity $\bm{w}$ of Eq.~(\ref{eq:O_Worldlines}) are
\begin{equation}
\left[ w^\mu \right]_\mathscr{O} = \begin{bmatrix} 1 \\ 0 \\ 0 \\ 0 \end{bmatrix}_\mathscr{O}
\label{eq:O_Worldlines_G}
\end{equation}
as measured by $\mathscr{O}$.
Unlike Minkowski spacetime, because there is no spacetime metric on $\mathbb{G}$, there is no one-form $\underline{\bm{w}}$ associated with $\bm{w}$ by metric duality.
However, the one-form $\bm{t}$ of Eq.~(\ref{eq:TimeForm}) may be regarded as a kind of `counterpart' to the vector $\bm{w}$.

The absence of a spacetime metric on $\mathbb{G}$ also means that $\bm{w}$ cannot be regarded as `normal' to $\mathcal{S}_t$: without a scalar product of four-vectors, there is no relation between $\bm{w}$ and the natural basis vectors $\bm{\partial}_x$, $\bm{\partial}_y$, and $\bm{\partial}_z$ tangent to the coordinate curves charting $\mathcal{S}_t$.
As previously mentioned, under Galilei boosts the hyperplanes of simultaneity $\mathcal{S}_t$ freely slide past one another, so that Figs.~\ref{fig:GalileiNewton_1} and \ref{fig:GalileiNewton_2} are equally valid.
Neither $\bm{w}$ of $\mathscr{O}$ nor $\bm{w}'$ of $\mathscr{O}'$ are normal to $\mathcal{S}_t$.
While for a different reason than was the case in Minkowski spacetime $\mathbb{M}$, it is nevertheless also the case here that the angle of these different frames' observer worldlines in a spacetime diagram relative to vertical has no invariant geometric meaning; the visual right angles relative to $\mathcal{S}_t$ made by $\bm{w}$ in Fig.~\ref{fig:GalileiNewton_1}, and by $\bm{w}'$ in Fig.~\ref{fig:GalileiNewton_2}, are merely a matter of convention for identifying the frame regarded as `fiducial' in a particular spacetime diagram.

The three-metric $\bm{\gamma}$ on $\mathcal{S}_t$ can be extended to a projection tensor on $\mathbb{G}$ with the same components according to $\mathscr{O}$ as shown in Eq.~(\ref{eq:Projection_0_2_M_Components}) on $\mathbb{M}$: 
\begin{equation}
\left[ \gamma_{\mu\nu} \right]_\mathscr{O} = \begin{bmatrix} 0 & 0_j \\ 0_i & \gamma_{ij} \end{bmatrix}_\mathscr{O} 
= \begin{bmatrix} 0 & 0 & 0 & 0 \\ 0 & 1 & 0 & 0 \\ 0 & 0 & 1 & 0 \\ 0 & 0 & 0 & 1 \end{bmatrix}_\mathscr{O}.
\label{eq:Projection_0_2_G_Components}
\end{equation}
It cannot be written in terms of a spacetime metric tensor, since there is none on $\mathbb{G}$; that is, there is no counterpart to the expressions in Eq.~(\ref{eq:Projection_0_2_M}).
Nevertheless it is always fair to extend a multilinear form on $\mathcal{S}_t$ to four dimensions, using the same symbol, since it behaves the same with respect to vectors on spacetime that happen to tangent to $\mathcal{S}_t$.
Moreover it is also possible to define a pair of $(1,1)$ projection tensors
\begin{eqnarray}
\overrightarrow{\bm{\gamma}} &=& \bm{\delta} - \bm{w} \otimes \bm{t},
\label{eq:ProjectorTensor_G} \\
\overleftarrow{\bm{\gamma}} &=& \bm{\delta} - \bm{t} \otimes \bm{w} ,
\label{eq:ProjectorTensor_G_L}
\end{eqnarray}
corresponding to the last equalities in Eqs.~(\ref{eq:Projection_1_1_M}) and (\ref{eq:Projection_1_1_M_L}), again with the same components 
\begin{equation}
\left[ {\left(\overrightarrow{\gamma} \right)^\mu}_\nu \right]_\mathscr{O} = \begin{bmatrix} 0 & 0_j \\ 0^i & {\delta^i}_j \end{bmatrix}_\mathscr{O} 
= \begin{bmatrix} 0 & 0 & 0 & 0 \\ 0 & 1 & 0 & 0 \\ 0 & 0 & 1 & 0 \\ 0 & 0 & 0 & 1 \end{bmatrix}_\mathscr{O}
\label{eq:Projection_1_1_G_Components}
\end{equation}
according to $\mathscr{O}$, and similarly for the components ${\left(\overleftarrow{\gamma}\right)_\mu}^\nu$ of $\overleftarrow{\bm{\gamma}}$.
The push-forward of the inverse three-metric $\Overrightarrow{\bm{\gamma}}$ on $\mathcal{S}_t$ onto $\mathbb{G}$ is yet another tensor with the same components according to $\mathscr{O}$: 
\begin{equation}
\left[ \left( \Overrightarrow{\gamma}\right)^{\mu\nu} \right]_\mathscr{O} = \begin{bmatrix} 0 & 0^j \\ 0^i & \gamma^{ij} \end{bmatrix}_\mathscr{O} 
= \begin{bmatrix} 0 & 0 & 0 & 0 \\ 0 & 1 & 0 & 0 \\ 0 & 0 & 1 & 0 \\ 0 & 0 & 0 & 1 \end{bmatrix}_\mathscr{O}
\label{eq:Projection_2_0_G_Components}.
\end{equation}
As discussed near the beginning of this subsection, strictly speaking the arrow notation is only operative for the tensors confined to $\mathcal{S}_t$; this is why the arrows are included as part of the tensor symbol itself in Eqs.~(\ref{eq:Projection_0_2_G_Components}), (\ref{eq:Projection_1_1_G_Components}), and (\ref{eq:Projection_2_0_G_Components}).
The necessity of regarding $\bm{\gamma}$, $\overrightarrow{\bm{\gamma}}$, $\overleftarrow{\bm{\gamma}}$, and $\Overrightarrow{\bm{\gamma}}$ as separate tensors on $\mathbb{G}$, in contrast to these being related by metric duality on $\mathbb{M}$, speaks to the comparative inelegance of Galilei/Newton spacetime.
As a final note, the properties of $\bm{\gamma}$, $\overrightarrow{\bm{\gamma}}$, $\overleftarrow{\bm{\gamma}}$, and $\Overrightarrow{\bm{\gamma}}$ under Galilei transformations are worked out in Appendix~\ref{sec:TransformationProperties}; of these, $\Overrightarrow{\bm{\gamma}}$ is special in that it has the same components shown in Eq.~(\ref{eq:Projection_2_0_G_Components}) in $\textit{any}$ frame related to $\mathscr{O}$ by a Galilei transformation (a property which will be relevant in Sec.~\ref{sec:StressInertia}; see also related remarks in Sec.~\ref{sec:ContrastComparison}).

The particle four-velocity $\bm{u}$ and four-momentum $\bm{p}$ can be expressed in terms of the $\mathscr{O}$ observer four-velocity $\bm{w}$.
Comparison of Eqs.~(\ref{eq:ThreeVelocity}), (\ref{eq:FourVelocity_G}), and (\ref{eq:O_Worldlines_G}) reveals
\begin{equation}
\bm{u} = \bm{w} + \bm{v}.
\label{eq:FourVelocityDecomposition_G}
\end{equation}
As for the four-momentum,
\begin{equation}
\bm{p} = m \,\bm{w} + \bm{q},
\label{eq:FourMomentumDecomposition_G}
\end{equation}
thanks to Eqs.~(\ref{eq:FourMomentum_G}), (\ref{eq:ThreeMomentum_G}), and (\ref{eq:O_Worldlines_G}).
In each case, contraction with $\overrightarrow{\bm{\gamma}}$ projects out the portion tangent to $\mathcal{S}_t$ (i.e. $\bm{v}$ and $\bm{q}$).

The absolute nature of particle inertia in $\mathbb{G}$---which ultimately stems from absolute time---means that inertia and energy are not linked; the particle four-momentum does not include the kinetic energy, which has to be introduced by hand as a separate concept (once again highlighting the lesser theoretical unity of Galilei/Newton spacetime).
From a four-dimensional perspective, one can feed $\bm{u}$ and Eq.~(\ref{eq:Force_G}) into the two slots of the four-dimensional $\bm{\gamma}$.
Alternatively, recognizing that Eq.~(\ref{eq:Force_G}) relates entities tangent to $\mathcal{S}_t$, one can take the scalar product of $\bm{v}$ (also tangent to $\mathcal{S}_t$) with Eq.~(\ref{eq:Force_G}) with respect to the three-metric $\bm{\gamma}$.
Either way, the result is
\begin{equation}
\frac{\mathrm{d} e_{\bm{q}}}{\mathrm{d}\tau} = \bm{\gamma} \left( \bm{v}, \bm{f} \right),
\label{eq:Power_G}
\end{equation}
where
\begin{equation}
e_{\bm{q}} = \frac{1}{2m} \, \bm{\gamma}\left( \bm{q}, \bm{q} \right)
\label{eq:KineticEnergy_G}
\end{equation}
defines the particle kinetic energy.

\subsection{Contrast and Comparison}
\label{sec:ContrastComparison}

Geometrically, the fiber bundle nature and associated causality structure of $\mathbb{G}$ are justifiably recognized as qualitatively different from the pseudo-Riemann manifold nature and associated causality structure of $\mathbb{M}$.
Note in particular that Eqs.~(\ref{eq:LineElement})-(\ref{eq:MinkowskiMatrix}) becomes senseless when $c \rightarrow \infty$; this corresponds to the non-existence of a spacetime metric on $\mathbb{G}$, and is central to the qualitative difference between $\mathbb{M}$ and $\mathbb{G}$.
[Note however that $\Overrightarrow{\bm{\gamma}}$ of Eq.~(\ref{eq:Projection_2_0_G_Components}) is the $c \rightarrow \infty$ limit of the inverse metric $\Overrightarrow{\bm{g}}$ on $\mathbb{M}$; see Eq.~(\ref{eq:MinkowskiInverseMetric}).
Thus $\Overrightarrow{\bm{\gamma}}$ is a kind of geometric remnant that survives the limit of $\mathbb{M}$ to $\mathbb{G}$, with the previously-noted invariance of Eq.~(\ref{eq:Projection_2_0_G_Components}) under Galilei transformations being analogous to the invariance of the inverse Minkowski metric under Lorentz transformations.]

But it is also worth pausing to reflect on the extent to which $\mathbb{G}$ is like $\mathbb{M}$ in the limit $c \rightarrow \infty$.
Begin with the visual appearance of spacetime diagrams.
In units with $c=1$, the null cones of Figs.~\ref{fig:Minkowski_1} and \ref{fig:Minkowski_2} are at $45^\circ$.
But in units more suited to ordinary human experience, for instance in which $c \approx 3 \times 10^8\ \mathrm{m} / \mathrm{s}$, the null cones become visually indistinguishable from horizontal.
The surfaces of simultaneity $\mathcal{S}_{t=0}$ and $\mathcal{S}'_{t'=0}$ in Figs.~\ref{fig:Minkowski_1} and \ref{fig:Minkowski_2} get squashed between the null cones pressing into the horizontal plane, visually appearing coincident as in Figs.~\ref{fig:GalileiNewton_1} and \ref{fig:GalileiNewton_2}.
Analytically, $\Lambda_{\bm{v}} \rightarrow 1$ and $\epsilon_{\bm{q}}/ c^2 \rightarrow m$, so that Eqs.~(\ref{eq:ProperTime_M}), (\ref{eq:FourVelocity_M}), and (\ref{eq:FourMomentum_M}) go over to (\ref{eq:ProperTime_G}), (\ref{eq:FourVelocity_G}), and (\ref{eq:FourMomentum_G}).
The evolution of particle kinetic energy takes the same form in Eqs.~(\ref{eq:Power_M}) and (\ref{eq:Power_G}), with Eq.~(\ref{eq:KineticEnergy_M}) reducing to Eq.~(\ref{eq:KineticEnergy_G}).
And of course, the Poincar\'e transformations with their familiar Lorentz boosts limit smoothly to inhomogeneous Galilei transformations with their familiar Galilei boosts.

\section{Fluid Dynamics}
\label{sec:FluidDynamics}

One way to motivate the geometric structures characterizing fluid dynamics on spacetime is to generalize particle mechanics to continuum mechanics via the formalism of kinetic theory.
For a simple fluid consisting of only a single particle type---and given a foliation of spacetime in which to study, for example, an initial value problem---interest focuses on macroscopic equations for particle number, momentum, and energy.
Some useful background can be found in, for instance, Refs.~\cite{de-Groot1962Non-equilibrium,Lindquist1966Relativistic-Tr,Ehlers1971General-Relativ,Israel1972The-Relativisti,Groot1980Relativistic-Ki,Cardall2013Conservative-31,Shibata2014Conservative-fo,Schafer2014Fluid-Dynamics-,Cardall2017Relativistic-an}.

\subsection{Kinetic Theory}
\label{sec:KineticTheory}

Kinetic theory is a link between microscopic particle dynamics and a macroscopic continuum description of a physical system.
It considers volume elements that are macroscopically small---in practical terms, infinitesimal---yet contain a very large number of particles, enabling a statistical description.
Consider a position space volume element $\bm{t}\, \mathrm{dV}_t$ in spacetime, where $\bm{t}$ is the one-form defined in Sec.~\ref{sec:GlobalInertialFrames} and $\mathrm{dV}_t$ is a three-volume in the hypersurface $\mathcal{S}_t$.
For a single particle type of mass $m$, consider also a thin tube of particle worldlines, with four-momenta such that their three-momenta fall within an invariant three-momentum range $\mathrm{dP}_m$ of the three-momentum determining the four-momentum $\bm{p}$.
The Lorentz-invariant three-momentum element associated with $\mathbb{M}$, i.e. restricted to the mass shell, can be represented schematically for example as
\begin{equation}
\mathrm{dP}_m = \frac{c^2}{(2\pi\hbar)^3} \frac{\mathrm{d}\bm{q}}{\epsilon_{\bm{q}}}
\label{eq:MomentumElement_M_q}
\end{equation} 
with reference to the decomposition of $\bm{p}$ in Eq.~(\ref{eq:FourMomentumDecomposition_M}).
The corresponding Galilei-invariant three-momentum element associated with $\mathbb{G}$ is
\begin{equation}
\mathrm{dP}_m = \frac{1}{(2\pi\hbar)^3} \frac{\mathrm{d}\bm{q}}{m}
\label{eq:MomentumElement_G_q}
\end{equation} 
with reference to Eq.~(\ref{eq:FourMomentumDecomposition_G}).
The number of such worldlines piercing $\bm{t}\, \mathrm{dV}_t$ is
\begin{equation}
f \, \left(\bm{t} \cdot \bm{p} \right) \, \mathrm{dP}_m \, \mathrm{dV}_t = f \, t_\alpha p^\alpha \, \mathrm{dP}_m \, \mathrm{dV}_t,
\label{eq:ParticleDistribution}
\end{equation}
which defines the single-particle distribution function $f$ as a scalar field on phase space (i.e. spacetime 
plus momentum space restricted to the mass shell).
Slice off a short section of the tube of wordlines whose timelike extent is $\bm{p} \, \mathrm{d}\lambda$, where $\lambda$ is the affine parameter of the worldlines.
The Liouville theorem and an assumption of uncorrelated particle collisions leads to the Boltzmann equation
\begin{equation}
\frac{\mathrm{d}f}{\mathrm{d}\lambda} = C( f ),
\label{eq:Boltzmann}
\end{equation}
expressing the change---due to microscopic interactions as given by the phase space density of collisions $C( f )$---in worldline occupancy along the length of the tube segment.
The left-hand side is a directional derivative along the phase flow:
\begin{equation}
\bm{L}_m( f ) = C(f),
\label{eq:BoltzmannLiouville}
\end{equation}
where $\bm{L}_m$ is the Liouville vector, restricted to the mass shell, arising from the equations determining the worldlines $\mathcal{X}(\lambda)$:
\begin{eqnarray}
\frac{\mathrm{d}\mathcal{X}(\lambda)}{\mathrm{d}\lambda} &=& \bm{p}, \label{eq:Worldline_1}\\
\frac{\mathrm{d}\bm{p}}{\mathrm{d}\lambda} &=& m \bm{f}. \label{eq:Worldline_2}
\end{eqnarray}
[The force vector $\bm{f}$ and its components $f^\mu$ are not to be confused with the distribution function $f$, a scalar.
The factor of $m$ appears in Eq.~(\ref{eq:Worldline_2}) so that $\bm{f}$ has the same definition as in Eqs.~(\ref{eq:Force_M}) and (\ref{eq:Force_G}), recalling that $\mathrm{d}\tau = m\, \mathrm{d}\lambda$.]
Clearly a plausible account in terms of kinetic theory requires a distinction between long-range forces $\bm{f}$, responsible for the deviation of particle worldlines from geodesics; and short-range forces responsible for particle collisions, over whose microscopic effective range particle worldlines can be regarded as rectilinear.

An explicit expression for the Liouville vector $\bm{L}_m$ based on Eqs.~(\ref{eq:Worldline_1}) and (\ref{eq:Worldline_2}), and the concomitant rendering of Eq.~(\ref{eq:BoltzmannLiouville}) into a partial differential equation, depends on a choice of phase space coordinates---four spacetime coordinates, and three momentum space coordinates on the mass shell.
Approach this task in a geometric spirit by choosing momentum space coordinates in accordance with a decomposition of $\bm{p}$ in terms of the four-velocity of observers associated with a reference frame in which the momentum components are to be reckoned.
That is, the \textit{coordinates} on the momentum space mass shell can be chosen to be the spacetime \textit{components} (with respect to some three-dimensional basis linearly independent of the observer four-velocity) of the three-momentum separated out by the chosen decomposition of $\bm{p}$. 

An example of such a decomposition of $\bm{p}$ is provided in Eqs.~(\ref{eq:FourMomentumDecomposition_M}) and (\ref{eq:FourMomentumDecomposition_G}), alluded to above in connection with invariant momentum space volume elements.
In this example, take both the spacetime and momentum space coordinates from a frame $\mathscr{O}$ with spacetime coordinates $\left( x^\mu \right) = \left( t, x^i \right)$ and observer four-velocity $\bm{w} = \bm{\partial}_t$.
In this frame $\bm{w}$ has the components displayed in Eqs.~(\ref{eq:O_Worldlines_M}) and (\ref{eq:O_Worldlines_G}), and the three-momentum $\bm{q}$ has the components displayed in Eqs.~
 (\ref{eq:ThreeMomentum_M}) and (\ref{eq:ThreeMomentum_G}); take, then, the momentum space coordinates to be $\left( q^i \right)$.
Then Eq.~(\ref{eq:Boltzmann}) reads 
\begin{equation}
\frac{\mathrm{d}x^\mu}{\mathrm{d}\lambda} \frac{\partial f}{\partial x^\mu} + \frac{\mathrm{d}q^i}{\mathrm{d}\lambda} \frac{\partial f}{\partial q^i} = C( f ),
\label{eq:Boltzmann_q}
\end{equation}
and Eqs.~(\ref{eq:Worldline_1}) and (\ref{eq:Worldline_2}) give
\begin{eqnarray}
\frac{\mathrm{d}x^\mu}{\mathrm{d}\lambda} &=& p^\mu, \label{eq:Worldline_1_q}\\
\frac{\mathrm{d}q^i}{\mathrm{d}\lambda} &=& m  f^i, \label{eq:Worldline_2_q}
\end{eqnarray}
assuming $\mathscr{O}$ is a global inertial frame on the affine spacetimes $\mathbb{M}$ or $\mathbb{G}$.  
Then comparison with Eq.~(\ref{eq:BoltzmannLiouville}) yields
\begin{equation}
\bm{L}_m = p^\alpha \bm{\partial}_{x^\alpha} + m  f^i \bm{\partial}_{q^i}
\end{equation} 
for the Liouville vector.

While the focus of this paper is on the affine spacetimes $\mathbb{M}$ and $\mathbb{G}$, indulge for a moment consideration of curved spacetime.
In this case Eq.~(\ref{eq:Worldline_2_q}) reads 
\begin{equation}
\frac{\mathrm{d}q^i}{\mathrm{d}\lambda} = -{\Gamma^i}_{\beta\alpha}\, p^\beta p^\alpha + m f^i, 
\end{equation}
so that the Liouville vector
\begin{equation}
\bm{L}_m = p^\alpha \bm{\partial}_{x^\alpha} - {\Gamma^i}_{\beta\alpha}\, p^\beta p^\alpha \bm{\partial}_{q^i} + m f^i \bm{\partial}_{q^i}
\label{eq:Liouville_q_Curved}
\end{equation} 
now includes the connection coefficients due to spacetime curvature.

Even on $\mathbb{M}$ or $\mathbb{G}$, connection coefficients appear in the Liouville vector if momentum space coordinates are chosen based on momentum components measured not with respect to a global inertial frame, but instead with respect to a frame field (tetrad) that varies across spacetime, as for instance resulting from use of curvilinear coordinates or accelerated observers.
Consider for instance a different decomposition of $\bm{p}$:
\begin{eqnarray}
\bm{p} &=& \frac{\epsilon_{\bm{s}} }{c^2} \,\bm{U} + \bm{s} \ \ \ (\mathrm{on\ } \mathbb{M}), 
\label{eq:FourMomentumDecomposition_U_M} \\
\bm{p} &=& m \,\bm{U} + \bm{s} \ \ \ (\mathrm{on\ } \mathbb{G}),
\label{eq:FourMomentumDecomposition_U_G}
\end{eqnarray}
where $\bm{U}$ is the four-velocity field of accelerated observers associated with some reference frame $\tilde{\mathscr{O}}$, not a global inertial frame; and $\bm{s}$ is the three-momentum in this frame.
With reference to $\tilde{\mathscr{O}}$, the invariant three-momentum volume element is
\begin{eqnarray}
\mathrm{dP}_m &=& \frac{c^2}{(2\pi\hbar)^3} \frac{\mathrm{d}\bm{s}}{\epsilon_{\bm{s}}} \ \ \ (\mathrm{for\ } \mathbb{M}), \label{eq:MomentumElement_M_s} \\
\mathrm{dP}_m &=& \frac{1}{(2\pi\hbar)^3} \frac{\mathrm{d}\bm{s}}{m} \ \ \ (\mathrm{for\ }\mathbb{G}). \label{eq:MomentumElement_G_s}
\end{eqnarray} 
The components of $\bm{s}$ are governed by 
\begin{equation}
\frac{\mathrm{d}s^{\tilde\imath}}{\mathrm{d}\lambda} = -{\Gamma^{\tilde\imath}}_{\tilde\beta \tilde\alpha}\, p^{\tilde\beta} p^{\tilde\alpha} + m f^{\tilde\imath}, 
\end{equation}
and the Liouville vector reads
\begin{equation}
\bm{L}_m = p^\alpha \bm{\partial}_{x^\alpha} -{\Gamma^{\tilde\imath}}_{\tilde\beta \tilde\alpha}\, p^{\tilde\beta} p^{\tilde\alpha} \bm{\partial}_{s^{\tilde\imath}} + m f^{\tilde\imath} \bm{\partial}_{s^{\tilde\imath}}.
\label{eq:Liouville_s}
\end{equation} 
In these expressions, unaccented indices represent components measured with respect to $\mathscr{O}$, while indices accented with a tilde $(\tilde{\ })$ are components measured with respect to $\tilde{\mathscr{O}}$.
The components of $\bm{U}$ and $\bm{s}$ according to $\tilde{\mathscr{O}}$ are
\begin{equation}
\begin{bmatrix} U^{\tilde\mu} \end{bmatrix}_{\tilde{\mathscr{O}}} =
\begin{bmatrix} 1 \\ 0 \\ 0 \\ 0 \end{bmatrix}_{\tilde{\mathscr{O}}}, \ \ \
\begin{bmatrix} s^{\tilde\mu} \end{bmatrix}_{\tilde{\mathscr{O}}} =
\begin{bmatrix} 0 \\ s^{\tilde\imath} \end{bmatrix}_{\tilde{\mathscr{O}}},
\end{equation}
so that indeed it is sensible to use the three components $\left( s^{\tilde\imath}\right)$ as coordinates on the mass shell in momentum space.
(Beware that the transformation relating ${\Gamma^{\tilde\imath}}_{\tilde\mu \tilde\nu}$ to ${\Gamma^\rho}_{\mu\nu}$ is more complicated than it would be if these were components of a tensor. In the case that $\tilde{\mathscr{O}}$ is a local Minkowski frame, the ${\Gamma^{\tilde\rho}}_{\tilde\mu \tilde\nu}$ are known as Ricci rotation coefficients.)

It is conceptually useful to combine the first two terms of Eqs.~(\ref{eq:Liouville_q_Curved}) or (\ref{eq:Liouville_s}) into a single operator acting on the distribution function $f$, representing the variation of $f$ in spacetime while holding the \textit{vector} $\bm{p}$ fixed; hence the presence of connection coefficients to account for the change in \textit{components} of $\bm{p}$.
This operator has been denoted by a variety of notations involving indices; see for instance Eqs.~(2.11) or (2.34) in Ref.~\cite{Lindquist1966Relativistic-Tr}, Eq.~(104) and associated footnote in Ref.~\cite{Ehlers1971General-Relativ}, and Eq.~(13) in Ref.~\cite{Israel1972The-Relativisti}.
Introduce instead a coordinate-free version of this operator, $\bm{\nabla}|_{\bm{p}}$, such that Eq.~(\ref{eq:BoltzmannLiouville}) reads
\begin{equation}
\bm{p} \cdot \bm{\nabla}|_{\bm{p}} f + m \bm{f} \cdot \bm{\mathscr{D}}|_m f = C(f).
\label{eq:BoltzmannGeometric}
\end{equation}
There are two covariant derivatives in this rendering of the Boltzmann equation: $\bm{\nabla}|_{\bm{p}}$, a spacetime derivative, with the vertical bar and its subscript indicating that the vector $\bm{p}$ (but not necessarily its components in an arbitrary coordinate system) is held fixed; and $\bm{\mathscr{D}}|_m$, a momentum space derivative, with the vertical bar and its subscript $m$ indicating restriction to the mass shell.
In this way the distracting connection coefficients multiplying momentum coordinate derivatives are subsumed into an operator associated with the spacetime geometry giving rise to them, leaving a long-range force $\bm{f}$ that makes worldlines differ from geodesics cleanly separated.
Because the use of a subscripted vertical bar collides with the common practice of using a vector subscript to convert a covariant derivative into a directional derivative, 
the dot operator ($\cdot$) is used in Eq.~(\ref{eq:BoltzmannGeometric}) instead.

A strength of the coordinate-free notation in Eq.~(\ref{eq:BoltzmannGeometric}) is that it immediately suggests a conservative formulation of the first term:
\begin{equation}
\bm{\nabla}|_{\bm{p}} \cdot \left( f \bm{p} \right) + m \bm{f} \cdot \bm{\mathscr{D}}|_m f = C(f).
\label{eq:BoltzmannGeometricConservative}
\end{equation}
Indeed such conservative formulations have been demonstrated in detail \cite{Cardall2003Conservative-fo,Cardall2013Conservative-31,Shibata2014Conservative-fo}; on $\mathbb{M}$, Eq.~(\ref{eq:BoltzmannGeometricConservative}) becomes
\begin{equation}
\nabla_\alpha \left( p^\alpha f \right) + \frac{\epsilon_{\bm{s}} }{c^2}\, \mathscr{D}_{\tilde\imath} \left(-\frac{c^2}{\epsilon_{\bm s}} {\Gamma^{\tilde\imath}}_{\tilde\beta \tilde\alpha}\, p^{\tilde\beta} p^{\tilde\alpha} f\right) + m f^{\tilde\imath} \, \mathscr{D}_{\tilde\imath} f = C(f)
\label{eq:BoltzmannConservative}
\end{equation}
when rendered, for instance, in terms of the coordinates used in Eq.~(\ref{eq:Liouville_s}). 
On $\mathbb{G}$, a similar equation obtains with $\epsilon_{\bm{s}} / c^2 \rightarrow m$.
The utility of a conservative formulation like Eq.~(\ref{eq:BoltzmannConservative}) 
is that the second term gives a vanishing surface integral when integrated over $\mathrm{dP}_m$ as given in Eqs.~(\ref{eq:MomentumElement_M_s})-(\ref{eq:MomentumElement_G_s}).

\subsection{Particle Number and Fluid Velocity Vectors}
\label{sec:ParticleNumber}

The first momentum moment of the particle distribution function provides a macroscopic particle number flux, the four-vector
\begin{equation}
\bm{N} = \int f \, \bm{p} \; \mathrm{dP}_m.
\label{eq:NumberVector}
\end{equation}
This can be used to define a fluid four-velocity $\bm{U}$, which can be regarded as the four-velocity of `comoving observers' that ride along with the fluid, associated with a reference frame $\tilde{\mathscr{O}}$, by writing also
\begin{equation}
\bm{N} = n \, \bm{U}
\label{eq:FluidVelocity}
\end{equation} 
and analyzing Eq.~(\ref{eq:NumberVector}) to define the particle number density $n$ as well as $\bm{U}$.
 
For this purpose begin with the decomposition of $\bm{p}$ as given in Eqs.~(\ref{eq:FourMomentumDecomposition_M}) and (\ref{eq:FourMomentumDecomposition_G}) on $\mathbb{M}$ and $\mathbb{G}$ respectively, regarding $\bm{w}$ as the four-velocity of observers---call them `fiducial observers'---associated with a global inertial frame $\mathscr{O}$.
Define the three-velocity $\bm{v}_{\bm{q}}$ associated with the three-momentum $\bm{q}$ measured by $\mathscr{O}$ as
\begin{eqnarray}
\bm{v}_{\bm{q}} &=& \frac{c^2 \,\bm{q}}{\epsilon_{\bm{q}}} \ \ \ (\mathrm{on\ } \mathbb{M}), 
\label{eq:ThreeVelocity_M_q} \\
\bm{v}_{\bm{q}} &=& \frac{\bm{q}}{m} \ \ \ (\mathrm{on\ } \mathbb{G}). 
\label{eq:ThreeVelocity_G_q}
\end{eqnarray}
Then, referring to Eqs.~(\ref{eq:MomentumElement_M_q}) and (\ref{eq:MomentumElement_G_q}), the number flux vector reads as
\begin{equation}
\bm{N} = \int f \, \left( \bm{w} + \bm{v}_{\bm{q}} \right) \, \frac{\mathrm{d}\bm{q}}{\left(2\pi\hbar\right)^3}
\label{eq:NumberVector_q}
\end{equation}
on both $\mathbb{M}$ and $\mathbb{G}$.
 
Next, recognize that just as the four-velocity $\bm{u}$ of an individual particle can be decomposed in terms of a three-velocity $\bm{v}$ measured by the fiducial observers of $\mathscr{O}$ as in Eqs.~(\ref{eq:FourVelocityDecomposition_M}) and (\ref{eq:FourVelocityDecomposition_G}) on $\mathbb{M}$ and $\mathbb{G}$ respectively, so also the fluid four-velocity $\bm{U}$ can be decomposed as
\begin{eqnarray}
\bm{U} &=& \Lambda_{\bm{V}} \left( \bm{w} + \bm{V} \right) \ \ \ (\mathrm{on\ }\mathbb{M}), 
\label{eq:FluidVelocity_M} \\
\bm{U} &=& \bm{w} + \bm{V} \ \ \ (\mathrm{on\ }\mathbb{G}),
\label{eq:FluidVelocity_G}
\end{eqnarray}
in which $\bm{V}$ is the fluid three-velocity, tangent to $\mathcal{S}_t$, measured in $\mathscr{O}$.
Using this to substitute for $\bm{w}$ in Eq.~(\ref{eq:NumberVector_q}) gives
\begin{equation}
\bm{N} = \frac{\bm{U}}{\Lambda_{\bm{V}}} \int \frac{f \,\mathrm{d}\bm{q}}{\left(2\pi\hbar\right)^3}
 - \bm{V} \int \frac{f \,\mathrm{d}\bm{q}}{\left(2\pi\hbar\right)^3}
 + \int \frac{f \,\bm{v}_{\bm{q}} \, \mathrm{d}\bm{q}}{\left(2\pi\hbar\right)^3}
\label{eq:NumberDefine}
\end{equation}
on $\mathbb{M}$, and similarly with $\Lambda_{\bm{V}} \rightarrow 1$ on $\mathbb{G}$.
Define, therefore, the particle number density $N$ and average three-velocity $\bm{V}$ measured by the fiducial observers of $\mathscr{O}$ as
\begin{eqnarray}
N &=& \int \frac{f \,\mathrm{d}\bm{q}}{\left(2\pi\hbar\right)^3}, 
\label{eq:NumberDensity_q} \\
\bm{V} &=& \frac{1}{N} \int \frac{f \,\bm{v}_{\bm{q}} \, \mathrm{d}\bm{q}}{\left(2\pi\hbar\right)^3}
\label{eq:FluidThreeVelocity_q}
\end{eqnarray}
on both $\mathbb{M}$ and $\mathbb{G}$.
Then the last two terms of Eq.~(\ref{eq:NumberDefine}) cancel, and comparison of the first term with 
Eq.~(\ref{eq:FluidVelocity}) yields
\begin{eqnarray}
n &=& \frac{N}{\Lambda_V} \ \ \ (\mathrm{on\ }\mathbb{M}), 
\label{eq:NumberDensityComoving_M} \\
n &=& N \ \ \ (\mathrm{on\ }\mathbb{G}).
\label{eq:NumberDensityComoving_G}
\end{eqnarray}
Note that the particle number vector decomposes as
\begin{equation}
\bm{N} = N \left(\bm{w} + \bm{V} \right)
\label{eq:ParticleNumberDecomposition}
\end{equation}
on both $\mathbb{M}$ and $\mathbb{G}$.

While $N$ has been recognized in Eq.~(\ref{eq:NumberDensity_q}) as the particle number density measured by the fiducial observers associated with $\mathscr{O}$, the meaning of $n$ as the number density measured by the comoving observers associated with $\tilde{\mathscr{O}}$ should be more directly shown.
Now that the fluid four-velocity $\bm{U}$ has been defined in Eqs.~(\ref{eq:FluidVelocity_M})-(\ref{eq:FluidVelocity_G}) and (\ref{eq:FluidThreeVelocity_q}), consider again Eq.~(\ref{eq:NumberVector}) defining $\bm{N}$ in light of the alternative decomposition of $\bm{p}$ given by Eqs.~(\ref{eq:FourMomentumDecomposition_U_M})-(\ref{eq:FourMomentumDecomposition_U_G}).
In this decomposition $\bm{s}$ is recognized as the particle three-momentum measured by the comoving observers, associated with $\tilde{\mathscr{O}}$, having four-velocity $\bm{U}$.
As above, define the associated particle three-velocity as
\begin{eqnarray}
\bm{v}_{\bm{s}} &=& \frac{c^2 \,\bm{s}}{\epsilon_{\bm{s}}} \ \ \ (\mathrm{on\ } \mathbb{M}), 
\label{eq:ThreeVelocity_M_s} \\
\bm{v}_{\bm{s}} &=& \frac{\bm{s}}{m} \ \ \ (\mathrm{on\ } \mathbb{G}). 
\label{eq:ThreeVelocity_G_s}
\end{eqnarray}
Then, referring to Eqs.~(\ref{eq:MomentumElement_M_s}) and (\ref{eq:MomentumElement_G_s}), the number flux vector reads as
\begin{eqnarray}
\bm{N} &=& \int f \, \left( \bm{U} + \bm{v}_{\bm{s}} \right) \, \frac{\mathrm{d}\bm{s}}{\left(2\pi\hbar\right)^3} \nonumber \\
&=& \bm{U} \int \frac{f \,\mathrm{d}\bm{s}}{\left(2\pi\hbar\right)^3} + \int \frac{f \,\bm{v}_{\bm{s}} \, \mathrm{d}\bm{s}}{\left(2\pi\hbar\right)^3}
\end{eqnarray}
on both $\mathbb{M}$ and $\mathbb{G}$.
Comparison with Eq.~(\ref{eq:FluidVelocity}) shows that
\begin{eqnarray}
n &=& \int \frac{f \,\mathrm{d}\bm{s}}{\left(2\pi\hbar\right)^3}, 
\label{eq:NumberDensity_s} \\
0 &=& \int \frac{f \,\bm{v}_{\bm{s}} \, \mathrm{d}\bm{s}}{\left(2\pi\hbar\right)^3},
\label{eq:FluidThreeVelocity_s}
\end{eqnarray}
so that $n$ is indeed the particle number density measured in $\tilde{\mathscr{O}}$; and that the average three-velocity measured in $\tilde{\mathscr{O}}$ vanishes, as expected in what is defined as the `comoving frame.'

Conservation of particles is expressed by the vanishing spacetime divergence of the particle number flux vector $\bm{N}$.
Take the divergence of Eq.~(\ref{eq:NumberVector}) and find
\begin{eqnarray}
\bm{\nabla} \cdot \bm{N} &=& \int \bm{p} \cdot \bm{\nabla}|_{\bm{p}} f \; \mathrm{dP}_m \nonumber \\
&=& \int  \bm{\nabla}|_{\bm{p}} \cdot \left( f \, \bm{p} \right) \; \mathrm{dP}_m \nonumber \\	
&=& \int C(f) \, \mathrm{dP}_m  -  \int m \bm{f} \cdot \bm{\mathscr{D}}|_m f \; \mathrm{dP}_m
\label{eq:NumberDivergence}
\end{eqnarray}
by virtue of Eqs.~(\ref{eq:BoltzmannGeometric}) and (\ref{eq:BoltzmannGeometricConservative}).
With only a single particle species, scattering is the only possible type of collision interaction; the integral of $C(f)$ over all momenta in the first term of the last line must vanish, as particles can be neither created nor destroyed.
As for the vanishing of the second term in the last line, consider only the paradigmatic forces on $\mathbb{M}$ and $\mathbb{G}$---the electromagnetic Lorentz force and Newton gravitational force respectively.

On $\mathbb{M}$, the Lorentz force on a particle of charge $e$ and four-momentum $\bm{p}$ is
\begin{equation}
m \bm{f}_\mathrm{Lorentz} = e \overrightarrow{\bm{F}} \cdot \bm{p},
\label{eq:LorentzForce}
\end{equation}
where $\bm{F}$ is an antisymmetric bilinear form, the electromagnetic field tensor. 
In examining its effect in the last term of Eq.~(\ref{eq:NumberDivergence}),
express the integral over the mass shell as an integral over all momentum space---with the four-dimensional volume element expressed schematically as $\mathrm{dP} = \mathrm{d}\bm{p} / (2\pi\hbar)^3$---and a Dirac delta function restricting the integrand:
\begin{equation}
- 2 \int \left( m \bm{f} \cdot \bm{\mathscr{D}} f \right) \theta( \bm{p})\, \delta\left( \bm{g}(\bm{p},\bm{p}) + m^2 c^2 \right) \mathrm{dP},
\end{equation}
in which $\theta(\bm{p}) = 1$ if $\bm{p}$ is future-directed and vanishes otherwise, and the momentum space derivative $\bm{\mathscr{D}}$ is no longer restricted to the mass shell.
In the case of the Lorentz force of Eq.~(\ref{eq:LorentzForce}), this can be expressed
\begin{equation}
- 2 \int \bm{\mathscr{D}} \cdot \left[ \theta( \bm{p})\, \delta\left( \bm{g}(\bm{p},\bm{p}) + m^2 c^2 \right)  f  \, m \bm{f}_\mathrm{Lorentz} \right] \mathrm{dP} = 0,
\end{equation}
yielding a vanishing surface integral if the integrand goes to zero for infinite momenta.
Both $\bm{f}_\mathrm{Lorentz}$ and the delta function could be brought inside the derivative because the antisymmetry of $\bm{F}$ implies both that $\bm{\mathscr{D}} \cdot \bm{f}_\mathrm{Lorentz} = 0$ and $\bm{g}(\bm{p},\bm{f}_\mathrm{Lorentz}) = 0$. 

On $\mathbb{G}$, the Newton gravitational force on a particle of mass $m$ is
\begin{equation}
\bm{f}_\mathrm{Newton} = - m \,\overrightarrow{\bm{D}} \Phi,
\label{eq:NewtonForce}
\end{equation}
where $\bm{D}$ is the covariant derivative operator on $\mathcal{S}_t$ and $\Phi$ is the Newton gravitational potential.
This is independent of particle momentum, so that the last term of Eq.~(\ref{eq:NumberDivergence}) is
\begin{equation}
-\bm{f}_\mathrm{Newton} \cdot \int \bm{\mathscr{D}}|_m f \; m\, \mathrm{dP}_m = 0,
\end{equation}
in which the perfect integral vanishes provided the distribution function goes to zero for infinite momenta.

Therefore, at least for the paradigmatic long-range forces on $\mathbb{M}$ and $\mathbb{G}$ respectively,
\begin{equation}
\bm{\nabla} \cdot \bm{N} = 0
\label{eq:ParticleConservation}
\end{equation}
expresses particle conservation in both cases.
[In fact the conditions on the force $\bm{f}$ required for the the last term of Eq.~(\ref{eq:NumberDivergence}) to vanish enter into the derivation of the Liouville theorem needed to obtain the Boltzmann equation in the first place.] 

\subsection{Stress-Inertia Tensor}
\label{sec:StressInertia}

The second momentum moment of the particle distribution function, the $(2,0)$ tensor
\begin{equation}
\bm{T} = \int f \, \bm{p} \otimes \bm{p}\; \mathrm{dP}_m,
\label{eq:StressInertiaTensor}
\end{equation}
is often called the stress-energy (or energy-momentum) tensor.
This is sensible in units with $c=1$; but recalling Eqs.~(\ref{eq:FourMomentum_M}) and (\ref{eq:FourMomentum_G}), for the purpose of addressing fluid dynamics on both $\mathbb{M}$ and $\mathbb{G}$ on as unified a footing as possible, it seems more sensible to call it the stress-inertia (or inertia-momentum) tensor.
Having defined the fluid four-velocity $\bm{U}$ associated with a comoving reference frame $\tilde{\mathscr{O}}$ in Sec.~\ref{sec:ParticleNumber}, use the decomposition of $\bm{p}$ in Eqs.~(\ref{eq:FourMomentumDecomposition_U_M}) and (\ref{eq:FourMomentumDecomposition_U_G}) to motivate definitions of an internal energy density $e$, pressure $p$, energy flux $\bm{J}$, and viscous stress $\bm{\Pi}$ measured in $\tilde{\mathscr{O}}$; then a comoving `fluid element' can be regarded as a system to which the the laws of thermodynamics can be applied to relate these quantities.

Dealing first with $\mathbb{M}$, Eqs.~(\ref{eq:FourMomentumDecomposition_U_M}), (\ref{eq:MomentumElement_M_s}), (\ref{eq:ThreeVelocity_M_s}), and (\ref{eq:StressInertiaTensor}) give
\begin{equation}
\bm{T} = \int f \; \frac{\epsilon_{\bm{s}}}{c^2} \left( \bm{U} + \bm{v}_{\bm{s}} \right) \otimes \left( \bm{U} + \bm{v}_{\bm{s}} \right) \, \frac{\mathrm{d}\bm{s}}{\left(2\pi\hbar\right)^3}.
\label{eq:StressInertiaTensor_M_1}
\end{equation}
Referring back to the particle kinetic energy defined in Eq.~(\ref{eq:KineticEnergy_M}), define similarly the particle energy without rest mass $e_{\bm{s}} = \epsilon_{\bm{s}} - m c^2$ with reference to $\tilde{\mathscr{O}}$ instead of $\mathscr{O}$, so that the coefficient of $\bm{U} \otimes \bm{U} / c^2$ can be expressed
\begin{equation}
\int   \frac{f  \, \epsilon_{\bm{s}} \, \mathrm{d}\bm{s}}{\left(2\pi\hbar\right)^3} = m c^2 \, n + e,
\end{equation}
where Eq.~(\ref{eq:NumberDensity_s}) has been used, and
\begin{equation}
e = \int  \frac{f \, e_{\bm{s}} \,\mathrm{d}\bm{s}}{\left(2\pi\hbar\right)^3}
\label{eq:InternalEnergyDensity}
\end{equation}
defines the internal energy density measured with respect to $\tilde{\mathscr{O}}$.
The energy flux relative to $\tilde{\mathscr{O}}$ is a vector making a tensor product with $\bm{U} / c^2$ in Eq.~(\ref{eq:StressInertiaTensor_M_1}):
\begin{equation}
\bm{J} =  \int   \frac{f  \, \epsilon_{\bm{s}} \, \bm{v}_{\bm{s}}\, \mathrm{d}\bm{s}}{\left(2\pi\hbar\right)^3} = \int   \frac{f  \, e_{\bm{s}} \, \bm{v}_{\bm{s}} \,\mathrm{d}\bm{s}}{\left(2\pi\hbar\right)^3},
\label{eq:InternalEnergyFlux} 
\end{equation}
where the last equality, due to Eq.~(\ref{eq:FluidThreeVelocity_s}), means that mass flux (times $c^2$) does not contribute to $\bm{J}$.
Finally, Eq.~(\ref{eq:StressInertiaTensor_M_1}) includes the pressure tensor
\begin{equation}
\bm{P} = \int  f  \, \frac{\epsilon_{\bm{s}}}{c^2} \left( \bm{v}_{\bm{s}} \otimes \bm{v}_{\bm{s}} \right)  \frac{\mathrm{d}\bm{s}}{\left(2\pi\hbar\right)^3}. 
\label{eq:PressureTensor_M}
\end{equation}
Because $\bm{v}_{\bm{s}}$ is orthogonal to $\bm{U}$, 
\begin{eqnarray}
\bm{0} &=& \bm{g}( \bm{U}, \bm{J} ) = \underline{\bm{U}} \cdot \bm{J}, 
\label{eq:Orthogonal_J} \\
\bm{0} &=& \underline{\bm{U}} \cdot \bm{P} = \bm{P} \cdot \underline{\bm{U}}.
\label{eq:Orthogonal_P}
\end{eqnarray}
It is useful to define the $(0,2)$ comoving frame projection tensor 
\begin{equation}
\bm{h} = \bm{g} + \frac{1}{c^2}\, \underline{\bm{U}} \otimes \underline{\bm{U}}
\label{eq:Projection_0_2_U}
\end{equation}
in $\tilde{\mathscr{O}}$, analogous to $\bm{\gamma}$ of Eq.~(\ref{eq:Projection_0_2_M}) for the fiducial inertial frame $\mathscr{O}$,
and satisfying 
\begin{equation}
\bm{0} = \bm{h} \cdot \bm{U} = \bm{U} \cdot \bm{h}.
\end{equation}
The pressure tensor decomposes as
\begin{equation}
\bm{P} = p \Overrightarrow{\bm{h}} + \bm{\Pi},
\label{eq:PressureDecomposition_M}
\end{equation}
in which the isotropic pressure $p$ and viscous stress tensor $\bm{\Pi}$ measured in $\tilde{\mathscr{O}}$ satisfy
\begin{eqnarray}
p &=& \frac{1}{3} \mathrm{Tr}\left( \bm{h} \cdot \bm{P} \right), \\
0 &=& \mathrm{Tr}\left( \bm{h} \cdot \bm{\Pi} \right).
\end{eqnarray}
Putting all this together,
\begin{equation}
\bm{T} = \left( m\, n + \frac{e}{c^2} \right) \bm{U}\otimes \bm{U} +  \bm{U}\otimes \frac{\bm{J}}{c^2}  + \frac{\bm{J}}{c^2}\otimes\bm{U} + p \Overrightarrow{\bm{h}} + \bm{\Pi}
\label{eq:StressInertiaTensor_M_2}
\end{equation}
exhibits the stress-inertia tensor on $\mathbb{M}$.

Turning next to $\mathbb{G}$, Eqs.~(\ref{eq:FourMomentumDecomposition_U_G}), (\ref{eq:MomentumElement_G_s}), (\ref{eq:ThreeVelocity_G_s}), and (\ref{eq:StressInertiaTensor}) give
\begin{equation}
\bm{T} = \int f \; m \left( \bm{U} + \bm{v}_{\bm{s}} \right) \otimes \left( \bm{U} + \bm{v}_{\bm{s}} \right) \, \frac{\mathrm{d}\bm{s}}{\left(2\pi\hbar\right)^3}.
\label{eq:StressInertiaTensor_G_1}
\end{equation}
With an inertia that does not depend on momentum, the coefficient of $\bm{U}\otimes\bm{U}$ integrates directly to $m\, n$ thanks to Eq.~(\ref{eq:NumberDensity_s}), and the terms mixed in $\bm{U}$ and $\bm{v}_{\bm{s}}$ vanish because of Eq.~(\ref{eq:FluidThreeVelocity_s}). 
There remains the pressure tensor
\begin{eqnarray}
\bm{P} = \int  f  \, m \left( \bm{v}_{\bm{s}} \otimes \bm{v}_{\bm{s}} \right)  \frac{\mathrm{d}\bm{s}}{\left(2\pi\hbar\right)^3}.
\label{eq:PressureTensor_G} 
\end{eqnarray}
Because there is no spacetime metric on $\mathbb{G}$, there is no direct analogue of Eq.~(\ref{eq:Projection_0_2_U}) to use in decomposing $\bm{P}$ into its isotropic and viscous parts.
However, while $\Overrightarrow{\bm{\gamma}}$ and $\Overrightarrow{\bm{h}}$ are distinct projection tensors on $\mathbb{M}$ for $\mathscr{O}$ and $\tilde{\mathscr{O}}$ respectively, on $\mathbb{G}$ it turns out that no separate $(2,0)$ projection operator is needed: $\mathscr{O}$ and $\tilde{\mathscr{O}}$ are related by a local Galilei transformation; and as discussed in Sec.~\ref{sec:GalileiNewtonSpacetime}, the components of the $(2,0)$ tensor $\Overrightarrow{\bm{\gamma}}$  do not change under such a transformation.
Therefore the pressure tensor decomposes as
\begin{equation}
\bm{P} = p \Overrightarrow{\bm{\gamma}} + \bm{\Pi},
\label{eq:PressureDecomposition_G}
\end{equation}
in which the isotropic pressure $p$ and viscous stress tensor $\bm{\Pi}$ measured in $\tilde{\mathscr{O}}$ satisfy
\begin{eqnarray}
p &=& \frac{1}{3} \mathrm{Tr}\left( \bm{\gamma} \cdot \bm{P} \right), \\
0 &=& \mathrm{Tr}\left( \bm{\gamma} \cdot \bm{\Pi} \right).
\end{eqnarray}
Putting all this together,
\begin{equation}
\bm{T} = m\, n  \, \bm{U}\otimes \bm{U} 
 + p \Overrightarrow{\bm{\gamma}} + \bm{\Pi}
\label{eq:StressInertiaTensor_G_2}
\end{equation}
exhibits the stress-inertia tensor on $\mathbb{G}$.
This is recognized as the $c \rightarrow \infty$ limit of Eq.~(\ref{eq:StressInertiaTensor_M_2}): the terms involving $e$ and $\bm{J}$ disappear; and from Eq.~(\ref{eq:Projection_0_2_U}), $\Overrightarrow{\bm{h}} \rightarrow \Overrightarrow{\bm{g}}$, and it has already been noted in Sec.~\ref{sec:ContrastComparison} that $\Overrightarrow{\bm{\gamma}}$ on $\mathbb{G}$ is the $c \rightarrow \infty$ limit of $\Overrightarrow{\bm{g}}$ on $\mathbb{M}$.

In order to find an equation satisfied by $\bm{T}$, take the divergence of Eq.~(\ref{eq:StressInertiaTensor}) and find
\begin{eqnarray}
\bm{\nabla} \cdot \bm{T} &=& \int \bm{p} \left( \bm{p} \cdot \bm{\nabla}|_{\bm{p}} f \right) \; \mathrm{dP}_m \nonumber \\
&=& \int  \bm{\nabla}|_{\bm{p}} \cdot \left( f \, \bm{p} \otimes \bm{p} \right) \; \mathrm{dP}_m \nonumber \\	
&=& \int \bm{p} \; C(f) \, \mathrm{dP}_m  -  \int \bm{p} \left( m \bm{f} \cdot \bm{\mathscr{D}}|_m f \right)  \mathrm{dP}_m, \nonumber \\
& & \label{eq:StressInertiaDivergence}
\end{eqnarray}
in which Eq.~(\ref{eq:BoltzmannGeometricConservative}) has been multiplied by $\bm{p}$.
With only a single particle species, the scattering that is the only possible interaction between point particles must, like colliding billiard balls, conserve four-momentum; thus the integral in the first term of the last line vanishes due to four-momentum conservation.
As in Sec.~\ref{sec:ParticleNumber}, evaluate the second term in terms of the paradigmatic forces on $\mathbb{M}$ and $\mathbb{G}$---the electromagnetic Lorentz force and Newton gravitational force respectively.

For the Lorentz force on $\mathbb{M}$, considerations similar to those in Sec.~\ref{sec:ParticleNumber} lead to 
\begin{equation}
- 2 \int \bm{p} \left(\bm{\mathscr{D}} \cdot \left[ \theta( \bm{p})\, \delta\left( \bm{g}(\bm{p},\bm{p}) + m^2 c^2 \right)  f  \, m \bm{f}_\mathrm{Lorentz} \right] \right) \mathrm{dP} = 0
\end{equation}
for the last term in Eq.~(\ref{eq:StressInertiaDivergence}).
After integration by parts, returning this to an integral over the mass shell gives
\begin{eqnarray}
\int f \, m \bm{f}_\mathrm{Lorentz} \; \mathrm{dP}_m &=& e \overrightarrow{\bm{F}} \cdot \int f \, \bm{p} \; \mathrm{dP}_m \nonumber \\
&=& e \overrightarrow{\bm{F}} \cdot \bm{N},
\end{eqnarray}
in which Eqs.~(\ref{eq:NumberVector}) and (\ref{eq:LorentzForce}) have been used.

For the Newton gravitational force on $\mathbb{G}$, upon integration by parts the last term of Eq.~(\ref{eq:StressInertiaDivergence}) becomes
\begin{eqnarray}
-  \int \bm{p} \left( m \bm{f}_\mathrm{Newton} \cdot \bm{\mathscr{D}}|_m f \right)  \mathrm{dP}_m &=& \bm{f}_\mathrm{Newton} \int f \; m \, \mathrm{dP}_m \nonumber \\
&=& - m \, N \,  \overrightarrow{\bm{D}} \Phi
\label{eq:StressInertiaSource_G}
\end{eqnarray}
when Eqs.~(\ref{eq:FourMomentumDecomposition_G}), (\ref{eq:MomentumElement_G_q}), (\ref{eq:NumberDensity_q}), and (\ref{eq:NewtonForce}) are taken into account. 

Therefore the divergence of the stress-inertia tensor $\bm{T}$ is 
\begin{eqnarray}
\bm{\nabla} \cdot \bm{T} &=& e \overrightarrow{\bm{F}} \cdot \bm{N} \ \ \ (\mathrm{on\ }\mathbb{M}), 
\label{eq:StressInertiaConservation_M} \\
\bm{\nabla} \cdot \bm{T} &=& - m \, N \,  \overrightarrow{\bm{D}} \Phi \ \ \ (\mathrm{on\ }\mathbb{G})
\label{eq:StressInertiaConservation_G}
\end{eqnarray}
for the paradigmatic long-range forces on $\mathbb{M}$ and $\mathbb{G}$ respectively.

\subsection{Three-Momentum Tensor}
\label{sec:ThreeMomentumTensor}

In contrast to the number vector of Eq.~(\ref{eq:NumberVector}) expressing the flux of a single quantity---particle number---the stress-inertia tensor of Eq.~(\ref{eq:StressInertiaTensor}) embodies the flux of four-momentum.
Thus, while the divergence of the number vector in Eq.~(\ref{eq:ParticleConservation}) will ultimately reduce to a single partial differential equation, Eqs.~(\ref{eq:StressInertiaConservation_M}) and (\ref{eq:StressInertiaConservation_G}) separate into four partial differential equations; and the divergence $\bm{\nabla} \cdot \bm{T}$ being a coordinate-free geometric expression, there is freedom in deciding what information one would like to extract, i.e. which partial differential equations one wishes to generate.

Begin the separation by defining a $(1,1)$ tensor $\bm{M}$ expressing the flux of three-momentum.
In particular, having chosen a fiducial frame $\mathscr{O}$---a global inertial frame, whose time coordinate $t$ yields hyperplanes of simultaneity $\mathcal{S}_t$, and whose associated observers have four-velocity $\bm{w}$---let it express the flux of the three momentum measured by $\mathscr{O}$:
\begin{equation}
\bm{M} = \int f \, \underline{\bm{q}} \otimes \bm{p}\; \mathrm{dP}_m
\label{eq:ThreeMomentumTensor}
\end{equation}
(note the first slot of this tensor is `index down').
Because $\underline{\bm{q}} = \bm{\gamma} \cdot \bm{p}$ [recall again Eqs.~(\ref{eq:FourVelocityDecomposition_M}) and (\ref{eq:FourVelocityDecomposition_G})], comparison with Eq.~(\ref{eq:StressInertiaTensor}) shows that
\begin{equation}
\bm{M} = \bm{\gamma} \cdot \bm{T}.
\label{eq:MomentumTensor}
\end{equation}
This applies to both $\mathbb{M}$ and $\mathbb{G}$; recall from the discussion in Sec.~\ref{sec:GalileiNewtonSpacetime} that the underbar notation, applied here to $\bm{q}$, is allowed on $\mathbb{G}$ because $\bm{q}$ is tangent to $\mathcal{S}_t$.

After having decomposed $\bm{T}$ in Eqs.~(\ref{eq:StressInertiaTensor_M_2}) and (\ref{eq:StressInertiaTensor_G_2}) in terms of quantities defined with respect to the comoving frame $\tilde{\mathscr{O}}$, 
the projection of momentum with respect to $\mathscr{O}$ in Eq.~(\ref{eq:MomentumTensor}) entails some algebraic complications associated with the relationship between the two frames.
(The more elegant geometry of $\mathbb{M}$ relative to $\mathbb{G}$ now begins to lead to messier practical consequences.)
To help manage this, group the various terms in $\bm{M}$ as
\begin{equation}
\bm{M} = \bm{M}_\mathrm{perfect} + \bm{M}_\mathrm{dissipative},
\label{eq:MomentumSplit}
\end{equation}
where $\bm{M}_\mathrm{perfect}$ includes terms involving $n$, $e$, and $p$, properties intrinsic to a fluid element; and $\bm{M}_\mathrm{dissipative}$ involves the quantities $\bm{J}$ and $\bm{\Pi}$ associated with energy and momentum exchange between fluid elements.

On $\mathbb{M}$, inserting Eq.~(\ref{eq:StressInertiaTensor_M_2}) in Eq.~(\ref{eq:MomentumTensor}) yields
\begin{equation}
\bm{M}_\mathrm{perfect} = \Lambda_{\bm{V}} \left( m\, n + \frac{e + p}{c^2} \right) \underline{\bm{V}} \otimes \bm{U} + p \overleftarrow{\bm{\gamma}},
\label{eq:MomentumPerfect_M}
\end{equation}
in which Eqs.~(\ref{eq:FluidVelocity_M}) and (\ref{eq:Projection_0_2_U}) have been used, and the left arrow in $\overleftarrow{\bm{\gamma}}$ denotes that the \textit{last} index of a multilinear form has been raised; and
\begin{equation}
\bm{M}_\mathrm{dissipative} = \Lambda_{\bm{V}} \underline{\bm{V}} \otimes \frac{\bm{J}}{c^2} + \frac{\bm{J}_{\bm{\gamma}}}{c^2} \otimes \bm{U} + \bm{\Pi}_{\bm{\gamma}},
\label{eq:MomentumDissipative_M}
\end{equation}
in which the linear form $\bm{J}_{\bm{\gamma}}$ and $(1,1)$ tensor $\bm{\Pi}_{\bm{\gamma}}$ 
\begin{eqnarray}
\bm{J}_{\bm{\gamma}} &=& \bm{\gamma} \cdot \bm{J}, 
\label{eq:J_Gamma} \\ 
\bm{\Pi}_{\bm{\gamma}} &=& \bm{\gamma} \cdot \bm{\Pi} 
\label{eq:Pi_Gamma}
\end{eqnarray}
have been defined.

On $\mathbb{G}$ things are simpler: inserting Eq.~(\ref{eq:StressInertiaTensor_G_2}) in Eq.~(\ref{eq:MomentumTensor}) yields
\begin{eqnarray}
\bm{M}_\mathrm{perfect} &=& m\, n \, \underline{\bm{V}} \otimes \bm{U} + p \overleftarrow{\bm{\gamma}}, 
\label{eq:MomentumPerfect_G} \\
\bm{M}_\mathrm{dissipative} &=& \bm{\Pi}_{\bm{\gamma}},
\label{eq:MomentumDissipative_G}
\end{eqnarray}
reflecting the decompositions in Eqs.~(\ref{eq:FluidVelocity_G}) and (\ref{eq:PressureDecomposition_G}). (Again, the underbar notation is allowed because $\bm{V}$ is tangent to $\mathcal{S}_t$.)

Because $\bm{\nabla} \bm{\gamma} = 0$ on affine $\mathbb{M}$ and $\mathbb{G}$ (as readily seen in $\mathscr{O}$), insertion of Eq.~(\ref{eq:MomentumTensor}) into Eqs.~(\ref{eq:StressInertiaConservation_M}) and (\ref{eq:StressInertiaConservation_G}) shows that $\bm{M}$ satisfies
\begin{eqnarray}
\bm{\nabla} \cdot \bm{M} &=& e\, \bm{F}_{\bm{\gamma}} \cdot \bm{N} \ \ \ (\mathrm{on\ }\mathbb{M}), 
\label{eq:MomentumConservation_M} \\
\bm{\nabla} \cdot \bm{M} &=& - m \, N \,  \bm{D} \Phi \ \ \ (\mathrm{on\ }\mathbb{G}),
\label{eq:MomentumConservation_G}
\end{eqnarray}
in which the bilinear form
\begin{equation}
\bm{F}_{\bm{\gamma}} = \bm{\gamma} \cdot \overrightarrow{\bm{F}}
\end{equation}
has been defined.
On $\mathbb{M}$, the Lorentz force density on a current density $e \bm{N}$ can be recognized with a suitable decomposition of the electromagnetic field tensor $\bm{F}$.
The Newton gravitational force density is apparent on $\mathbb{G}$.

\subsection{Energy Vector}
\label{sec:EnergyVector}

Having defined a $(1,1)$ tensor $\bm{M}$ (first index down) embodying the flux of three-momentum as measured by the fiducial---and, on $\mathbb{M}$ and $\mathbb{G}$, inertial---frame $\mathscr{O}$, define also a vector $\bm{E}$ expressing the flux of total kinetic energy measured by $\mathscr{O}$, that is, the particle energy that excludes its rest mass:
\begin{equation}
\bm{E} = \int f \, e_{\bm{q}} \, \bm{p}\; \mathrm{dP}_m,
\label{eq:EnergyVector}
\end{equation}
recalling that $e_{\bm{q}}$ is defined by Eqs.~(\ref{eq:KineticEnergy_M}) and (\ref{eq:KineticEnergy_G}) on $\mathbb{M}$ and $\mathbb{G}$ respectively.

On $\mathbb{M}$, the relation $e_{\bm{q}} + m c^2 = \epsilon_{\bm{q}} = - \underline{\bm{w}}\cdot \bm{p}$ gives
\begin{eqnarray}
\bm{E} &=& \int f \left( - \underline{\bm{w}}\cdot \bm{p} - m c^2 \right) \bm{p} \;  \mathrm{dP}_m \nonumber \\
&=& - \underline{\bm{w}}\cdot \bm{T} - m c^2 \bm{N},
\label{eq:EnergyVector_M}
\end{eqnarray}
thanks to Eqs.~(\ref{eq:NumberVector}) and (\ref{eq:StressInertiaTensor}).
Because energy is a form of inertia, it is not surprising that $\bm{E}$ can be expressed in terms of the stress-inertia tensor $\bm{T}$.
Insertion of Eqs.~(\ref{eq:FluidVelocity}) and (\ref{eq:StressInertiaTensor_M_2}) into Eq.~(\ref{eq:EnergyVector_M}) gives rise to several contractions.
Most simply, Eq.~(\ref{eq:FluidVelocity_M}) gives
\begin{equation}
- \underline{\bm{w}}\cdot \bm{U} = \Lambda_{\bm{V}} c^2.
\end{equation}
Together with the fact that $\bm{J}$ and $\bm{\Pi}$ are orthogonal to $\bm{U}$ [see Eqs.~(\ref{eq:Orthogonal_J}), (\ref{eq:Orthogonal_P}), (\ref{eq:Projection_0_2_U}), and (\ref{eq:PressureDecomposition_M})], Eq.~(\ref{eq:FluidVelocity_M}) also implies that
\begin{eqnarray}
- \underline{\bm{w}}\cdot \bm{J} &=& \underline{\bm{V}} \cdot \bm{J} = \bm{V}\cdot \bm{J}_{\bm{\gamma}}, \\ 
- \underline{\bm{w}}\cdot \bm{\Pi} &=& \underline{\bm{V}} \cdot \bm{\Pi} = \bm{V}\cdot \bm{\Pi}_{\bm{\gamma}},
\end{eqnarray}
where $\bm{J}_{\bm{\gamma}}$ and $\bm{\Pi}_{\bm{\gamma}}$ are defined in Eqs.~(\ref{eq:J_Gamma}) and (\ref{eq:Pi_Gamma}).
Breaking $\bm{E}$ into two pieces
\begin{equation}
\bm{E} = \bm{E}_\mathrm{perfect} + \bm{E}_\mathrm{dissipative}
\label{eq:EnergySplit}
\end{equation}
as was done with the momentum tensor $\bm{M}$,
\begin{eqnarray}
\bm{E}_\mathrm{perfect} &=& \left[ \left( \Lambda_{\bm{V}} - 1 \right) mc^2 \, n  + \Lambda_{\bm V} \left( e + p \right) \right]\bm{U} \nonumber \\
& & - \; p\, \bm{w}, 
\label{eq:EnergyPerfect_M} \\
\bm{E}_\mathrm{dissipative} &=& \Lambda_{\bm{V}} \bm{J} +  \bm{V}\cdot \bm{\Pi}_{\bm{\gamma}} + \frac{1}{c^2} \left( \bm{V}\cdot \bm{J}_{\bm{\gamma}} \right) \bm{U}
\label{eq:EnergyDissipative_M}
\end{eqnarray}
collect the terms arising from Eq.~(\ref{eq:EnergyVector_M}).

On $\mathbb{G}$, use of Eq.~(\ref{eq:KineticEnergy_G}) in Eq.~(\ref{eq:EnergyVector}) yields
\begin{equation}
\bm{E} = \frac{1}{2m}  \int f \, \bm{\gamma}(\bm{q},\bm{q} ) \, \bm{p}\; \mathrm{dP}_m.
\label{eq:EnergyVector_G_1}
\end{equation}
Because of the absolute nature of inertia on $\mathbb{G}$, it does not contain energy or momentum; thus Eq.~(\ref{eq:EnergyVector_G_1}) is not readily expressed (at least \textit{in toto}) in terms of the stress-inertia tensor $\bm{T}$.
Indeed, Eq.~(\ref{eq:EnergyVector_G_1}) contains a third moment in $\bm{q}$, while $\bm{T}$ is only a second moment in $\bm{q}$ on $\mathbb{G}$; thus expression of $\bm{E}$ in terms of quantities measured by $\tilde{\mathscr{O}}$ (in order to leverage thermodynamics) will have to proceed more directly.
The three-momentum $\bm{q}$ according to $\mathscr{O}$ can be expressed, as expected, as
\begin{equation}
\bm{q} = m \left(\bm{V} + \bm{v}_{\bm{s}} \right)
\label{eq:ThreeMomentum_q_s}
\end{equation}
thanks to Eqs.~(\ref{eq:FourMomentumDecomposition_G}), (\ref{eq:FourMomentumDecomposition_U_G}), (\ref{eq:FluidVelocity_G}), and (\ref{eq:ThreeVelocity_G_s}).
This gives three cross terms in the scalar product in Eq.~(\ref{eq:EnergyVector_G_1}).
The term with a scalar product of $\bm{V}$ with itself is
\begin{eqnarray}
\frac{m}{2}\, \bm{\gamma}(\bm{V},\bm{V}) \int f \, \bm{p}\; \mathrm{dP}_m  
&=& \frac{m}{2}\, \bm{\gamma}(\bm{V},\bm{V}) \, \bm{N} \nonumber \\
&=& \frac{1}{2}m \,n \, \bm{\gamma}(\bm{V},\bm{V}) \, \bm{U}
\end{eqnarray}
by Eqs.~(\ref{eq:NumberVector}) and (\ref{eq:FluidVelocity}).
The cross terms with a scalar product of $\bm{V}$ with $\bm{v}_{\bm{s}}$ give
\begin{eqnarray}
\underline{\bm{V}} \cdot \int f \,m \,\bm{v}_{\bm{s}} \otimes \bm{p} \; \mathrm{dP}_m &=& \underline{\bm{V}} \cdot \int f \,m \,\bm{v}_{\bm{s}} \otimes ( \bm{U} + \bm{v}_{\bm{s}} ) \frac{\mathrm{d}\bm{s}}{(2\pi\hbar)^3} \nonumber \\
&=& \underline{\bm{V}} \cdot \bm{P} \nonumber \\
&=& p \,\bm{V} + \bm{V}\cdot \bm{\Pi}_{\bm{\gamma}},
\end{eqnarray}
in which Eqs.~(\ref{eq:FourMomentumDecomposition_U_G}), (\ref{eq:MomentumElement_G_s}), (\ref{eq:FluidThreeVelocity_s}), (\ref{eq:PressureTensor_G}), (\ref{eq:PressureDecomposition_G}), and (\ref{eq:Pi_Gamma}) have been employed.
Finally, the term arising from the scalar product of $\bm{v}_{\bm{s}}$ with itself is
\begin{eqnarray}
\int f \, \frac{m}{2} \, \bm{\gamma}(\bm{v}_{\bm{s}}, \bm{v}_{\bm{s}})\,  \bm{p} \; \mathrm{dP}_m &=&
\int f \, e_{\bm{s}} \, ( \bm{U} + \bm{v}_{\bm{s}} ) \frac{\mathrm{d}\bm{s}}{(2\pi\hbar)^3} \nonumber \\
&=& e \, \bm{U} + \bm{J},
\end{eqnarray}
where Eqs.~(\ref{eq:FourMomentumDecomposition_U_G}) and (\ref{eq:MomentumElement_G_s}) have been used, and the definitions in Eqs.~(\ref{eq:InternalEnergyDensity}) and (\ref{eq:InternalEnergyFlux}) have been adopted.
Putting all this together, the expressions
\begin{eqnarray}
\bm{E}_\mathrm{perfect} &=& \left[ \frac{1}{2}m \,n \, \bm{\gamma}(\bm{V},\bm{V}) + e \right] \bm{U} + p\, \bm{V}, 
\label{eq:EnergyPerfect_G} \\
\bm{E}_\mathrm{dissipative} &=& \bm{J} + \bm{V}\cdot \bm{\Pi}_{\bm{\gamma}}
\label{eq:EnergyDissipative_G}
\end{eqnarray}
comprise the terms in $\bm{E}$ on $\mathbb{G}$.

On $\mathbb{M}$, an equation obeyed by $\bm{E}$ follows quickly from Eqs.~(\ref{eq:ParticleConservation}), (\ref{eq:StressInertiaConservation_M}), and (\ref{eq:EnergyVector_M}):
\begin{equation}
\bm{\nabla} \cdot \bm{E} = e\, N \,\bm{V} \cdot \bm{F} \cdot \bm{w} \ \ \ (\mathrm{on\ }\mathbb{M}), 
\label{eq:EnergyConservation_M}
\end{equation}
in which Eq.~(\ref{eq:ParticleNumberDecomposition}) and the antisymmetry of the electromagnetic field tensor $\bm{F}$ have been used.
With a suitable decomposition of $\bm{F}$ this is recognized as Joule heating.

On $\mathbb{G}$, the fact that $\bm{E}$ is not expressible solely in terms of the first and second momentum moments means that its governing equation cannot be justified in terms of those previously obtained for $\bm{N}$ and $\bm{T}$.
Instead, a short calculation like that in Eq.~(\ref{eq:StressInertiaSource_G})---but with an extra factor of momentum and a contraction---can be undertaken directly, with the result
\begin{equation}
\bm{\nabla} \cdot \bm{E} = - m \, N \,\bm{V} \cdot \bm{D} \Phi  \ \ \ (\mathrm{on\ }\mathbb{G}), 
\label{eq:EnergyConservation_G}
\end{equation}
featuring a `gravitational power' source term.

\subsection{3+1 Fluid Dynamics}
\label{sec:3_1_FluidDynamics}

To summarize the current section to this point: the previous subsections have motivated three principal geometric objects characterizing a simple fluid on both Minkowski spacetime $\mathbb{M}$ and Galilei/Newton spacetime $\mathbb{G}$---the particle number vector $\bm{N}$, the $(1,1)$ three-momentum tensor $\bm{M}$ (first index down), and the energy vector $\bm{E}$.
These names derive from their defining expressions in Eqs.~(\ref{eq:NumberVector}), (\ref{eq:ThreeMomentumTensor}), and (\ref{eq:EnergyVector}), which exhibit the quantities whose spacetime flux is generated by the flow of particles with four-momenta $\bm{p}$.
Note that separate mention of three-momentum and energy already presupposes some chosen fiducial frame---in this paper, on $\mathbb{M}$ and $\mathbb{G}$, a global inertial frame $\mathscr{O}$---with which are associated a family of observers with constant four-velocity $\bm{w}$, and a foliation of spacetime (whether $\mathbb{M}$ or $\mathbb{G}$) into hyperplanes of simultaneity $\mathcal{S}_t$.

In addition to these spacetime fluxes of quantities defined with respect to fiducial observers, crucial for allowing the equations governing them to form a closed system are the definitions of a family of comoving observers and quantities measured by them.
This begins with a specification of a `fluid velocity.'
Associated with this is a notion of macroscopically small but microscopically large `fluid elements' of a fixed number of particles, which can be regarded as a system to which thermodynamics can be applied in order to provide the necessary closures.

The average flow of particle number defines a fluid velocity vector $\bm{U}$, taken to be the four-velocity of comoving observers, whose reference frame is denoted $\tilde{\mathscr{O}}$.
The fluid four-velocity $\bm{U}$, three-velocity $\bm{V}$, and particle density $N$ are defined by Eqs.~(\ref{eq:FluidVelocity_M})-(\ref{eq:FluidVelocity_G}) and (\ref{eq:NumberDensity_q})-(\ref{eq:FluidThreeVelocity_q}).
(The three-velocity $\bm{V}$ can be regarded as a four-vector on $\mathbb{M}$ or $\mathbb{G}$ that happens to be tangent to $\mathcal{S}_t$.)
Even in coordinate-free notation, these definitions reference the fiducial frame $\mathscr{O}$, in the following sense: they rely on a decomposition, in Eqs.~(\ref{eq:FourMomentumDecomposition_M}), (\ref{eq:FourMomentumDecomposition_G}) and (\ref{eq:ThreeVelocity_M_q})-(\ref{eq:ThreeVelocity_G_q}), of particle four-momentum $\bm{p}$ that separates the portion tangent to $\mathcal{S}_t$ from $\bm{w}$, i.e. the particle three-momentum $\bm{q}$ and associated particle velocity $\bm{v}_{\bm{q}}$. 

The definition of fluid velocity $\bm{U}$ allows quantities measured by comoving observers to be defined. 
In particular, an alternative decomposition of $\bm{p}$ can be given, see Eqs.~(\ref{eq:FourMomentumDecomposition_U_M})-(\ref{eq:FourMomentumDecomposition_U_G}) and (\ref{eq:ThreeVelocity_M_s})-(\ref{eq:ThreeVelocity_G_s}), furnishing the particle three-momentum $\bm{s}$ and velocity $\bm{v}_{\bm{s}}$ according to $\tilde{\mathscr{O}}$.
(Notice that reference to a different set of observers does not require use of components and a change of coordinates; use of coordinate-free notation is fully viable, and even conceptually simplifying, for this purpose.)
Then the particle density $n$ according to $\tilde{\mathscr{O}}$ can be given [Eq.~(\ref{eq:NumberDensity_s})], and the vanishing average velocity in the comoving frame confirmed [Eq.~(\ref{eq:FluidThreeVelocity_s})].
The internal energy density $e$ [Eq.~(\ref{eq:InternalEnergyDensity})] and pressure $p$ [Eqs.~(\ref{eq:PressureTensor_M}), (\ref{eq:PressureDecomposition_M}) and (\ref{eq:PressureTensor_G}), (\ref{eq:PressureDecomposition_G})] 
round out the quantities characterizing a fluid element.
The quantities responsible for dissipation---which are non-zero only if the distribution function $f$ is asymmetric as a function of the comoving frame rectangular components of $\bm{v}_{\bm{s}}$---are the internal energy flux $\bm{J}$ [Eq.~(\ref{eq:InternalEnergyFlux})] and viscous stress $\bm{\Pi}$ [Eqs.~(\ref{eq:PressureTensor_M}), (\ref{eq:PressureDecomposition_M}) and (\ref{eq:PressureTensor_G}), (\ref{eq:PressureDecomposition_G})], representing internal energy and momentum exchange \textit{between} fluid elements.
(Ultimately by definition---and with the caveat that it is not the only possible convention, see e.g. Ref.~\cite{Groot1980Relativistic-Ki}---in the present treatment of a simple fluid there is no net particle number exchange between fluid elements.) 
Again, the assumption of fluid dynamics is that, given some microscopic model, all of these quantities defined in the comoving frame are related by (non-equilibrium, if necessary) thermodynamics.

In the course of defining both the primary geometric objects associated with fiducial observers---the spacetime fluxes $\bm{N}$, $\bm{M}$, and $\bm{E}$---and the fluid velocity characterizing comoving observers and the physical quantities measured by them, relationships between them have also been given in the preceding subsections.
For $\bm{N}$, these include Eqs.~(\ref{eq:NumberDensityComoving_M})-(\ref{eq:ParticleNumberDecomposition}).
For $\bm{M}$, the relevant equations are Eq.~(\ref{eq:MomentumSplit}), followed by Eqs.~(\ref{eq:MomentumPerfect_M})-(\ref{eq:MomentumDissipative_M}) on $\mathbb{M}$, and by Eqs.~(\ref{eq:MomentumPerfect_G})-(\ref{eq:MomentumDissipative_G}) on $\mathbb{G}$.
For $\bm{E}$, the relationships are given in Eq.~(\ref{eq:EnergySplit}), followed by Eqs.~(\ref{eq:EnergyPerfect_M})-(\ref{eq:EnergyDissipative_M}) on $\mathbb{M}$, and by Eqs.~(\ref{eq:EnergyPerfect_G})-(\ref{eq:EnergyDissipative_G}) on $\mathbb{G}$.
For both $\bm{M}$ and $\bm{E}$, it has seemed worthwhile to separate contributions labeled `perfect' (pertaining to quantities characterizing individual fluid elements) and `dissipative' (involving quantities expressing `leakage' between fluid elements)---both to help manage the algebraic complications associated with relating two sets of observers, and because the limit of a perfect fluid is commonly used in practice. 

Having completed this summary view to this point, the next step is to convert the balance equations obeyed by $\bm{N}$, $\bm{M}$, and $\bm{E}$ in Eqs.~(\ref{eq:ParticleConservation}), (\ref{eq:MomentumConservation_M})-(\ref{eq:MomentumConservation_G}), and (\ref{eq:EnergyConservation_M})-(\ref{eq:EnergyConservation_G}), involving spacetime divergences, into $3+1$ equations suitable for initial value problems.
As alluded to above, an important step towards a $3+1$ perspective has already been taken in the separation of momentum and energy embodied in $\bm{M}$ and $\bm{E}$, defined with reference to the fiducial observer four-velocity $\bm{w}$ and hyperplanes of simultaneity $\mathcal{S}_t$; what remains is to similarly decompose the spacetime divergences.
On the affine spacetimes $\mathbb{M}$ and $\mathbb{G}$, and for the inertial fiducial observers $\mathscr{O}$, rectangular coordinates give an instant $3+1$ separation 
\begin{equation}
\left[ \nabla_\mu \right]_\mathscr{O} = \left[ \frac{\partial}{\partial t}, \frac{\partial}{\partial x^i} \right]_\mathscr{O}.
\end{equation}
But in more geometric terms, the time derivative can be expressed (thanks to the constancy of $\bm{w}$) as the Lie derivative $\mathcal{L}_{\bm w} = \bm{w} \cdot \bm{\nabla}$, and the spatial derivative in terms of a covariant derivative $\bm{D}$ on $\mathcal{S}_t$.
These operate on the corresponding elements of a decomposition of the slots of $\bm{N}$, $\bm{M}$, and $\bm{E}$ which are contracted with the spacetime operator $\bm{\nabla}$.

Such a decomposition of $\bm{N}$ is given already in Eq.~(\ref{eq:ParticleNumberDecomposition}), such that Eq.~(\ref{eq:ParticleConservation}) takes the $3+1$ form
\begin{equation}
\mathcal{L}_{\bm w} N + \bm{D} \cdot \bm{F}_N = 0
\label{eq:Number_3_1}
\end{equation}
on both $\mathbb{M}$ and $\mathbb{G}$, where
\begin{equation}
\bm{F}_N = N \bm{V}
\label{eq:Flux_N}
\end{equation}
is the `spatial' flux, i.e. the portion of the spacetime particle flux $\bm{N}$ tangent to $\mathcal{S}_t$, and the fiducial particle density $N$ in $\mathscr{O}$ is given in terms of the comoving density $n$ in $\tilde{\mathscr{O}}$ by Eqs.~(\ref{eq:NumberDensityComoving_M})-(\ref{eq:NumberDensityComoving_G}).

Similar decompositions of $\bm{M}$ and $\bm{E}$ remain to be given.
For this purpose the decompositions of the fluid four-velocity $\bm{U}$ in Eqs.~(\ref{eq:FluidVelocity_M})-(\ref{eq:FluidVelocity_G}) are needed.
Also needed are decompositions of the internal energy flux $\bm{J}$:
\begin{eqnarray}
\bm{J} &=& \frac{1}{c^2} \left( \bm{V} \cdot \bm{J}_{\bm{\gamma}} \right) \bm{w} + \bm{J}_{\overleftarrow{\bm{\gamma}}} \ \ \ (\mathrm{on\ }\mathbb{M}), \\
\bm{J} &=&  \bm{J}_{\overleftarrow{\bm{\gamma}}} \ \ \ (\mathrm{on\ }\mathbb{G}), 
\end{eqnarray}
in which the linear form $\bm{J}_{\bm{\gamma}}$ defined in Eq.~(\ref{eq:J_Gamma}) and the vector 
\begin{equation}
\bm{J}_{\overleftarrow{\bm{\gamma}}} = \bm{J} \cdot \overleftarrow{\bm{\gamma}}
\end{equation}
are both tangent to $\mathcal{S}_t$.
Similarly, 
\begin{eqnarray}
\bm{\Pi}_{\bm{\gamma}} &=& \frac{1}{c^2} \left( \bm{V} \cdot \bm{\Pi}_{\bm{\gamma}\bm{\gamma}} \right) \otimes \bm{w} + \bm{\Pi}_{\bm{\gamma}\overleftarrow{\bm{\gamma}}} \ \ \ (\mathrm{on\ }\mathbb{M}), \\
\bm{\Pi}_{\bm{\gamma}} &=&  \bm{\Pi}_{\bm{\gamma}\overleftarrow{\bm{\gamma}}} \ \ \ (\mathrm{on\ }\mathbb{G})
\end{eqnarray}
where
\begin{eqnarray}
\bm{\Pi}_{\bm{\gamma}\bm{\gamma}} &=& \bm{\Pi}_{\bm{\gamma}} \cdot \bm{\gamma}, \\
\bm{\Pi}_{\bm{\gamma}\overleftarrow{\bm{\gamma}}} &=& \bm{\Pi}_{\bm{\gamma}} \cdot \overleftarrow{\bm{\gamma}}
\end{eqnarray}
further decompose the viscous stress $\bm{\Pi}$, already partially decomposed as $\bm{\Pi}_{\bm{\gamma}}$ in Eq.~(\ref{eq:Pi_Gamma}).
(The decompositions on $\mathbb{M}$ utilize the fact that $\bm{J}$ and $\bm{\Pi}$ are orthogonal to $\bm{U}$).

These decompositions allow the three-momentum Eqs.~(\ref{eq:MomentumConservation_M})-(\ref{eq:MomentumConservation_G}) to take the $3+1$ form
\begin{eqnarray}
\mathcal{L}_{\bm w} \bm{S} + \bm{D} \cdot \bm{F}_{\bm{S}} &=& e \bm{F}_{\bm{\gamma}} \cdot \bm{N} \ \ \ (\mathrm{on\ }\mathbb{M}), 
\label{eq:Momentum_3_1_M} \\
\mathcal{L}_{\bm w} \bm{S} + \bm{D} \cdot \bm{F}_{\bm{S}} &=&  - m \, N \,  \bm{D} \Phi \ \ \ (\mathrm{on\ }\mathbb{G}).
\label{eq:Momentum_3_1_G}
\end{eqnarray}
Here the momentum density $\bm{S}$ is a linear form that can be expressed as
\begin{equation}
\bm{S} = \bm{S}_\mathrm{perfect} +  \bm{S}_\mathrm{dissipative},
\label{eq:Density_S}
\end{equation}
with a corresponding breakdown of the spatial flux $\bm{F}_{\bm{S}}$.
In particular,
\begin{eqnarray}
\bm{S}_\mathrm{perfect} &=& \Lambda_{\bm{V}}^2 \left( m\, n + \frac{e + p}{c^2} \right) \underline{\bm{V}} \ \ \ (\mathrm{on\ }\mathbb{M}), \label{eq:Density_S_P_M} \\
\bm{S}_\mathrm{perfect} &=& m\, n \, \underline{\bm{V}} \ \ \ (\mathrm{on\ }\mathbb{G}),
\label{eq:Density_S_P_G}
\end{eqnarray}
and
\begin{eqnarray}
\bm{S}_\mathrm{dissipative} &=& \frac{1}{c^2} \left[ \left(\bm{V}\cdot \bm{\Pi}_{\bm{\gamma}\bm{\gamma}} \right) + \Lambda_{\bm{V}} \bm{J}_{\bm{\gamma}} \right] \; + \; \frac{\Lambda_{\bm{V}}}{c^4}  \left( \bm{V} \cdot \bm{J}_{\bm{\gamma}} \right) \underline{\bm{V}} \ \ \ (\mathrm{on\ }\mathbb{M}), 
\label{eq:Density_S_D_M}  \\
\bm{S}_\mathrm{dissipative} &=& \bm{0} \ \ \ (\mathrm{on\ }\mathbb{G}). \label{eq:Density_S_D_G}
\end{eqnarray}
Meanwhile 
\begin{equation}
\bm{F}_{\bm{S}_\mathrm{perfect}} = \bm{S}_\mathrm{perfect} \bm{V} + p \overleftarrow{\bm{\gamma}}
\label{eq:Flux_S_P}
\end{equation}
is the perfect flux on both $\mathbb{M}$ and $\mathbb{G}$, and
\begin{eqnarray}
\bm{F}_{\bm{S}_\mathrm{dissipative}} &=& \frac{1}{c^2} \left[ \Lambda_{\bm{V}} \underline{\bm{V}} \otimes \bm{J}_{\overleftarrow{\bm{\gamma}}} + \Lambda_{\bm{V}} \bm{J}_{\bm{\gamma}} \otimes \bm{V} \right] \;+ \; \bm{\Pi}_{\bm{\gamma}\overleftarrow{\bm{\gamma}}} \ \ \ (\mathrm{on\ }\mathbb{M}), 
\label{eq:Flux_S_D_M} \\
\bm{F}_{\bm{S}_\mathrm{dissipative}} &=& \bm{\Pi}_{\bm{\gamma}\overleftarrow{\bm{\gamma}}} \ \ \ (\mathrm{on\ }\mathbb{G})
\label{eq:Flux_S_D_G}
\end{eqnarray}
are the dissipative fluxes.

The above-referenced decompositions also allow the energy Eqs.~(\ref{eq:EnergyConservation_M})-(\ref{eq:EnergyConservation_G}) to take the $3+1$ form
\begin{eqnarray}
\mathcal{L}_{\bm w} E + \bm{D} \cdot \bm{F}_{E} &=& e\, N \,\bm{V} \cdot \bm{F} \cdot \bm{w}  \ \ \ (\mathrm{on\ }\mathbb{M}), 
\label{eq:Energy_3_1_M} \\
\mathcal{L}_{\bm w} E + \bm{D} \cdot \bm{F}_{E} &=&  - m \, N \,\bm{V} \cdot \bm{D} \Phi \ \ \ (\mathrm{on\ }\mathbb{G}).
\label{eq:Energy_3_1_G}
\end{eqnarray}
Here the energy density $E$ can be expressed as
\begin{equation}
E = E_\mathrm{perfect} + E_\mathrm{dissipative},
\label{eq:Density_E}
\end{equation}
with a corresponding breakdown of the spatial flux $\bm{F}_{E}$.
In particular,
\begin{eqnarray}
E_\mathrm{perfect} &=& \Lambda_{\bm{V}} \left[ \left( \Lambda_{\bm{V}} - 1 \right) mc^2 \, n  + \Lambda_{\bm V} \left( e + p \right) \right] \; - \; p  \ \ \ (\mathrm{on\ }\mathbb{M}), 
\label{eq:Density_E_P_M}\\
E_\mathrm{perfect} &=&   \frac{1}{2}m \,n \, \bm{\gamma}(\bm{V},\bm{V}) + e \ \ \ (\mathrm{on\ }\mathbb{G}),
\label{eq:Density_E_P_G}
\end{eqnarray}
and
\begin{eqnarray}
E_\mathrm{dissipative} &=& \frac{2 \Lambda_{\bm{V}} }{c^2} \left( \bm{V} \cdot \bm{J}_{\bm{\gamma}}\right) \; + \; \frac{1}{c^2} \bm{V} \cdot \bm{\Pi}_{\bm{\gamma}\bm{\gamma}} \cdot \bm{V} \ \ \ (\mathrm{on\ }\mathbb{M}), 
\label{eq:Density_E_D_M}\\
E_\mathrm{dissipative} &=& 0 \ \ \ (\mathrm{on\ }\mathbb{G}).
\label{eq:Density_E_D_G}
\end{eqnarray}
Meanwhile 
\begin{equation}
\bm{F}_{E_\mathrm{perfect}} = \left( E_\mathrm{perfect} + p \right) \bm{V}
\label{eq:Flux_E_P}
\end{equation}
is the perfect flux on both $\mathbb{M}$ and $\mathbb{G}$, and
\begin{eqnarray}
\bm{F}_{E_\mathrm{dissipative}} &=& \Lambda_{\bm{V}} \bm{J}_{\bm{\gamma}} + \bm{V} \cdot \bm{\Pi}_{\bm{\gamma}\overleftarrow{\bm{\gamma}}} \; + \; \frac{\Lambda_{\bm{V}}}{c^2} \left( \bm{V} \cdot \bm{J}_{\bm{\gamma}} \right) \bm{V} \ \ \ (\mathrm{on\ }\mathbb{M}), 
\label{eq:Flux_E_D_M}\\
\bm{F}_{E_\mathrm{dissipative}} &=& \bm{J}_{\bm{\gamma}} + \bm{V} \cdot \bm{\Pi}_{\bm{\gamma}\overleftarrow{\bm{\gamma}}} \ \ \ (\mathrm{on\ }\mathbb{G})
\label{eq:Flux_E_D_G}
\end{eqnarray}
are the dissipative fluxes.

In the dissipative densities and fluxes there are a number of terms on $\mathbb{M}$ that vanish when $c\rightarrow \infty$, resulting in expressions that reduce (as expected) to those obtained on $\mathbb{G}$. 
Such terms are typically noted and labeled as `inherently relativistic' dissipative effects. 
The present discussion clarifies their origin: comoving frame quantities are required to close the system via thermodynamics, and the terms in question ultimately arise due to the mismatch in hyperplanes of simultaneity between the fiducial and comoving frames on $\mathbb{M}$. 

\section{Conclusion}
\label{sec:Conclusion}

Taking fluid dynamics on Minkowski spacetime as a reference with which comparisons might be made and against which contrasts can be drawn, the purpose of this paper is to consider Galilei/Newton fluid dynamics in geometric terms from a spacetime perspective.
In a different spirit than the remarks in the concluding section of Ref.~\cite{Cardall2017Relativistic-an}, which emphasize distinct principles that might be taken to underlie non-relativistic and relativistic fluid dynamics, the goal here has been to start with a common framework---a generic kinetic theory of classical particles---in order to explore the extent to which concepts and geometric objects pertaining to fluid dynamics can be developed in parallel on both Minkowski spacetime $\mathbb{M}$ and Galilei/Newton spacetime $\mathbb{G}$.

\begin{table}
\caption{Geometric entities characterizing $3+1$ Minkowski and Galilei/Newton spacetimes.}
\centering
\begin{tabular}{lll}
\toprule
\textbf{Symbol}	& \textbf{Description}	& \textbf{References}\\
\midrule
$\mathcal{M}$		& Generic spacetime 			& Sec.~\ref{sec:GlobalInertialFrames}	\\
				& \qquad (four-dimensional manifold) 			&					\\
$\mathbb{M}$		& Minkowski spacetime			& Secs.~\ref{sec:CausalStructure}, 															\ref{sec:MinkowskiSpacetime} \\
				& \qquad (affine; pseudo-Riemann metric embodying 	&				\\
				 &\qquad \qquad absolute light cones)	&							\\
$\mathbb{G}$		& Galilei/Newton spacetime		& Secs.~\ref{sec:CausalStructure}, 															\ref{sec:GalileiNewtonSpacetime} \\
				& \qquad (affine; fiber bundle embodying absolute time)			&		\\
$\mathscr{O}$		& Fiducial global inertial reference frame		& 																		Sec.~\ref{sec:GlobalInertialFrames} \\
$\left( x^\mu \right)$	& Coordinates on $\mathbb{M}$, $\mathbb{G}$ according to $\mathscr{O}$ 	& 												Sec.~\ref{sec:GlobalInertialFrames}	\\		
				& \qquad (global inertial spacetime coordinates)	&					\\
$t$				& Time coordinate $x^0$ according to $\mathscr{O}$	&
						 Sec.~\ref{sec:GlobalInertialFrames}	\\
				& \qquad (relative on $\mathbb{M}$, absolute on $\mathbb{G}$) &			\\
$\mathcal{S}_t$	& Hypersurfaces of simultaneity	& Sec.~\ref{sec:GlobalInertialFrames}	\\
				& \qquad (level surfaces of $t$)		&								\\
				& \qquad (Euclid `position space' on $\mathbb{M}$ and $\mathbb{G}$)	&	\\
				& \qquad (relative on $\mathbb{M}$, absolute on $\mathbb{G}$)	&		\\
$( x^i )$			& Coordinates on $\mathcal{S}_t$ according to $\mathscr{O}$ 	& 															Sec.~\ref{sec:GlobalInertialFrames}	\\		
				& \qquad (rectangular position space coordinates)	&					\\
$\bm{t}$			& Time one-form				& Eqs.~(\ref{eq:TimeForm_0}), 																(\ref{eq:TimeForm}) 				\\
				& \qquad (relative on $\mathbb{M}$, absolute on $\mathbb{G}$) &			\\
				& \qquad (Invariant under 	Galilei transformations on $\mathbb{G}$) &		\\
$\bm{w}$			& Four-velocity vector of $\mathscr{O}$ observers	& 					
											Sec.~\ref{sec:GlobalInertialFrames}; 	\\					&	&				Eq.~(\ref{eq:O_Worldlines_M}) on $\mathbb{M}$,	\\
				&	&				Eq.~(\ref{eq:O_Worldlines_G}) on $\mathbb{G}$ 	\\
$\bm{\gamma}$, $\Overrightarrow{\bm{\gamma}}$	& Metric and inverse on $\mathcal{S}_t$		& 
											Eq.~(\ref{eq:ThreeMetric})			\\
				& \qquad (flat on $\mathbb{M}$ and $\mathbb{G}$)	&					\\
$\bm	{g}$, $\Overrightarrow{\bm{g}}$		& Metric and inverse on $\mathbb{M}$		
					& Eqs.~(\ref{eq:MinkowskiMetric})-(\ref{eq:MinkowskiInverseMetric})		\\
				& \qquad (Invariant under Lorentz transformations)	&					\\
$\bm{\delta}$		& Identity tensor on $\mathbb{M}$ and $\mathbb{G}$ 	&				\\
				& \qquad ( $(1,1)$; Kronecker delta)		&								\\
$\bm{\gamma}$, $\overrightarrow{\bm{\gamma}}$, $\overleftarrow{\bm{\gamma}}$, 
$\Overrightarrow{\bm{\gamma}}$	& Projection tensors to $\mathcal{S}_t$ 
	on $\mathbb{M}$ and $\mathbb{G}$		&		
	Eqs.~(\ref{eq:Projection_0_2_M})-(\ref{eq:Projection_2_0_M_Components}) on $\mathbb{M}$, \\
				& \qquad ( $(0,2)$, $(1,1)$, $(1,1)$, $(2,0)$ )	&
	Eqs.~(\ref{eq:Projection_0_2_G_Components})-(\ref{eq:Projection_2_0_G_Components}) on $\mathbb{G}$ \\	
				& \qquad ($\Overrightarrow{\bm{\gamma}}$ invariant under 
						Galilei transformations on $\mathbb{G}$) &		\\
$\tilde{\mathscr{O}}$		& Accelerated reference frame	& Sec.~\ref{sec:KineticTheory}		\\
					& \quad (Often associated with the fluid) 	& Sec.~\ref{sec:ParticleNumber} \\
$\bm{U}$		& Four-velocity vector of $\tilde{\mathscr{O}}$ observers	& 
									Sec.~\ref{sec:KineticTheory}		\\
					& \quad (Often the four-velocity of the fluid) 	& 
									Sec.~\ref{sec:ParticleNumber} \\
$\bm{h}$		& Projector orthogonal to $\bm{U}$ on $\mathbb{M}$	&
				 Eq.~(\ref{eq:Projection_0_2_U}) on $\mathbb{M}$					\\
			& \quad ( $\Overrightarrow{\bm{\gamma}}$ on $\mathbb{G}$ plays a role 
					analogous to $\Overrightarrow{\bm{h}}$ on $\mathbb{M}$)		\\
\bottomrule
\end{tabular}
\label{tab:SpacetimeEntities}
\end{table}

Comparison and contrast begin with the spacetimes themselves, discussed in Sec.~\ref{sec:Spacetime}.
See Table~\ref{tab:SpacetimeEntities}.
Both $\mathbb{M}$ and $\mathbb{G}$ are four-dimensional differentiable manifolds;
more particularly, they are both affine spaces admitting global inertial reference frames.
Selection of a particular inertial frame $\mathscr{O}$ as `fiducial' provides an associated family of observers with constant four-velocity $\bm{w}$ and hyperplanes of simultaneity $\mathcal{S}_t$.
Minkowski spacetime $\mathbb{M}$ is a pseudo-Riemann manifold with spacetime metric $\bm{g}$ embodying an invariant null cone structure, with the pull-back $\bm{\gamma}$ of $\bm{g}$ onto each `position space' hyperplane $\mathcal{S}_t$ constituting the flat three-metric thereon. 
In contrast, $\mathbb{G}$ has no spacetime metric; it is a fiber bundle with a one-dimensional base manifold, `time,' and `position space' fibers constituting invariant hyperplanes of simultaneity $\mathcal{S}_t$, upon which $\bm{\gamma}$ is the flat three-metric.
While a pseudo-Riemann manifold and a fiber bundle are qualitatively different, the correspondence between the invariant structures of $\mathbb{M}$ and $\mathbb{G}$ can be pictured in a spacetime diagram as the limit $c \rightarrow \infty$ pressing a null cone of $\mathbb{M}$ down onto a fiducial observer position space hyperplane (visualized as the horizontal axis), squeezing the distinct hyperplanes of simultaneity of inertial observers in relative motion in $\mathbb{M}$ into an absolute hyperplane of simultaneity of $\mathbb{G}$.
Moreover, the (push-forward of the) inverse three-metric $\Overrightarrow{\bm{\gamma}}$ on $\mathbb{G}$ is the $c \rightarrow \infty$ limit of the inverse spacetime metric $\Overrightarrow{\bm{g}}$ on $\mathbb{M}$, with the Galilei invariance of the components of $\Overrightarrow{\bm{\gamma}}$ recognized as a geometric remnant of the Lorentz invariance of the components of $\Overrightarrow{\bm{g}}$.

A classical particle worldline is a curve on spacetime, whose suitably parametrized tangent vector is the particle four-momentum $\bm{p}$.
See Table~\ref{tab:ParticleEntities}.
The zeroth component $p^0$ is the particle inertia; on $\mathbb{M}$ the inertia $\epsilon_{\bm{q}} / c^2$ depends on the particle three-momentum $\bm{q}$, while on $\mathbb{G}$ the inertia is simply the particle rest mass $m$---an absolute inertia resulting from absolute simultaneity.
Because kinetic energy $e_{\bm{q}}$ contributes to inertia on $\mathbb{M}$, it can readily be extracted from $\bm{p}$.
In contrast, the absolute nature of inertia on $\mathbb{G}$ means that inertia and energy are not linked, such that the notion of kinetic energy must be introduced by hand as a separate concept.
However, once defined on $\mathbb{G}$ in such a way that it is the $c \rightarrow \infty$ limit of $e_{\bm{q}}$ on $\mathbb{M}$, the kinetic energy turns out to obey the same geometric equation on both $\mathbb{M}$ and $\mathbb{G}$ [see Eqs.~(\ref{eq:Power_M}) and (\ref{eq:Power_G})].

\begin{table}
\caption{Geometric entities characterizing particles on Minkowski and Galilei/Newton spacetimes.}
\centering
\begin{tabular}{lll}
\toprule
\textbf{Symbol}	& \textbf{Description}	& \textbf{References}\\
\midrule
$\lambda$				& Affine parameter 			& Sec.~\ref{sec:GlobalInertialFrames}	\\
					& \quad (Units of time/mass)	&								\\
$\mathcal{X}(\lambda)$	& Worldline				& Sec.~\ref{sec:GlobalInertialFrames}	\\
$\bm{p}$				& Particle four-momentum vector 		& 
					Eqs.~(\ref{eq:FourMomentum}), (\ref{eq:FourMomentumCoordinates})	\\
$m$					& Particle rest mass					& Sec.~\ref{sec:GlobalInertialFrames}	\\
$\tau$				& Proper time measured by particle	& Secs.~\ref{sec:GlobalInertialFrames},
						\ref{sec:MinkowskiSpacetime}, \ref{sec:GalileiNewtonSpacetime}	\\
					& \quad ($\mathrm{d}\tau = m\, \mathrm{d}\lambda$)  	& 			\\
$\bm{u}$				& Particle four-velocity vector	& 
						Eqs.~(\ref{eq:FourVelocity_0}), (\ref{eq:FourVelocity}),			\\
					& 	&  Eqs.~(\ref{eq:FourVelocity_M}), 															(\ref{eq:FourVelocityDecomposition_M}) on $\mathbb{M}$,	\\
					& 	&  Eqs.~(\ref{eq:FourVelocity_G}), 															(\ref{eq:FourVelocityDecomposition_G}) on $\mathbb{G}$		\\
$\bm{v}$, $\bm{v}_{\bm{q}}$	& Particle three-velocity vector measured by $\mathscr{O}$ & 
						Eqs.~(\ref{eq:FourVelocity})-(\ref{eq:ThreeVelocity})			\\
					& \quad (Tangent to $\mathcal{S}_t$)	&
						Eq.~(\ref{eq:ThreeVelocity_M_q}) on $\mathbb{M}$,			\\
					&	& Eq.~(\ref{eq:ThreeVelocity_G_q}) on $\mathbb{G}$		\\
$\Lambda_{\bm{v}}$		& Particle Lorentz factor on $\mathbb{M}$ measured by $\mathscr{O}$	&
								Eq.~(\ref{eq:LorentzFactor})						\\
$\bm{q}$				& Particle three-momentum vector measured by $\mathscr{O}$ & 
						Eqs.~(\ref{eq:FourMomentum_M}), (\ref{eq:ThreeMomentum_M}),
						(\ref{eq:FourMomentumDecomposition_M}) on $\mathbb{M}$,		\\	
					& \quad (Tangent to $\mathcal{S}_t$)	&						
						Eqs.~(\ref{eq:FourMomentum_G}), (\ref{eq:ThreeMomentum_G}),
						(\ref{eq:FourMomentumDecomposition_G}) on $\mathbb{G}$		\\
$\epsilon_{\bm{q}}$		& Particle total energy on $\mathbb{M}$ measured by $\mathscr{O}$
						& Eq.~(\ref{eq:FourMomentum_M})							\\
$e_{\bm{q}}$			& Particle kinetic energy measured by $\mathscr{O}$
						& Eq.~(\ref{eq:KineticEnergy_M}) on $\mathbb{M}$,				\\
					&	& Eq.~(\ref{eq:KineticEnergy_G}) on $\mathbb{G}$				\\
$\bm{s}$				& Particle three-momentum vector measured by $\tilde{\mathscr{O}}$ & 
						Eq.~(\ref{eq:FourMomentumDecomposition_U_M}) on $\mathbb{M}$, \\	
					&	& Eq.~(\ref{eq:FourMomentumDecomposition_U_G}) on $\mathbb{G}$ \\	
$\epsilon_{\bm{s}}$		& Particle total energy on $\mathbb{M}$ measured by $\tilde{\mathscr{O}}$
						& Eq.~(\ref{eq:FourMomentumDecomposition_U_M})			\\
$\bm{v}_{\bm{s}}$	& Particle three-velocity vector measured by $\tilde{\mathscr{O}}$ & 
						Eq.~(\ref{eq:ThreeVelocity_M_s}) on $\mathbb{M}$,			\\
					&	& Eq.~(\ref{eq:ThreeVelocity_G_s}) on $\mathbb{G}$		\\
$e_{\bm{s}}$			& Particle kinetic energy measured by $\tilde{\mathscr{O}}$
						& Sec.~(\ref{sec:StressInertia}) on $\mathbb{M}$,			\\
					&	& Sec.~(\ref{sec:EnergyVector}) on $\mathbb{G}$				\\
\bottomrule
\end{tabular}
\label{tab:ParticleEntities}
\end{table}

A kinetic theory of classical particles motivates certain geometric objects pertaining to fluid dynamics, as summarized more fully in Sec.~\ref{sec:3_1_FluidDynamics}.
See also Table~\ref{tab:FluidEntities}.
Two types of reference frames must be invoked.
First, a fiducial frame $\mathscr{O}$ is necessary to cover an extended region of spacetime with a single coordinate system.
(On $\mathbb{M}$ and $\mathbb{G}$, as addressed in this paper, the fiducial frame is a global inertial one in which the paradigmatic forces of electromagnetism and gravity can be defined respectively.)  
Second, a comoving frame $\tilde{\mathscr{O}}$ is also necessary, in order to define quantities related by thermodynamics, so that a closed system of equations can be obtained.
Defining and working with quantities related to these two frames does not require use of components and changes of coordinates; use of coordinate-free geometric notation is fully viable, and even conceptually simplifying, for this purpose.

\begin{table}
\caption{Geometric entities characterizing a simple fluid on Minkowski and Galilei/Newton spacetimes.}
\centering
\begin{tabular}{lll}
\toprule
\textbf{Symbol}	& \textbf{Description}	& \textbf{References}\\
\midrule
$\bm{t}\, \mathrm{dV}_t$	& Position space volume element associated with $\mathcal{S}_t$	& 												Sec.~(\ref{sec:KineticTheory})	\\
$\mathrm{dP}_m$		& Invariant three-momentum volume element &
					 			Eqs.~(\ref{eq:MomentumElement_M_q}) and 												(\ref{eq:MomentumElement_M_s}) on $\mathbb{M}$, 	\\
				&	&	Eqs.~(\ref{eq:MomentumElement_G_q}) and 													(\ref{eq:MomentumElement_G_s}) on $\mathbb{G}$			\\
$f$					& Particle distribution function		& Eq.~(\ref{eq:ParticleDistribution}) 	\\
$\bm{N}$				& Particle number four-vector 		& Eq.~(\ref{eq:NumberVector})		\\
					& \quad (First momentum moment of $f$)		&					\\
$\bm{T}$				& Stress-inertia tensor			& Eq.~(\ref{eq:StressInertiaTensor}) 	\\
					& \quad ( $(2,0)$ Second momentum moment of $f$)	&			\\
$\bm{M}$				& Three-momentum tensor		& Eq.~(\ref{eq:ThreeMomentumTensor})\\
					& \quad ( $(1,1)$ Spatial projection of $\bm{T}$)	&
						Eq.~(\ref{eq:MomentumTensor})			\\
$\bm{E}$				& Energy vector four-vector	 	& Eq.~(\ref{eq:EnergyVector})		\\
					& \quad (Related to $\bm{T}$ and $\bm{N}$ on $\mathbb{M}$)	
						& Eq.~(\ref{eq:EnergyVector_M}) on $\mathbb{M}$		\\
					& \quad (Not simply related to $\bm{T}$ on $\mathbb{G}$)	
						& Eq.~(\ref{eq:EnergyVector_G_1}) on $\mathbb{G}$		\\
$\bm{U}$				& Fluid four-velocity vector		& 
						Eqs.~(\ref{eq:FluidVelocity_M})-(\ref{eq:FluidVelocity_G}),			\\
					& \quad (Obtained together with $\bm{V}$ and $N$)	&
						Eqs.~(\ref{eq:NumberDensity_q})-(\ref{eq:FluidThreeVelocity_q})	\\
$\bm{V}$				& Fluid three-velocity vector measured by $\mathscr{O}$		& 
						Eqs.~(\ref{eq:FluidVelocity_M})-(\ref{eq:FluidVelocity_G}),			\\
					& \quad (Obtained together with $\bm{U}$ and $N$)	&
						Eqs.~(\ref{eq:NumberDensity_q})-(\ref{eq:FluidThreeVelocity_q})	\\
					& \quad (Tangent to $\mathcal{S}_t$)			&				\\
$N$					& Particle number density measured by $\mathscr{O}$		& 
						Eqs.~(\ref{eq:FluidVelocity_M})-(\ref{eq:FluidVelocity_G}),			\\
					& \quad (Obtained together with $\bm{U}$ and $\bm{V}$)	&
						Eqs.~(\ref{eq:NumberDensity_q})-(\ref{eq:FluidThreeVelocity_q})	\\
$n$					& Particle number density measured by $\tilde{\mathscr{O}}$		& 
						Eq.~(\ref{eq:NumberDensity_s})			\\
$e$					& Internal energy density measured by $\tilde{\mathscr{O}}$		& 
						Eq.~(\ref{eq:InternalEnergyDensity})			\\
$p$					& Pressure measured by $\tilde{\mathscr{O}}$		& 
		Eqs.~(\ref{eq:PressureTensor_M}), (\ref{eq:PressureDecomposition_M}) on $\mathbb{M}$ \\
& &		Eqs.~(\ref{eq:PressureTensor_G}), (\ref{eq:PressureDecomposition_G}) on $\mathbb{G}$ \\
$\bm{J}$				& Internal energy flux vector measured by $\tilde{\mathscr{O}}$	& Eq.~(\ref{eq:InternalEnergyFlux})	\\
$\bm{\Pi}$				& Viscous stress tensor measured by $\tilde{\mathscr{O}}$		& 
		Eqs.~(\ref{eq:PressureTensor_M}), (\ref{eq:PressureDecomposition_M}) on $\mathbb{M}$ \\
	& \quad ( $(2,0)$ ) & Eqs.~(\ref{eq:PressureTensor_G}), (\ref{eq:PressureDecomposition_G})
		on $\mathbb{G}$ \\
\bottomrule
\end{tabular}
\label{tab:FluidEntities}
\end{table}

A geometric spacetime perspective capable of embracing both Minkowski and Galilei/Newton fluid dynamics on a somewhat common footing must of necessity be at least partly a $3+1$ formulation, as opposed to being fully four-dimensional.
This is not because of the particle number four-vector $\bm{N}$---the first $\bm{p}$ moment of the particle distribution function $f$---which is of essentially the same character on $\mathbb{M}$ and $\mathbb{G}$, as discussed in Sec.~\ref{sec:ParticleNumber}.
The important difference appears in connection with the second $\bm{p}$ moment of $f$, the tensor $\bm{T}$ addressed in Sec.~\ref{sec:StressInertia}.
This is normally known as the stress-energy (or energy-momentum) tensor, but the present comparison with the Galilei/Newton case brings into sharp focus the realization that it is more properly understood as a stress-\textit{inertia} (or \textit{inertia}-momentum) tensor.
This has some impact on the nature of the $(1,1)$ three-momentum tensor $\bm{M}$, obtained in Sec.~\ref{sec:ThreeMomentumTensor} on both $\mathbb{M}$ and $\mathbb{G}$ as the projection of the first slot of $\bm{T}$ onto the hyperplanes of simultaneity $\mathcal{S}_t$. 
But the much larger difference is in connection with an energy four-vector $\bm{E}$ introduced in Sec.~\ref{sec:EnergyVector}, due to the conceptual difference in particle kinetic energy $e_{\bm{q}}$ discussed in the paragraph before last: on $\mathbb{M}$, because kinetic energy contributes to inertia, the vector $\bm{E}$ can readily be obtained in terms of the projection of $\bm{T}$ along the inertial observer four-velocity $\bm{w}$; but on $\mathbb{G}$, the absolute nature of inertia implies that a comparable projection of $\bm{T}$ (via the time one-form $\bm{t}$) would be effectively redundant with the particle number vector $\bm{N}$.
Thus there is no simple relation between $\bm{E}$ and $\bm{T}$ on $\mathbb{G}$, despite the fact that $\bm{E}$ on $\mathbb{G}$ turns out to be the $c \rightarrow \infty$ limit of $\bm{E}$ on $\mathbb{M}$.
 
While this \textit{ab initio} distinction between three-momentum and energy is necessary to accommodate the Galilei/Newton case, the geometric objects $\bm{N}$, $\bm{M}$, and $\bm{E}$---the spacetime fluxes of particle number, three-momentum, and energy---nevertheless obey balance equations involving spacetime divergences on both $\mathbb{M}$ and $\mathbb{G}$:
\begin{eqnarray}
\bm{\nabla} \cdot \bm{N} &=& 0, \label{eq:Divergence_4_N} \\
\bm{\nabla} \cdot \bm{M} &=& \bm{Q}_{\bm{S}}, \label{eq:Divergence_4_S}\\
\bm{\nabla} \cdot \bm{E} &=& \bm{Q}_{E}, \label{eq:Divergence_4_E}
\end{eqnarray} 
where the sources $\bm{Q}_{\bm{S}}$ and $\bm{Q}_{E}$ are for instance those given in Secs.~\ref{sec:ThreeMomentumTensor} and \ref{sec:EnergyVector} for electromagnetism on $\mathbb{M}$ and gravity on $\mathbb{G}$.

However, a fully $3+1$ formulation is desirable for practical solution of initial value problems in any case, and this also can be expressed in a coordinate-free geometric way, as given in Sec.~\ref{sec:3_1_FluidDynamics}. 
See also Table~\ref{tab:DensityFlux}.
On the affine spacetimes $\mathbb{M}$ and $\mathbb{G}$,
\begin{eqnarray}
\mathcal{L}_{\bm w} N + \bm{D} \cdot \bm{F}_N &=& 0, \label{eq:Divergence_3_1_N} \\
\mathcal{L}_{\bm w} \bm{S} + \bm{D} \cdot \bm{F}_{\bm{S}} &=& \bm{Q}_{\bm{S}}, \label{eq:Divergence_3_1_S}\\
\mathcal{L}_{\bm w} E + \bm{D} \cdot \bm{F}_{E} &=& \bm{Q}_{E}, \label{eq:Divergence_3_1_E}
\end{eqnarray}
where $N$, $\bm{S}$, and $E$ are are volume densities of particle number, three-momentum, and internal-plus-kinetic energy measured by the fiducial observers of $\mathscr{O}$, with $\bm{F}_{N}$, $\bm{F}_{\bm{S}}$, $\bm{F}_{E}$ being the corresponding spatial fluxes; and $\mathcal{L}_{\bm w}$ and $\bm{D}$ are respectively the Lie derivative along the fiducial observer four-velocity $\bm{w}$ and the three-covariant derivative on the hyperplanes of simultaneity $\mathcal{S}_t$.
In coordinates associated with the fiducial (and, on $\mathbb{M}$ and $\mathbb{G}$, global inertial) observers, these equations reduce immediately to the familiar conservative formulations of special relativistic and non-relativistic hydrodynamics 
(see Appendix~\ref{sec:PartialDifferentialEquations}).
And in this geometric form, they will provide a useful conceptual bridge to arbitrary-Lagrange-Euler and general relativistic formulations.

\begin{table}
\caption{Densities and fluxes measured in the fiducial frame $\mathscr{O}$ in terms of the fluid three-velocity $\bm{V}$ measured by $\mathscr{O}$ and quantities determined by thermodynamics and constitutive relations in the comoving frame $\tilde{\mathscr{O}}$.}
\centering
\begin{tabular}{lll}
\toprule
\textbf{Symbol}	& \textbf{Description}	& \textbf{References}\\
\midrule
$N$		& Particle number density	& Eq.~(\ref{eq:NumberDensityComoving_M})	on $\mathbb{M}$, \\
	&	& Eq.~(\ref{eq:NumberDensityComoving_G})	on $\mathbb{G}$	\\
$\bm{S}$	& Three-momentum density	& Eq.~(\ref{eq:Density_S}),	\\
	&	& 	Eqs.~(\ref{eq:Density_S_P_M}), (\ref{eq:Density_S_D_M}) on $\mathbb{M}$,	\\
	&	& 	Eqs.~(\ref{eq:Density_S_P_G}), (\ref{eq:Density_S_D_G}) on $\mathbb{G}$	\\
$E$	& Internal/kinetic energy density	& Eq.~(\ref{eq:Density_E}),	\\
	&	& 	Eqs.~(\ref{eq:Density_E_P_M}), (\ref{eq:Density_E_D_M}) on $\mathbb{M}$,	\\
	&	& 	Eqs.~(\ref{eq:Density_E_P_G}), (\ref{eq:Density_E_D_G}) on $\mathbb{G}$	\\
$\bm{F}_N$ 	& Particle number flux    	& Eq.~(\ref{eq:Flux_N})	\\
$\bm{F}_{\bm{S}}$	& Three-momentum flux	&	Eq.~(\ref{eq:Flux_S_P}), \\
	&	&	Eq.~(\ref{eq:Flux_S_D_M}) on $\mathbb{M}$, \\	
	&	&	Eq.~(\ref{eq:Flux_S_D_G}) on $\mathbb{G}$ \\	
$\bm{F}_{E}$	& Internal/kinetic energy flux	&	Eq.~(\ref{eq:Flux_E_P}), \\
	&	&	Eq.~(\ref{eq:Flux_E_D_M}) on $\mathbb{M}$, \\	
	&	&	Eq.~(\ref{eq:Flux_E_D_G}) on $\mathbb{G}$ \\	
\bottomrule
\end{tabular}
\label{tab:DensityFlux}
\end{table}

\vspace{6pt} 

\funding{This work was supported by the U.S. Department of Energy, Office of Science, Office of Nuclear Physics under contract number DE-AC05-00OR22725.}


\conflictsofinterest{The author declares no conflict of interest. The funders had no role in the design of the study; in the collection, analyses, or interpretation of data; in the writing of the manuscript, or in the decision to publish the results.}

%

%

\appendixtitles{yes}
\appendixsections{multiple} 
\appendix

\section{Transformation properties of the projector in Galilei/Newton spacetime}
\label{sec:TransformationProperties}

In Galilei/Newton spacetime, consider the manner in which the tensors $\bm{\gamma}$, $\overrightarrow{\bm{\gamma}}$, $\overleftarrow{\bm{\gamma}}$, and $\Overrightarrow{\bm{\gamma}}$---discussed in the paragraph containing Eq.~(\ref{eq:Projection_0_2_G_Components})---behave under a Galilei transformation $G$ between a frame $\mathscr{O}$ and a frame $\mathscr{O}'$. 
Given components $\gamma_{\mu\nu}$ of $\bm{\gamma}$ in $\mathscr{O}$, the components in $\mathscr{O}'$ are
\begin{equation}
\gamma_{\mu'\nu'} = {G^\alpha}_{\mu'} {G^\beta}_{\nu'} \gamma_{\alpha\beta}.
\label{eq:Projection_0_2_Transform}
\end{equation}
Expressing Eq.~(\ref{eq:Projection_0_2_G_Components}) in block matrix form as
\begin{equation}
\begin{bmatrix} \gamma_{\mu\nu} \end{bmatrix} = 
\begin{bmatrix} 0 & 0 \\ 0 & 1 \end{bmatrix}, 
\end{equation}
and given a Galilei transformation matrix
\begin{equation}
\begin{bmatrix} {G^\mu}_{\mu'} \end{bmatrix} =
\begin{bmatrix} 1 & 0 \\ v & R \end{bmatrix}
\end{equation}
in which $v$ is a $3\times 1$ matrix of Galilei boost parameters and $R$ is a $3\times 3$ orthogonal rotation matrix ($R^T = R^{-1}$), Eq.~(\ref{eq:Projection_0_2_Transform}) can be expressed in block matrix form as
\begin{eqnarray}
\begin{bmatrix} \gamma_{\mu'\nu'} \end{bmatrix} &=& 
\begin{bmatrix} 1 & v^T \\ 0 & R^T \end{bmatrix} 
\begin{bmatrix} 0 & 0 \\ 0 & 1 \end{bmatrix} 
\begin{bmatrix} 1 & 0 \\ v & R \end{bmatrix} \nonumber \\
&=&
\begin{bmatrix} v^T v & v^T R \\ R^T v & 1 \end{bmatrix}.
\end{eqnarray}
The inverse Galilei transformation matrix is
\begin{equation}
\begin{bmatrix} {G^{\mu'}}_\mu \end{bmatrix} =
\begin{bmatrix} 1 & 0 \\ -R^{-1} v & R^{-1} \end{bmatrix},
\end{equation}
used to find ${\left( \overrightarrow{\gamma} \right)^{\mu'}}_{\nu'} = {G^{\mu'}}_{\alpha} {G^\beta}_{\nu'} {\left( \overrightarrow{\gamma} \right)^{\alpha}}_{\beta}$:
\begin{eqnarray}
\begin{bmatrix} {\left( \overrightarrow{\gamma} \right)^{\mu'}}_{\nu'} \end{bmatrix} &=& 
\begin{bmatrix} 1 & 0 \\ -R^{-1} v & R^{-1} \end{bmatrix}
\begin{bmatrix} 0 & 0 \\ 0 & 1 \end{bmatrix} 
\begin{bmatrix} 1 & 0 \\ v & R \end{bmatrix} \nonumber \\
&=&
\begin{bmatrix} 0 & 0 \\ R^{-1} v & 1 \end{bmatrix}.
\end{eqnarray}
A similar calculation for ${\left( \overleftarrow{\gamma} \right)_{\mu'}}^{\nu'} = {G^{\alpha}}_{\mu'} {G^{\nu'}}_{\beta} {\left( \overleftarrow{\gamma} \right)_{\alpha}}^{\beta}$ yields
\begin{eqnarray}
\begin{bmatrix} {\left( \overleftarrow{\gamma} \right)_{\mu'}}^{\nu'} \end{bmatrix} &=&
\begin{bmatrix} 1 & v^T \\ 0 & R^T \end{bmatrix}
\begin{bmatrix} 0 & 0 \\ 0 & 1 \end{bmatrix} 
\begin{bmatrix} 1 & - v^T R \\ 0 & R \end{bmatrix} \nonumber \\
&=& 
\begin{bmatrix} 0 & v^T R \\ 0 & 1 \end{bmatrix}.
\end{eqnarray}
While the components of $\bm{\gamma}$, $\overrightarrow{\bm{\gamma}}$, and $\overleftarrow{\bm{\gamma}}$ all change, the case of $\Overrightarrow{\bm{\gamma}}$ is special, for the components $\left(\Overrightarrow{\gamma}\right)^{\mu'\nu'} = {G^{\mu'}}_{\alpha} {G^{\nu'}}_{\beta} \left(\Overrightarrow{\gamma}\right)^{\alpha\beta}$ are
\begin{eqnarray}
\begin{bmatrix} \left(\Overrightarrow{\gamma}\right)^{\mu'\nu'} \end{bmatrix} &=& 
\begin{bmatrix} 1 & 0 \\ -R^{-1} v & R^{-1} \end{bmatrix}
\begin{bmatrix} 0 & 0 \\ 0 & 1 \end{bmatrix} 
\begin{bmatrix} 1 & - v^T R \\ 0 & R \end{bmatrix} \nonumber \\ 
&=& 
\begin{bmatrix} \left(\Overrightarrow{\gamma}\right)^{\mu\nu} \end{bmatrix},
\end{eqnarray}
remaining unchanged under Galilei transformations.

\section{Reduction to partial differential equations}
\label{sec:PartialDifferentialEquations}

The $3+1$ geometric formulation of the equations of fluid dynamics in Eqs.~(\ref{eq:Divergence_3_1_N})-(\ref{eq:Divergence_3_1_E}) is closely tied to partial differential equations in conservative form useful for numerical simulations.
See for instance Ref.~\cite{Gourgoulhon201231-Formalism-in} for an accessible introduction to the 
necessary differential geometry---in this case, Lie derivatives, and covariant derivatives on the position space slices $\mathcal{S}_t$, which are in themselves three-dimensional flat Riemann manifolds endowed with a metric $\bm{\gamma}$.
 
As a simple example, use a reference frame $\bar{\mathscr{O}}$ that differs from our fiducial frame $\mathscr{O}$, a global inertial frame, only in the use of curvilinear spatial coordinates $( x^{\bar\imath} )$ on the position space slices $\mathcal{S}_t$ instead of the rectilinear $( x^i )$. 
Because $\bm{w}$ is a constant vector (being the four-velocity of inertial observers), and thanks to Eqs.~(\ref{eq:O_Worldlines_M}) and (\ref{eq:O_Worldlines_G}) on $\mathbb{M}$ and $\mathbb{G}$ respectively, the Lie derivative in (for instance) Eq.~(\ref{eq:Divergence_3_1_N}) becomes
\begin{equation}
\mathcal{L}_{\bm w} N = w^{\bar\mu} \frac{\partial N}{\partial x^{\bar\mu}} = \frac{\partial N}{\partial t},
\end{equation}
and similarly 
\begin{equation}
\mathcal{L}_{\bm w} {S_{\bar\jmath}} = \frac{\partial S_{\bar\jmath}}{\partial t}
\end{equation}
for the components of Eq.~(\ref{eq:Divergence_3_1_S}).
Using the natural basis associated with the coordinates $( x^{\bar\imath} )$, and for the moment writing $\bm{F}_N = \bm{F}$, the position space divergence in (for instance) Eq.~(\ref{eq:Divergence_3_1_N}) is
\begin{eqnarray}
\bm{D} \cdot \bm{F} &=& D_{\bar\imath} F^{\bar\imath} 
	= \frac{\partial F^{\bar\imath}}{\partial x^{\bar\imath}}  
		+  {\Gamma^{\bar\imath}}_{\bar\jmath \bar\imath} F^{\bar\jmath} \\
&=& \frac{\partial F^{\bar\imath}}{\partial x^{\bar\imath}} 
		+ \frac{\partial \ln{\sqrt{\bar\gamma}}}{\partial x^{\bar\imath}} F^{\bar\imath}  
	=  \frac{1}{\sqrt{\bar\gamma}} \frac{\partial}{\partial x^{\bar\imath}} \left( \sqrt{\bar\gamma} F^{\bar\imath} \right).
\end{eqnarray}
Here the Christoffel symbols 
\begin{equation}
{\Gamma^{\bar\imath}}_{\bar\jmath \bar k} = \frac{1}{2} \gamma^{\bar\imath \bar\ell} \left( 
	\frac{\partial \gamma_{\bar\ell \bar\jmath}}{\partial x^{\bar k}} 
	+ \frac{\partial \gamma_{\bar\ell \bar k}}{\partial x^{\bar\jmath}}  
	- \frac{\partial \gamma_{\bar\jmath \bar k}}{\partial x^{\bar\ell}} \right)
\end{equation} 
account for the variation of the non-constant curvilinear basis vectors.
From this results the identity ${\Gamma^{\bar\imath}}_{\bar\jmath \bar\imath} = {\Gamma^{\bar\imath}}_{\bar\imath \bar\jmath} = \partial \ln{\sqrt{\bar\gamma}} / \partial x^{\bar\jmath}$ for a contracted Christoffel symbol (where $\bar\gamma = \det\left[ \gamma_{\bar\imath\bar\jmath} \right]$ is the determinant of the matrix of metric components) found in divergences.
But each free index gets a Christoffel symbol correction term; for a $(1,1)$ tensor as in Eq.~(\ref{eq:Divergence_3_1_S}), it turns out that  
\begin{equation}
D_{\bar\imath} {F_{\bar\jmath}}^{\bar\imath}  
	= \frac{1}{\sqrt{\bar\gamma}} \frac{\partial}{\partial x^{\bar\imath}} \left( \sqrt{\bar\gamma} {F_{\bar\jmath}}^{\bar\imath} \right) 
	- \frac{1}{2} F^{\hat\imath \hat k} \, \frac{\partial \gamma_{\bar\imath\bar k}}{\partial x^{\bar\jmath}}
\end{equation}
 for the symmetric case $F^{\bar\imath \bar\jmath} = F^{\bar\jmath \bar\imath}$.
 Now a metric
 \begin{equation}
 \begin{bmatrix} \gamma_{\bar\imath\bar\jmath} \end{bmatrix}_{\bar{\mathscr{O}}}
 = 
 \begin{bmatrix}
 1 & 0 & 0 \\
 0 & a (x^{\bar 1}) & 0 \\
 0 & 0 & b ( x^{\bar 1} ) \, c ( x^{\bar 2} )
 \end{bmatrix}
 \end{equation}
with $\sqrt{\bar\gamma} = \sqrt{a b c}$ is sufficiently general to cover rectangular, cylindrical, and spherical coordinates on a flat three-dimensional manifold; see Table~\ref{tab:CurvilinearMetric}.

\begin{table}
\caption{Coordinates, metric functions, and metric derivatives for common curvilinear coordinate systems.}
\centering
\begin{tabular}{cccccccccc}
\toprule
\textbf{System}	& $x^{\bar 1}$ & $x^{\bar 2}$ &	$x^{\bar 3}$ & $a$ & $b$ & $c$ & $\frac{1}{a} \frac{\partial a}{\partial x^1}$ & $\frac{1}{b} \frac{\partial b}{\partial x^1}$ & $\frac{1}{c} \frac{\partial c}{\partial x^2}$ \\
\midrule
Rectangular & $x$ & $y$ & $z$ & $1$ & $1$ & $1$ & $0$ & $0$ & $0$ \\
Cylindrical & $\varrho$ & $z$ & $\phi$ & $1$ & $\varrho$ & $1$ & $0$ & $\frac{1}{\varrho}$ & 0 \\
Spherical & $r$ & $\theta$ & $\phi$ & $r$ & $r$ & $\sin\theta$ & $\frac{1}{r}$ & $\frac{1}{r}$ & $\frac{\cos\theta}{\sin\theta}$ \\
\bottomrule
\end{tabular}
\label{tab:CurvilinearMetric}
\end{table}

Assuming a perfect fluid and vanishing source terms $\bm{Q}_{\bm{S}}$ and $\bm{Q}_E$ (no external forces), Eqs.~(\ref{eq:Divergence_3_1_N})-(\ref{eq:Divergence_3_1_E}) are
\begin{eqnarray}
\frac{\partial N}{\partial t} + \frac{1}{\sqrt{\bar\gamma}} \frac{\partial}{\partial x^{\bar\imath}} \left[ \sqrt{\bar\gamma} \left(F_N\right)^{\bar\imath} \right] &=& 0, \\
\frac{\partial S_{\bar 1}}{\partial t} + \frac{1}{\sqrt{\bar\gamma}} \frac{\partial}{\partial x^{\bar\imath}} \left[ \sqrt{\bar\gamma} {\left(F_S\right)_{\bar 1}}^{\bar\imath} \right] 
&=& \frac{1}{2} \left[ \frac{ {\left(F_S\right)_{\bar 2}}^{\bar 2} }{a} \frac{\partial a}{\partial x^{\bar 1}}
	+ \frac{ {\left(F_S\right)_{\bar 3}}^{\bar 3} }{b} \frac{\partial b}{\partial x^{\bar 1}} \right], \\
\frac{\partial S_{\bar 2}}{\partial t} + \frac{1}{\sqrt{\bar\gamma}} \frac{\partial}{\partial x^{\bar\imath}} \left[ \sqrt{\bar\gamma} {\left(F_S\right)_{\bar 2}}^{\bar\imath} \right] 
&=& \frac{1}{2} \frac{ {\left(F_S\right)_{\bar 3}}^{\bar 3} }{c} \frac{\partial c}{\partial x^{\bar 2}}, \\
\frac{\partial S_{\bar 3}}{\partial t} + \frac{1}{\sqrt{\bar\gamma}} \frac{\partial}{\partial x^{\bar\imath}} \left[ \sqrt{\bar\gamma} {\left(F_S\right)_{\bar 3}}^{\bar\imath} \right] 
&=& 0, \\
\frac{\partial E}{\partial t} + \frac{1}{\sqrt{\bar\gamma}} \frac{\partial}{\partial x^{\bar\imath}} 
\left[ \sqrt{\bar\gamma} \left(F_E\right)^{\bar\imath} \right] &=& 0.
\end{eqnarray}
From Table~\ref{tab:DensityFlux}, the perfect fluid densities are
\begin{eqnarray}
N &=& n, \\
S_{\bar\jmath} &=& m \, n\, V_{\bar\jmath} \\
E &=&   \frac{1}{2}m \,n \, \gamma_{\bar\imath \bar\jmath} V^{\bar\imath}V^{\bar\jmath} + e
\end{eqnarray}
on $\mathbb{G}$ and
\begin{eqnarray}
N &=& \Lambda_V \, n, \\
S_{\bar\jmath} &=& \Lambda_V^2 \left( m \, n + \frac{e+p}{c^2} \right) V_{\bar\jmath} \\
E &=&  \Lambda_V \left[ \left( \Lambda_V - 1 \right) mc^2 \,n + \Lambda_V \left( e + p \right) \right] - p, \\
\end{eqnarray}
on $\mathbb{M}$, and the fluxes are
\begin{eqnarray}
\left(F_N\right)^{\bar\imath} &=& N V^{\bar\imath}, \\
{\left(F_S\right)_{\bar\jmath}}^{\bar\imath} &=& S_{\bar\jmath} V^{\bar\imath} + p\, {\delta_{\bar\jmath}}^{\bar\imath}, \\
\left(F_E\right)^{\bar\imath} &=& \left( E + p \right) V^{\bar\imath}
\end{eqnarray}
on both $\mathbb{M}$ and $\mathbb{G}$.
The system is closed with an equation of state $p = p(n, e)$.


\def \prd {Phys. Rev. D}

\reftitle{References}

\end{document}